\documentclass[preprint,tightenlines,eqsecnum,floats,aps,amsmath,amssymb,nofootinbib,prd,showpacs]{revtex4}

\usepackage{amsmath,amssymb,amsfonts}
\usepackage{graphicx}
\usepackage{enumerate} 

\usepackage{hyperref} 

\def\be{\begin{equation}}
\def\ee{\end{equation}}
\def\ba{\begin{eqnarray}}
\def\ea{\end{eqnarray}}
\def\nn{\nonumber}

\def\a{\alpha}
\def\b{\beta}
\def\g{\gamma}
\def\d{\delta}
\def\e{\epsilon}

\def\k{\kappa}
\def\l{\lambda}
\def\m{\mu}
\def\n{\nu}
\def\x{\xi}
\def\p{\pi}

\def\r{\rho}
\def\vr{\varrho}
\def\s{\sigma}

\def\t{\tau}

\def\ph{\phi}

\def\ps{\psi}
\def\o{\omega}

\def\G{\Gamma}
\def\D{\Delta}
\def\Th{\Theta}
\def\L{\Lambda}

\def\Ps{\Psi}

\newcommand{\cyl}{{\rm Cyl}} 
\newcommand{\hilbert}{{\cal H}} 
\newcommand{\abs}[1]{{\left|{#1}\right|}} 
\newcommand{\Abs}[1]{{\Big|{#1}\Big|}} 
\newcommand{\inner}[2]{{\langle {#1}\vert {#2} \rangle}} 
\newcommand{\ket}[1]{\vert{#1}\rangle} 
\newcommand{\bra}[1]{\langle{#1}\vert} 

\newcommand{\triad}[2]{e^{#1}_{#2}} 
\newcommand{\cotriad}[2]{e_{#1}^{#2}} 
\newcommand{\ftriad}[2]{{}^o\! e^{#1}_{#2}} 
\newcommand{\fcotriad}[2]{{}^o\!\o_{#1}^{#2}} 
\newcommand{\fq}{{}^o\!q} 
\newcommand{\tE}{\tilde{E}} 

\newcommand{\Pl}{\ell_{\rm Pl}} 
\newcommand{\sgn}{{\rm sgn}} 
\newcommand{\muzero}{{}^o\!\m} 
\newcommand{\Lzero}{{}^o\!L} 
\newcommand{\mubar}{{\bar \m}} 

\newcommand{\secref}[1]{Section~\ref{#1}}

\newcommand{\eqnref}[1]{(\ref{#1})}
\newcommand{\figref}[1]{Figure~\ref{#1}}
\newcommand{\appref}[1]{Appendix~\ref{#1}}

\begin{document}


\title{Loop Quantum Cosmology in Bianchi Type I Models:\\
Analytical Investigation}
\author{Dah-Wei Chiou}
\email{dwchiou@socrates.berkeley.edu}
\affiliation{${}^{1}$Department of Physics,
University of California at Berkeley,\\
Berkeley, CA 94720,
U.S.A.\\
\\
${}^{2}$Institute for Gravitational Physics and Geometry,
Physics Department, The Pennsylvania State University, University Park, PA 16802,
U.S.A.}

\begin{abstract}
The comprehensive formulation for loop quantum cosmology in the
spatially flat, isotropic model was recently constructed. In this
paper, the methods are extended to the anisotropic Bianchi I
cosmology. Both the precursor and the improved strategies are
applied and the expected results are established: (i) the scalar
field again serves as an internal clock and is treated as emergent
time; (ii) the total Hamiltonian constraint is derived by imposing
the fundamental discreteness and gives the evolution as a difference
equation; and (iii) the physical Hilbert space, Dirac observables
and semi-classical states are constructed rigorously. It is also
shown that the state in the kinematical Hilbert space associated
with the classical singularity is decoupled in the difference
evolution equation, indicating that the big bounce may take place
when any of the area scales undergoes the vanishing behavior. The
investigation affirms the robustness of the framework used in the
isotropic model by enlarging its domain of validity and provides
foundations to conduct the detailed numerical analysis.
\end{abstract}

\maketitle


\tableofcontents

\section{Introduction}
The loop quantum cosmology (LQC) in the spatially flat, isotropic
model was recently investigated in detail \cite{Ashtekar:2006uz}.
(For a brief review, see \cite{Ashtekar:2006rx}.) That investigation
introduced a conceptual framework to construct the \emph{physical
sector} and gave the technical tools to systematically explore the
effects of quantum geometry both on the gravitational and matter
sectors. With a massless scalar field as the matter source, the
analysis led to the significant results: (i) the scalar field was
shown to serve as an internal clock, thereby providing an explicit
realization of the ``emergent time'' idea; (ii) the physical Hilbert
space, Dirac observables and semi-classical states were constructed
rigorously; and (iii) the Hamiltonian constraint was solved
numerically, showing that in the backward evolution of states which
are semi-classical at late times, \emph{the big bang is replaced by
a quantum bounce}. Furthermore, thanks to the non-perturbative,
background independent methods, the quantum evolution is
\emph{deterministic across the deep Planck regime}.

These results are attractive and affirm the assertion that the
quantum geometry effects of loop quantum gravity (LQG) hold a key to
many long standing questions
\cite{Bojowald:2001xe,Bojowald:2002gz,Bojowald:2003md,Ashtekar:2003hd,Bojowald:2006da}.
The detailed investigation enriches our understanding of quantum
geometry effects both in semi-classical and Planck regimes. However,
it has a serious drawback: the critical value $\r_{\rm crit}$ of the
matter density at which the bounce occurs can be made arbitrarily
small by increasing the momentum $p_\ph$ of the scalar filed (which
is a constant of motion). This is seriously problematic since large
values of $p_\ph$ are permissible and even preferred for
semi-classical states at late times and it is physically
unreasonable to expect that quantum corrections significantly modify
the classical predictions at low matter densities.

A careful consideration urges that this unpleasant feature is just
the artifact of approximation schemes and should be overcome by
modifying the definition of the constraint operator in a more
sophisticated way. In \cite{Ashtekar:2006wn}, the new Hamiltonian
constraint was constructed, based on the same principles used in
\cite{Ashtekar:2006uz,Ashtekar:2006rx} but improved by a more direct
implementation of the underlying physical ideas. The new method
retains all the attractive features while resolves the major
weakness: in the improved dynamics, \emph{the big bounce occurs
precisely when the matter density enters the Planck regime},
regardless of the value of $p_\ph$ and other initial conditions.

The investigations \cite{Ashtekar:2006uz} and \cite{Ashtekar:2006wn}
altogether provide a solid foundation for analyzing physical issues
of LQC. However, one of the limitations in this formulation is that
it is obtained in a specific mini-superspace model and thus the
domain of validity is restricted. In order to test the robustness of
this framework, it is necessary to apply the methodology to more
general situations. The next step is to consider Bianchi I model,
the simplest case of homogeneous, anisotropic models. In classical
general relativity, the generic approach to the initial singularity
is usually understood in terms of the Belinsky-Khalatnikov-Lifshitz
(BKL) scenario. In this scenario, along with the Bianchi IX model,
the Kasner solution of vacuum Bianchi I model plays a pivotal role
for the infinite fragmentation of the homogeneous patches close to
the singularity \cite{Belinsky:1982pk,Rendall:2005nf}. Therefore, a
detailed construction of Bianchi I model (with or without matter
source) in quantum theory could provide a tool to explore the
quantum effect near the classical singularity in generic cases.
Moreover, a better understanding of anisotropic Bianchi I models may
also shed light on how the isotropy appears in anisotropy. For
example, we can ask what kinds of semi-classical (coherent) states
in Bianchi I model give rise to the isotropic symmetry, either
exactly or approximately. Since so far there is no systematic
procedure which leads us to the dynamics of the symmetry reduced
theory from that of the full theory of LQG, understanding the
isotropy in the anisotropic framework could give a useful insight
into symmetry reduction.

In anisotropic models, the main structural difference from the
isotropic case is that the operator $\Th$ used in
\cite{Ashtekar:2006uz,Ashtekar:2006rx,Ashtekar:2006wn} is no longer
positive definite. However, a detailed analysis shows that what
matters is just the operator $\Th_+$, obtained by projecting the
action of $\Th$ to the positive eigenspace in its spectral
decomposition. Therefore, most analytical structures should go
through without a major modification.

The purpose of this paper is to give a detailed analytical
investigation of LQC with a massless scalar field in Bianchi I
model. Both the precursor strategy (``\,$\muzero$-scheme'') used in
\cite{Ashtekar:2006uz} and the improved strategy
(``$\mubar$-scheme'') used in \cite{Ashtekar:2006wn} are applied and
reconstructed for the Bianchi I model. The analytical issues are
investigated in detail and the parallel results are reproduced. In
particular, the scalar field again serves as an internal clock, the
physical Hilbert space with the inner product is constructed such
that Dirac observables are all self-adjoint and thereby the forms of
expectation values and uncertainties of Dirac observables are
formulated. Meanwhile, the corresponding Wheeler-DeWitt (WDW) theory
is studied and by analogy the semi-classical states for LQC are
constructed in both schemes, which match the semi-classical states
for WDW at late times.

However, since Bianchi I models have more variety, there are new
features significantly different from those in isotropic models and
more careful consideration is needed. In the classical dynamics,
apart from the solutions which have the similar characteristics to
those in isotropic models (i.e., expanding or contracting in all
three directions), the Bianchi I model with a massless scalar also
admits ``Kasner-like'' solutions (namely, expanding in two
directions while contracting in the other direction or the opposite
way around). The Kasner-like solution with two expanding and one
contracting directions eventually encounters the ``Kasner-like
singularity'' (a given regular cubical cell stretches as an
infinitely long line) in the backward evolution. When the quantum
correction is taken into account, according to the implications from
the decoupling of the problematic state in the difference evolution
equation, we anticipate that the singularity (Kasner-like or not) is
resolved and might be replaced by the \emph{big bounce}. In the
Kasner-like case, the big bounce could occur as many as \emph{three
times}, not just once, whenever the corresponding classical solution
undergoes vanishing behavior of the area scale factors. Furthermore,
in $\mubar$-scheme, the bounce is expected to happen at the moment
when the \emph{directional factor} $\vr_I:=p_\ph^{2}/\abs{p_I}^3$
(which plays the same role as the matter density does in the
isotropic case) in any of the three diagonal directions ($p_I$ is
the area scale factor in the direction $I$) approaches the Planck
regime of ${\cal O}(\hbar\,\Pl^{-4})$ ($\Pl$ is the Planck length),
irrespective of the value of $p_\ph$. In order to conclusively show
that this anticipation is indeed the case, we need to conduct a
thorough numerical investigation to reveal more phenomenological
ramifications due to quantum geometry effects in the Bianchi I
model. With the analytical foundation at hand, we are ready to
investigate the effective dynamics and numerics, which we wish to
study in more detail in a separate paper \cite{in progress}. (Some
results for the effective dynamics have been studied in \cite{to
appear}.)

This paper is organized as follows. In \secref{sec:classical
dynamics} we summarize the results of the classical dynamics of the
anisotropic Bianchi I cosmology with a massless scalar source. In
\secref{sec:Ashtekar variables} the Bianchi models are expressed in
terms of Ashtekar variables and the phase space of the diagonal
Bianchi models is constructed. In particular, the Hamiltonian
constraint in Bianchi I model is given in terms of the diagonal
variables $c^I$ and $p_I$ and serves as the staring point of the
whole formulation. \secref{sec:kinematics} deals with the
kinematical structure, highlighting the construction using
holonomies to replace connection variables. The gravitational part
of Hamiltonian constraint is derived in \secref{sec:dynamics} with
the quantum discreteness imposed by $\muzero$-scheme. The WDW theory
is studied in \secref{sec:WDW theory} to show that the singularity
is not resolved with WDW equation and to offer a systematic method
to construct the physical sector, which then can be heuristically
modified for the context of LQC. In \secref{sec:analytical issues},
the analytical issues in LQC are studied in detail: the total
Hamiltonian constraint is derived and gives the evolution as a
difference equation; the state in the kinematical Hilbert space
associated with the classical singularity is completely decoupled in
the difference evolution equation; and finally the Dirac observables
are recognized and the physical Hilbert space is constructed with
the physical inner product. \secref{sec:kinematics} --
\secref{sec:analytical issues} complete the analytical investigation
for the precursor strategy ($\muzero$-scheme). The organization of
this part deliberately parallels that of \cite{Ashtekar:2006uz} and
the notations are also used in the similar manner. The procedures in
\secref{sec:dynamics} -- \secref{sec:analytical issues} are then
repeated all over again with appropriate modifications for the
improved strategy ($\mubar$-scheme) in \secref{sec:improved
dynamics}. Finally, the results of this paper and outlooks for
further research are discussed in \secref{sec:discussion}.

Some technical details and related topics are supplemented in the
appendices. The detailed calculation for the WDW semi-classical
states is shown in \appref{sec:coherent states}. The issues to
obtain the formulation for the isotropic model by direct inspection
from the anisotropic results are discussed in \appref{sec:isotropic
model}. The comparison with the results of
\cite{Ashtekar:2006uz,Ashtekar:2006rx,Ashtekar:2006wn} and their
subtle differences are also remarked.

\section{Classical Dynamics of the Bianchi I Model}\label{sec:classical dynamics}
The Bianchi I model without matter source (Kasner model) has been
quantized in the context of LQC in \cite{Bojowald:2003md},
as has the isotropic model with massless scalar source in
\cite{Ashtekar:2006uz,Ashtekar:2006rx,Ashtekar:2006wn}.
In order to study the LQC in Bianchi I model
in the presence of massless scalar field, we first have to figure
out its classical dynamics and solutions. This has been done in
\cite{Salisbury:2005sa}, and the results are briefly outlined as
follows.

The Hilbert action for the cosmology with a massless scalar filed $\ph$ minimally coupled
to the gravity is given by
\ba\label{eqn:Hilbert action}
S&=&S_{\rm grav}+S_{\ph}\nn\\
&=&\frac{1}{8\p G}\int d^4x\sqrt{-g}\,R-\frac{1}{2}\int d^4x\sqrt{-g}\,g^{\m\n}\ph_{,\m}\ph_{,\n}.
\ea
In the Bianchi I model, we can take the spacetime metric to be the form
\be\label{eqn:metric}
g_{\m\n}=
\left(
\begin{array}{cccc}
-n^2(t)&0&0&0\\
0&a_1^2(t)&0&0\\
0&0&a_2^2(t)&0\\
0&0&0&a_3^2(t)
\end{array}
\right)
\ee
and the homogeneous field $\ph({\vec x},t)=\ph(t)$ to be independent of the spatial coordinates.
We then have the Hamiltonian constraint
\ba\label{eqn:H}
H&=&C_{\rm grav}+C_{\ph}\nn\\
&=&
\p G\left(-\frac{2\p_1\p_2}{a_3}-\frac{2\p_1\p_3}{a_2}-\frac{2\p_2\p_3}{a_1}
+\frac{a_1\p_1^2}{a_2a_3}+\frac{a_2\p_2^2}{a_1a_3}+\frac{a_3\p_3^2}{a_1a_2}\right)
+\frac{p_\ph^2}{2a_1a_2a_3}=0,
\ea
where $\p_i$ are the conjugate momenta of $a_i$ and $p_\ph$ is the conjugate momentum of $\ph$,
which satisfy the canonical relations
\be\label{eqn:Poisson}
\{a_i,\p_j\}=\d_{ij} \quad\text{and}\quad \{\ph,p_\ph\}=1.
\ee

The general solution is given by
\be
a_i(t)=a_{i,o}\left(\frac{\t}{\t_o}\right)^{\k_i}=
a_{i,o}\left(\frac{n_ot+\int_0^t dt_2
\int_0^{t_2}dt_1\l(t_1)+\t_o}{\t_o}\right)^{\k_i},
\ee
and
\be
\ph(t)=\ph_o+\frac{\k_\phi}{\sqrt{4\p
G}}\ln\left(\frac{n_ot+\int_0^t dt_2
\int_0^{t_2}dt_1\l(t_1)+\t_o}{\t_o}\right),
\ee
where $\t$ is the
proper time defined through $d\t=n(t)dt$, $\l:=\dot{n}\equiv \frac{d n}{dt}$ is a
positive but otherwise arbitrary function of $t$ and the constants
$\k_i$, $\k_\ph$ satisfy the constraint:
\be\label{eqn:parameter constraint}
\k_1+\k_2+\k_3=1 \quad\text{and}\quad
\k_1^2+\k_2^2+\k_3^2+\k_\ph^2=1.
\ee
Since the Hamiltonian is
independent of $\ph$, the momentum $p_{\ph}$ is a constant of
motion, which is given by
\be\label{eqn:momentum-phi}
p_{\ph}=\frac{a_{1,o}a_{2,o}a_{3,o}}{\t_o}\sqrt{\frac{1}{8\p G}}\
\k_{\ph} :=\hbar\sqrt{8\p G}\,{\cal K}\k_\ph.
\ee
The complete set
of solutions will be over-counted unless we fix the constant ratio
$a_{1,0}a_{2,0}a_{3,0}/\t_o$; thereby we introduce the dimensionless
constant ${\cal K}$, which is to be fixed.\footnote{The choice of
${\cal K}$ is in fact a gauge fixing. Change of $\cal{K}$ is
equivalent to the rescaling: $\k_i\rightarrow{\cal K}\k_i$ and
$\k_\ph\rightarrow{\cal K}\k_\ph$ with the new constraint:
$\k_1+\k_2+\k_3={\cal K}$ and $\k_1^2+\k_2^2+\k_3^2+\k_\ph^2={\cal
K}^2$. The rescaled $\k_i$ and $\k_\ph$ still yield the same
evolution equation with respect to the internal time $\ph$ as the
rescaling does not change \eqnref{eqn:classical sol a-phi}.}
Note that $\ph$ is a monotonic function of $t$ and thus it can be used as the internal time.
Expressed in terms of the internal time $\ph$, the solution reads as
\be\label{eqn:classical sol a-phi}
a_i(\ph)=a_{i,o}\ e^{\sqrt{8\p G}\frac{\k_i}{\k_\ph}(\ph-\ph_o)},
\ee
where the dependence on the coordinate time $t$ is completely
eliminated and only the correlation between the dynamical variables
$a_i$ and $\ph$ is truly physical! This realizes the idea of
\emph{emergent time} \cite{Isham:1992ms} and shows that one can
employ some subset or combination of dynamical fields as an internal
clock \cite{Rovelli:2001bz}. [Later in \secref{subsec:WDW general
solutions}, \secref{subsec:LQC general solutions} and
\secref{subsec:improved general solutions}, it will be shown that
$\phi$ is well-suited to be emergent time also in the quantum theory
(for both WDW theory and LQC).]

\ \newline
\small \textbf{Remark:} If without any matter sources, the classical
dynamics of vacuum Bianchi I model yields the standard \emph{Kasner
solution}:
\be\label{eqn:Kasner sol}
ds^2=-dt^2+t^{2\k_1}dx_1^2+t^{2\k_2}dx_2^2+t^{2\k_3}dx_3^2,
\ee
where $\k_1,\k_2,\k_3$ are constants subject to the constraint:
\be\label{eqn:Kasner constraint}
\k_1+\k_2+\k_3=1
\quad\text{and}\quad \k_1^2+\k_2^2+\k_3^2=1.
\ee
With the exception
of the trivial Minkowskian solutions (where two of $\k_i$ vanish), all solutions
have two of $\k_i$ positive and the other negative, which therefore
lead to the ``Kasner singularity''\footnote{At the ``Kasner(-like) singularity'',
a given regular cubical cell is stretched as a linear line,
infinitely long in one direction and infinitely thin in the other two.}
at $t=0$ and the ``planar
collapse''\footnote{``Planar collapse'' means that a given
regular cubical cell is stretched as a planar piece,
infinitely large in two directions and infinitely thin in the other one.
This is however not a spacetime singularity, since the spacetime curvature
goes to zero (instead of infinity) when Kanser-like solutions evolve
to the far future.}at $t=\infty$.
Now with the massless scalar source introduced, the solution
\eqnref{eqn:classical sol a-phi} with the constraint
\eqnref{eqn:parameter constraint} are the modifications of
\eqnref{eqn:Kasner sol} and \eqnref{eqn:Kasner constraint} and it
also admits ``Kasner-like'' solutions (which thereby give the
``Kasner-like'' singularity and planar collapse) except that the
role of $t$ is replaced by the internal time $\ph$. On the other
hand, because of $\k_\ph$, we also have nontrivial solutions with
$\k_i$ all positive, which are not allowed for standard Kasner
solutions. These ``Kasner-unlike'' solutions have only a
(Kasner-unlike) singularity\footnote{At the ``Kasner-unlike
singularity'', a given regular cubical cell contracts to a point.}
but no planar collapse and behave quite the same as the isotropic
solution. In particular, the constraint with $\k_1=\k_2=\k_3=1/3$
reproduces the isotropic solution. \hfill $\Box$\normalsize

\section{Bianchi Models in Ashtekar Variables}\label{sec:Ashtekar variables}
In the construction of LQC, so far, quantum gravity has mostly been
analyzed in mini- or midi-superspace models which are obtained by
ignoring an infinite number of degrees of freedom of the full
theory. That is, the canonical dynamical variables are enormously
reduced by the symmetry and only a finite number of relevant degrees
of freedom are preserved and quantized. This strategy has been used
in \cite{Bojowald:2002gz} and formulated rigorously in
\cite{Ashtekar:2003hd} for the isotropic models; it was also
extended to diagonal Bianchi class A models (the subclass of Bianchi
types I, II, VIII and IX, or referred as ``type D'') in
\cite{Bojowald:2003md}. In this section, we follow and outline the
ideas of \cite{Ashtekar variables} to describe the Bianchi models in
Ashtekar variables and end up with the Hamiltonian
constraint for the Bianchi I model in terms of the diagonal variables $c^I$
and $p_I$.

\subsection{The Description with Ashtekar Variables}
Bianchi cosmologies are homogeneous cosmological models, in which there
is a foliation of spacetime: $M=\Sigma\times\!R$ such that $\Sigma$ is
space-like and there is a transitive isometry group freely acting on
$\Sigma$. This means that we have the Killing fields on $\Sigma$. We
can choose a fiducial triad of vectors $\ftriad{a}{i}$ and a
fiducial co-triad of covectors $\fcotriad{a}{i}$ that are left
invariant by the action of the Killing fields.
The fiducial 3-metric is given by the co-triad $\fcotriad{a}{i}$:
\be
\fq_{ab}=\fcotriad{a}{i}\,\fcotriad{b}{j}\,\d_{ij}.
\ee
These bases $\ftriad{a}{i}$ and $\fcotriad{a}{i}$ satisfy
the relations:
\be
[\,\ftriad{}{i},\ftriad{}{j}]^a={C_{ij}}^k\,\ftriad{a}{k},
\ee
\be
2D_{[a}\fcotriad{b]}{i}=-{C_{jk}}^i\,\fcotriad{a}{j}\,\fcotriad{b}{k},
\ee
where ${C_{ij}}^k$ are the structure constants of the above
mentioned isometry group. The classification of these structure constants leads
to a classification of the homogeneous spaces. The case that we study
in this paper is Bianchi I type, in which the structure constants ${C_{ij}}^k$ all vanish.

The Ashtekar variables consist of the canonical pairs
${\mbox{$\tE$}^a}_i$ and ${A_a}^i$. The variables ${\mbox{$\tE$}^a}_i$ are
densitized triads and the connections ${A_a}^i$ are defined by
\be
{A_a}^i=\G_a^i-\g
{K_a}^i,
\ee
where $\G^i_a=\e^i{{}_j}^k\G^j_{ak}$ with
$\G^j_{ak}=\frac{1}{2}\triad{b}{k}(\partial_a\cotriad{b}{j}-\partial_b\cotriad{a}{j}+
e^{cj}e_{al}\partial_b\cotriad{c}{l})$ the spin connection of the
physical triad $\triad{a}{i}$; ${K_a}^i=K_{ab}E^{bi}=q^{-1/2}K_{ab}\tE^{bi}$ with $K_{ab}$ the
extrinsic curvature of the 3-surface and $q$ the determinant of the
3-metric; and $\g$ is the Barbero-Immirzi parameter.
In the context of Bianchi models, it is more convenient to
consider the variables:
\ba
{E^i}_j&:=&\fq^{-1/2}\,\fcotriad{a}{i}\,{\mbox{$\tE$}^a}_j,\nn\\
{A_i}^j&:=&{A_a}^j\,\ftriad{a}{i}.
\ea
They are essentially the same as the above canonical variables,
but stripped of all the complicated but fixed spatial dependence enforced by
the Bianchi symmetry.
Conversely, we have
\ba
{\mbox{$\tE$}^a}_i&=&\sqrt{\fq}\,{E^j}_i\,\ftriad{a}{j}\,,\nn\\
{A_a}^i&=&{A_j}^i\,\fcotriad{a}{j}.
\ea
These matrices ${E^i}_j$ and ${A_i}^j$ depend only on time (due to spatial homogeneity)
and are our \emph{reduced canonical variables}.
Thus as expected, the number of degrees of freedom (in the configuration space)
has been reduced from 9 per space point to just 9 globally.
The Gauss, vector and Hamiltonian constraints in terms of these reduced variables are expressed
respectively as
\ba
\mathcal{G}^i&=&{C_{jk}}^jE^{ki}+G\,\e^{ijk}A_{mj}{E^m}_k=0,\label{eqn:Gauss constraint}\\
\mathcal{V}_k&=&-{E^j}_i{A_m}^i{C_{jk}}^m+G\,\e^{imn}{E_i}^jA_{jm}A_{kn}=0\label{eqn:vector constraint},\\
\mathcal{S}&=&\e^{ijk}{C_{mn}}^p{E^m}_i{E^n}_jA_{pk}+G{E^m}_i{E^n}_j({A_m}^i{A_n}^j-{A_n}^i{A_m}^j)=0
\label{eqn:Hamiltonian constraint},
\ea
where $\g=i$ is used in the last line for the Hamiltonian constraint.

\subsection{Phase Space in the Diagonal Bianchi Models}\label{subsec:phase space}
For the diagonal Bianchi class A models, one can fix the gauge of the $SO(3)$ rotation so
that ${A_i}^j$ and ${E^i}_j$ are diagonalized:
\be
\tilde{c}^I={\L^i}_{(I)}{A_i}^j{\L_j}^{(I)},
\qquad
\tilde{p}_I={\L_i}^{(I)}{E^i}_j\,{\L^j}_{(I)},
\ee
where ${\L^i}_I$ is the $SO(3)$ rotation matrix and ${\L_i}^I$ is its inverse.\footnote{In this paper,
the Einstein convention is adopted: the indices
repeated on both upper and lower scripts are summed.
When Einstein convention is not wanted,
those indices which are repeated on both scripts but not to be summed are parenthesized.}
In this gauge,
\eqnref{eqn:Gauss constraint} and \eqnref{eqn:vector constraint} are
identically satisfied and the phase space is reduced to
6-dimensional and labeled by $\tilde{c}^I$ (the diagonalized entries of
${A_{i}}^{j}$) and $\tilde{p}_I$ (the diagonalized entries of
${E^{i}}_{j}$) with the single constraint $\mathcal{S}=0$. In the
diagonal components the symplectic structure is given by
\be
\{\tilde{c}^I,\tilde{p}_J\}=8\p\g\,G V_o\,\d^I_J,
\ee
where
$V_o=\Lzero_1\Lzero_2\Lzero_3$ is the volume of the ``unit cell''
$\cal{V}$ adapted to the fiducial triad and $\Lzero_I$ are the
lengths of the edges of this cell with respect to $\fq_{ab}$. (For
simplicity, we assume that this cell is rectangular with respect to
$\fq_{ab}$.)

In the passage from the full to the reduced theory, we introduced a fiducial metric $\fq_{ab}$.
There is a freedom in rescaling the diagonal entries of this metric by constants:
$\fq_{ii}\mapsto k_i^2\,\fq_{ii}$ for $i=1,2,3$. Under this rescaling, the variables $\tilde{c}^I$,
$\tilde{p}_I$ transform correspondingly via $\tilde{c}^I\mapsto k_{(I)}^{-1}\,\tilde{c}^{(I)}$
and $\tilde{p}_1\mapsto k_2^{-1}k_3^{-1}\tilde{p}_1$ and so on.
Since rescaling of the fiducial metric does not change physics, $\tilde{c}^I$ and $\tilde{p}_I$
do not have direct physical meaning and it is convenient to introduce the new variables
\be
c^I=\Lzero_{(I)}\,\tilde{c}^{(I)}
\quad\text{and}\quad
p_1=\Lzero_2\Lzero_3\tilde{p}_1,
\quad
p_2=\Lzero_1\Lzero_3\tilde{p}_2,
\quad
p_3=\Lzero_1\Lzero_2\tilde{p}_3,
\ee
which is independent of the choice of the fiducial metric $\fq_{ab}$.
This is equivalent to the relation
\be
\sqrt{\fq}\,\tilde{c}^{(I)}\tilde{p}_{(I)}=\sqrt{q}\,c^{(I)}p_{(I)}.
\ee
In terms of the new variables, the canonical relation is
\be\label{eqn:Poisson2}
\{{c}^I,{p}_J\}=8\p\g\,G \,\d^I_J.
\ee

The relation between the densitized triad and the 3-metric is given by
$qq^{ab}=\d^{ij}{\mbox{$\tE$}^a}_i{\mbox{$\tE$}^b}_j$, which leads to
\be\label{eqn:p and a}
p_1=(a_2a_3)\,{\sgn}(a_1a_2a_3),
\quad
p_2=(a_1a_3)\,{\sgn}(a_1a_2a_3),
\quad
p_3=(a_1a_2)\,{\sgn}(a_1a_2a_3),
\ee
where $a_i$ are the scale factors in the metric of \eqnref{eqn:metric}.
Comparing the canonical relations \eqnref{eqn:Poisson} and \eqnref{eqn:Poisson2} with
the help of \eqnref{eqn:p and a}, we have
\ba\label{eqn:c pi a}
c^1&=&4\p\g\,G \left(\frac{a_1}{a_2a_3}\p_1-\frac{1}{a_3}\p_2-\frac{1}{a_2}\p_3\right){\sgn}(a_1a_2a_3),\nn\\
c^2&=&4\p\g\,G \left(-\frac{1}{a_3}\p_1+\frac{a_2}{a_1a_3}\p_2-\frac{1}{a_1}\p_3\right){\sgn}(a_1a_2a_3),\nn\\
c^3&=&4\p\g\,G \left(-\frac{1}{a_2}\p_1-\frac{1}{a_1}\p_2+\frac{a_3}{a_1a_2}\p_3\right){\sgn}(a_1a_2a_3).
\ea

\subsection{The Hamiltonian Constraint in Bianchi I Models}\label{subsec:Hamiltonian constraint}
The gravitational part of the Hamiltonian constraint is given by
\be
C_{\rm grav}=q^{-1/2}G_{ab}n^an^b=\frac{1}{2\sqrt{q}}\left({}^{(3)}\!R-K_{ab}K^{ab}+({K^a}_a)^2\right).
\ee
For Bianchi I models, the spatial slice is flat and thus the intrinsic curvature ${}^{(3)}R=0$
and the spin connection of the triad $\G^i_a=0$, which follows ${A_a}^i=-\g {K_a}^i$.
Consequently, we have
\be
{K_a}^b={K_a}^i{E^b}_i=-\frac{1}{\g\sqrt{q}}{A_a}^i{\mbox{$\tE$}^b}_i
=-\frac{1}{\g}\sqrt{\frac{\fq}{q}}{A_j}^i{E^k}_i\,\fcotriad{a}{j}\,\ftriad{b}{k}
\ee
and
\be
C_{\rm grav}=\frac{1}{\g^2}\frac{\fq}{q^{3/2}}
\left({E^m}_i{E^n}_j{A_m}^i{A_n}^j-{E^m}_i{E}_{mj}{A_n}^i{A}^{nj}\right).
\ee
In the gauge that ${A_i}^j$ and ${E^i}_j$ are diagonal, the Hamiltonian reads as
\ba\label{eqn:Cgrav}
C_{\rm grav}&=&\frac{2}{\g^2}\frac{\fq}{q^{3/2}}
\left(\tilde{c}^1\tilde{p}_1\tilde{c}^2\tilde{p}_2
+\tilde{c}^1\tilde{p}_1\tilde{c}^3\tilde{p}_3
+\tilde{c}^2\tilde{p}_2\tilde{c}^3\tilde{p}_3\right)\nn\\
&=&\frac{2}{\g^2\sqrt{q}}
\left(c^1p_1c^2p_2+c^1p_1c^3p_3+c^2p_2c^3p_3\right).
\ea
[Note that $C_{\rm grav}$ in \eqnref{eqn:Cgrav} is proportional to $\mathcal{S}$
in \eqnref{eqn:Hamiltonian constraint} up to the factor $q^{-3/2}$ with ${C_{ij}}^k=0$.]

By the identities \eqnref{eqn:p and a}, \eqnref{eqn:c pi a} and $\sqrt{q}=\abs{a_1a_2a_3}$,
\eqnref{eqn:Cgrav} becomes
\be\label{eqn:Cgrav2}
C_{\rm grav}=16\p^2G^2
\left(\frac{2\p_1\p_2}{a_3}+\frac{2\p_1\p_3}{a_2}+\frac{2\p_2\p_3}{a_1}
-\frac{a_1\p_1^2}{a_2a_3}-\frac{a_2\p_2^2}{a_1a_3}-\frac{a_3\p_3^2}{a_1a_2}\right)
{\sgn}(a_1a_2a_3),
\ee
which is exactly the same as the gravitational part in \eqnref{eqn:H} up to an overall constant factor.
Fixing the difference of the overall factors for $C_{\rm grav}$
used in \eqnref{eqn:H} and in \eqnref{eqn:Cgrav2}
and referring to \eqnref{eqn:Cgrav},
we end up with the Hamiltonian constraint in terms of the diagonal variables $c^I$ and $p_I$:
\ba\label{eqn:Hamiltonian}
H&=&C_{\rm grav}+C_{\ph}\nn\\
&=&-\frac{2}{\g^2}
\left(c^1p_1c^2p_2+c^1p_1c^3p_3+c^2p_2c^3p_3\right)
\frac{1}{\sqrt{\abs{p_1p_2p_3}}}
+8\p G\,\frac{1}{\sqrt{\abs{p_1p_2p_3}}}p_{\ph}^2.
\ea
In the new variables, the classical solution \eqnref{eqn:classical sol a-phi}
now reads as
\ba\label{eqn:classical sol p-phi}
p_1(\ph)&=&p_{1,o}\ e^{\sqrt{8\p G}\big(\frac{\k_2+\k_3}{\k_\ph}\big)(\ph-\ph_o)}
=p_{1,o}\ e^{\sqrt{8\p G}\big(\frac{1-\k_1}{\k_\ph}\big)(\ph-\ph_o)},\nn\\
p_2(\ph)&=&p_{1,o}\ e^{\sqrt{8\p G}\big(\frac{\k_1+\k_3}{\k_\ph}\big)(\ph-\ph_o)}
=p_{2,o}\ e^{\sqrt{8\p G}\big(\frac{1-\k_2}{\k_\ph}\big)(\ph-\ph_o)},\nn\\
p_3(\ph)&=&p_{1,o}\ e^{\sqrt{8\p G}\big(\frac{\k_1+\k_2}{\k_\ph}\big)(\ph-\ph_o)}
=p_{3,o}\ e^{\sqrt{8\p G}\big(\frac{1-\k_3}{\k_\ph}\big)(\ph-\ph_o)}.
\ea

\ \newline \small \textbf{Remark}: The Hamiltonian constraint
\eqnref{eqn:Hamiltonian} reduces to that of the isotropic model (as
given in \cite{Ashtekar:2003hd}) if we further impose the isotropy
condition $c^1=c^2=c^3=c$ and $p_1=p_2=p_3=p$ with the canonical
structure $\{c,p\}=8\p\g\,G/3$. \hfill $\Box$\normalsize

\section{Kinematics: Quantization}\label{sec:kinematics}
In this section, we follow the lines in \cite{Ashtekar:2003hd} and
generalize the procedures to seek the quantum kinematics for
Bianchi I models. Elementary variables and the representation of
their algebra are obtained in \secref{subsec:elementary
variables} and \secref{subsec:rep of algebra}. In
\secref{subsec:triad operators}, the triad operators are constructed
in terms of holonomies; the fact that the eigenvalues of the triad
operators are bounded above is discussed, suggesting that the
classical singularity could be resolved by the quantum geometry.

\subsection{Elementary Variables}\label{subsec:elementary variables}
Before quantization, we should single out the ``elementary
functions'' on the classical phase space which are to be lifted to
their quantum counterparts. In the full theory, the configuration
variables are constructed from holonomies $h_e(A)$ associated with
edges $e$ and momentum variables $E(S,f)$, which are momenta $E$
smeared with test fields $f$ on 2-surfaces. In Bianchi I models,
however, because of homogeneity, we do not need all edges $e$ and
surfaces $S$. Symmetric connection $A$ in $\mathcal{A}_S$ can be
recovered by knowing holonomies $h_I$ along the edges $e_I$, which
lie along straight lines in the ``diagonal'' directions (for
$I=1,2,3$) in $\Sigma$. Similarly, it is sufficient to smear triads
only on the rectangular surfaces of the elementary cell
$\mathcal{V}$ (with respect to $\fq_{ab}$).

The $SU(2)$ holonomy along $e_I$ in the diagonal direction $I$ is given by:
\ba\label{eqn:hI}
h_I(A,\m_I)&:=&{\cal P}\exp\int_{e_I}A
=\exp(-\frac{i}{2}\m_{(I)}c^{(I)}\s_{(I)})\nn\\
&=&
\cos\Big(\frac{\m_{(I)}c^{(I)}}{2}\Big)
+2\sin\Big(\frac{\m_{(I)}c^{(I)}}{2}\Big)\t_{(I)},
\ea
where $2\,i\t_I=\s_I$ with $\s_I$ the Pauli matrices and $\m_I\in(-\infty,\infty)$
(and $\m_{I}\Lzero_{I}$ is the oriented length of the edge with respect to $\fq_{ab}$).
According to Peter-Weyl theorem, a basis on the Hilbert space of $L_2$ functions on $SU(2)$
is given by the matrix elements of the irreducible representations of the group.
Therefore, the \emph{cylindrical functions}
of the symmetric connections $A$ are obtained by the sums of
the matrix elements of the irreducible representations of the holonomies $h_I$, that is
\be\label{eqn:fA}
f(A)=
\sum_{\vec \m}
\sum_{{\vec j},{\vec \a},{\vec \b}}
{{\x^{({\vec j}){\vec \b}}_{\vec \m}}_{\vec \a}}\,
{R^{(j_1)\a_1}}_{\b_1}\!\left(h_1(A,\m_1)\right)
{R^{(j_2)\a_2}}_{\b_2}\!\left(h_2(A,\m_2)\right)
{R^{(j_3)\a_3}}_{\b_3}\!\left(h_3(A,\m_3)\right).
\ee
As mentioned in the beginning of \secref{subsec:phase space},
in terms of the diagonal variables,
the vector constraint \eqnref{eqn:vector constraint} is identically satisfied and thus
the symmetric theory is invariant under the $SU(2)$ gauge transformation.
By \eqnref{eqn:hI}, on the other hand, the holonomy $h_I(A,\m_I$) is given by
$\exp(-\frac{i}{2}\m_{(I)}c^{(I)}\s_{(I)})$,
which is not invariant if we perform a global $SU(2)$ rotation by
$\s_I\mapsto\s_I'={U_I}^J\s_J$.
Therefore, in order to have the cylindrical functions $SU(2)$-invariant,
the only possibility is that all the three
irreducible representations in \eqnref{eqn:fA} must be trivial
(i.e. $j_1=j_2=j_3=0$) and in the trivial
representation, $R^{(j=0)}\!\left(h_I(A,\m_I)\right)$
is simply $\exp(-\frac{i}{2}\m_{(I)}c^{(I)})$.

Consequently, the algebra generated by sums of products of the matrix elements of these holonomies
is just the algebra of \emph{almost periodic functions} of $c^I$. A typical element of the
almost periodic function is written as:
\be
g(c^1,c^2,c^3)=\sum_{j}\x_{j}\,e^{\frac{i}{2}{\m_{jI}c^I}},
\ee
where $j$ runs over a finite number of integers, $\m_{jI}\in \mathbb{R}$
and $\x_{j}\in \mathbb{C}$.
As in the terminology used in the full theory, we can regard a finite number of edges labeled
by $(j,I)$ as a ``graph'' and the function $g(c_I)$ as a cylindrical function with respect to
that graph. Analogous to the space $\cyl$ of cylindrical functions on ${\cal A}$ in the full theory,
the vector space of the almost periodic functions is called the cylindrical functions of \emph{symmetric}
connections and denoted by $\cyl_S$.

Similarly, the momentum functions $E(S,f)$ in the full theory is replaced by $E(S_I,f_I)$ in the homogeneous case,
which are obtained by smearing ${\mbox{$\tE$}^a}_i$ with a \emph{constant}-valued function $f_I$
on the rectangular face $S_I$ of the elementary cell ${\cal V}$ normal to the diagonal direction $I$
(with respect to $\fq_{ab}$). That is
\be\label{eqn:area operator}
E(S_I,f_I)=\int_{S_I}\Sigma^{(I)}_{ab}f_{(I)}dx^adx^b=p_{I}\frac{\Lzero_{I}}{V_o}A_{S_I,f_I},
\ee
where $\Sigma^I_{ab}=\e_{abc}{\mbox{$\tE$}^c}_i\,{\L^i}_I
=\e_{abc}\sqrt{\fq}\,p_{I}\ftriad{c}{I}$
and $A_{S_I,f_I}$ is the area of $S_I$ measured by $\fq_{ab}$ times the orientation factor
(which depends on $f_I$).
In terms of classical geometry, $p_I$ is related to the ``physical'' length of the edges and the volume
of the elementary cell ${\cal V}$ via
\be\label{eqn:L and V}
L_I=\sqrt{\abs{p_I}}
\quad\text{and}\quad
V=\sqrt{\abs{p_1p_2p_3}}.
\ee

According to \eqnref{eqn:Poisson2}, the only non-vanishing Poisson brackets among the elementary functions are:
\be
\{g(c^1,c^2,c^3),p_I\}={4\p\g\,G}
\sum_j(i\,\m_{jI}\,\x_{j})\,e^{\frac{i}{2}{\m_{jK}\,c^{K}}}.
\ee
The right hand side is again in $\cyl_S$ and thus the space of the elementary variables is
closed under the Poisson bracket. Note that, contrary to the full theory, the smeared momenta
$E(S_I,f_I)$ now commute with one another since they are all proportional to $p_I$ due to homogeneity.
Therefore, the \emph{triad representation} in Bianchi I models
does exist and this fact will be exploited in \secref{subsec:triad operators}.

\subsection{Representation of the Algebra of Elementary Variables}\label{subsec:rep of algebra}
In the previous subsection, we have seen that the algebra of the
elementary variables in Bianchi I models is in a perfect parallel to
the case of the isotropic model studied in \cite{Ashtekar:2003hd}.
The elementary variables in different diagonal directions are
independent of one another and each direction behaves as a copy of
the isotropic case. Consequently, the result in
\cite{Ashtekar:2003hd} can be generalized to the Bianchi I model in
the obvious way, when we construct quantum kinematics and seek a
representation of the algebra of the elementary variables.

Therefore, the gravitational part of the Hilbert space for Bianchi I models is
$\hilbert^S_{\rm grav}=L^2(\mathbb{R}^3_{\rm Bohr},d^3\!\m_{\rm Bohr})$,
the concrete construction of which
is the Cauchy completion of the space $\cyl_S$ of almost periodic functions of $c^I$ with respect
to the inner product:
\be\label{eqn:inner product}
\inner{e^{\frac{i}{2}{\vec \m_1}\cdot{\vec c}}}{e^{\frac{i}{2}{\vec \m_2}\cdot{\vec c}}}=\d_{{\vec \m_1}{\vec \m_2}}.
\ee
(Here, the right hand side is the Kronecker delta, not the Dirac distribution.)
Thus, the almost periodic functions ${\cal N}_{\vec \m}({\vec c})
\equiv{\cal N}_{\m_1,\m_2,\m_3}(c^1,c^2,c^3):=e^{i{\vec \m}\cdot{\vec c}/2}$
constitute an orthonormal basis in $\hilbert^S_{\rm grav}$.
For $g_1$ and $g_2$ in $\cyl_S$, we have
\be
({\hat g}_1g_2)({\vec c})=g_1({\vec c})g_2({\vec c})
\ee
and by \eqnref{eqn:Poisson2} the momentum operator is represented via
\be
{\hat p}_I=-8\p i{\g\Pl^2}\frac{\partial}{\partial c^I}
\quad\text{and thus}\quad
({\hat p}_Ig)({\vec c})={4\p\g\Pl^2}\sum_j(\x_j\m_{jI}){\cal N}_{{\vec \m}_j},
\ee
where the Planck length square is defined as $\Pl^2=G\hbar$.
The basis vectors ${\cal N}_{\vec \m}$ are normalized eigenstates of ${\hat p}_I$. Writing
${\cal N}_{\vec \m}({\vec c})=\inner{{\vec c}}{{\vec \m}}$, we have
\be
{\hat p}_I\ket{{\vec \m}}={4\p\g\Pl^2}\m_I\ket{{\vec \m}}\equiv p_{\m_I}\ket{{\vec \m}}.
\ee
Meanwhile, \eqnref{eqn:L and V} also gives us the operators for the physical lengths of the edges and the volume of
the cell ${\cal V}$:
\ba
{\hat L}_I\ket{{\vec \m}}&&=
\left({4\p\g\abs{\m_I}}\right)^\frac{1}{2}\Pl\ket{{\vec \m}}
\equiv
L_{\m_I}\ket{{\vec \m}},\label{eqn:length operator}\\
{\hat V}\ket{{\vec \m}}&&=
\left({4\p\g}\right)^\frac{3}{2}\sqrt{\abs{\m_1\m_2\m_3}}\ \Pl^3\ket{{\vec \m}}
\equiv
V_{{\vec \m}}\ket{{\vec \m}}.\label{eqn:volume operator}
\ea

\subsection{Inverse Triad Operators}\label{subsec:triad operators}
In the case that the Hamiltonian has a matter sector minimally
coupled to the gravity, when the quantum geometry is taken into
account, the inverse triad operators play a key role
\cite{Bojowald:2002gz,Bojowald:2001vw}. For Bianchi I models, the
inverse triad operators can be obtained by directly generalizing the
result of the isotropic case in \cite{Ashtekar:2003hd}. The triad
coefficients ${\sgn}(p_I)\abs{p_I}^{-1/2}$ can be expressed as the
Poisson brackets $\{c^{(I)},L_{(I)}\}$, which are then to be
replaced by the commutator (divided by $i\hbar$) in quantum theory. However,
as the operator corresponding to the connection does not exist in
the full LQG \cite{Ashtekar:2004eh,Rovelli:2004tv,Thiemann:2001yy},
there is no operator $\hat{c}^I$ on $\hilbert^S_{\rm grav}$
corresponding to $c^I$ and we have to further re-express the Possion
bracket in terms of \emph{holonomies} \`{a} la Thiemann's trick.

On the reduced phase space, the triad coefficients can be written as:
\be\label{eqn:triad coefficient}
\frac{{\sgn}(p_I)}{\sqrt{\abs{p_I}}}=\frac{1}{2\p\g\,G\,\muzero_I}{\rm tr}
\left(\t_{I}h_{I}^{(\muzero_I)}\{h_{I}^{(\muzero_I)-1},L_{I}\}\right),
\ee
where
\ba\label{eqn:hI muzero}
h_I^{(\muzero_I)}&:=&{\cal P}\exp\left(\int_0^{\muzero_I\Lzero_I}\!\!{A_a}^{i}\,\t_i \,dx^a\right)
=\exp\left(\muzero_{(I)}c^{(I)}\t_{(I)}\right)\nn\\
&=&\cos\Big(\frac{\muzero_{(I)}c^{(I)}}{2}\Big)+2\sin\Big(\frac{\muzero_{(I)}c^{(I)}}{2}\Big)\,\t_{I}
\ea
is the holonomy of the connection ${A_a}^i$ evaluated along a line of length $\muzero_I\Lzero_I$
(with respect to $\fq_{ab}$)
parallel to the edge of the elementary cell ${\cal V}$ in the diagonal direction $I$.
Note that the expression of \eqnref{eqn:triad coefficient} is independent
of the choice of $\muzero_I$.
In quantum theory, the Poisson brackets are replaced by the commutators.
This leads to the inverse triad operators:
\ba\label{eqn:triad operator}
&&\widehat{\left[{\frac{1}{\sqrt{\abs{p_I}}}}\right]}
=-\frac{i\,{\sgn}(p_I)}{2\p\g\Pl^2\muzero_I}
{\rm tr}\left(\t_{I}{\hat h}_{I}^{(\muzero_I)}[{\hat h}_{I}^{\muzero_I-1},{\hat L}_{I}]\right)\nn\\
=&&-\frac{i\,{\sgn}(p_I)}{2\p\g\Pl^2\muzero_I}\!
\left[\sin\!\Big(\frac{\muzero_{(I)}{c}^{(I)}}{2}\Big){\hat
L}_{I}\cos\!\Big(\frac{\muzero_{(I)}{c}^{(I)}}{2}\Big) \!\!-\!
\cos\!\Big(\frac{\muzero_{(I)}{c}^{(I)}}{2}\Big){\hat
L}_{I}\sin\!\Big(\frac{\muzero_{(I)}{c}^{(I)}}{2}\Big)\right].
\qquad
\ea
As in the strategy used for isotropic models, to
construct inverse triad operators, there are two ambiguities, labeled by an
half integer $j$ and a real number $\ell$ ($0<\ell<1$)
\cite{Bojowald:2006da,Bojowald:2002ny}. The general considerations
in \cite{Vandersloot:2005kh,Perez:2005fn} urge one to set $j=1/2$. For
$\ell$, a general selection criterion is not available and
$\ell=1/2$ and $\ell=3/4$ have been used in the literature. The
difference on this choice does not affects any qualitative features
and here we use $\ell=1/2$ since it gives a simpler expression.\footnote{Note that
$\ell=3/4$ was used in \cite{Ashtekar:2006uz} while $\ell=1/2$ was used in \cite{Ashtekar:2006wn}.}

The eigenstates of the inverse triad operator are $\ket{{\vec \m}}$ with the eigenvalues given by
\be\label{eqn:triad operator on states}
\widehat{\left[{\frac{1}{\sqrt{\abs{p_I}}}}\right]}\ket{{\vec \m}}
=\frac{1}{4\p\g\Pl^2\muzero_I}\Abs{L_{\m_I+\muzero_I}-L_{\m_I-\muzero_I}}\,\ket{{\vec \m}},
\ee
where $L_{\m_I}$ are the eigenvalues of the length operators defined in \eqnref{eqn:length operator}
and the equations
\ba\label{eqn:e cos sin}
e^{\pm\frac{i}{2}\muzero_1c^1}\ket{\m_1,\m_2,\m_3}&=&\ket{\m_1\pm\muzero_1,\m_2,\m_3},\nn\\
\cos(\frac{1}{2}\muzero_{1}c^{1})\ket{\m_1,\m_2,\m_3}
&=&\frac{1}{2}\left(\ket{\m_1+\muzero_1,\m_2,\m_3}+\ket{\m_1-\muzero_1,\m_2,\m_3}\right),\nn\\
\sin(\frac{1}{2}\muzero_{1}c^{1})\ket{\m_1,\m_2,\m_3}
&=&\frac{1}{2i}\left(\ket{\m_1+\muzero_1,\m_2,\m_3}-\ket{\m_1-\muzero_1,\m_2,\m_3}\right)
\ea
as well as the corresponding ones for $c^2$ and $c^3$ are used.
While the choice of $\muzero_I$ gives no difference in classical level,
it changes the eigenvalues of the inverse triad operators. The difference due to $\muzero_I$
is negligible when the operators act on the states $\ket{\vec{\m}}$ with $\abs{\m_I}\gg\muzero_I$
but becomes significant when the values of $\m_I$ are comparable to $\muzero_I$.
In $\muzero$-scheme, we will impose $\muzero_I=k\equiv\sqrt{3}/2$ as the imprint of
the fundamental discreteness in the full theory (see \secref{subsec:grav constraint operator}),
while in $\mubar$-scheme, the inverse triad operators are redefined in a refined way such
that $\muzero_I$ are replaced by $\mubar_I$, which vary adaptively on different states $\ket{\vec{\m}}$
(see \secref{subsec:improved total hamiltonian}).

Since the inverse triad operator is diagonal with real eigenvalues in
the basis of $\ket{\vec{\m}}$, it is self-adjoint in $\hilbert_{\rm grav}^S$.
It can also be shown that the inverse triad operator commutes with $\hat{p}_I$
despite the presence of $c^I$ in \eqnref{eqn:triad operator}.
A key property of the inverse triad operators is that they are
all bounded above. At the value $\m_I=\pm\muzero_I$, the upper bound
is obtained:
\be
\abs{p_I}^{-\frac{1}{2}}_{\max}={(2\p\g)}^{-1/2}\,
\Pl^{-1}.
\ee
The fact that the eigenvalue of $\widehat{{1}/{\sqrt{p_I}}}$ goes to zero (instead of infinity)
when $\m_I\rightarrow0$ commands the matter part of the Hamiltonian constraint
$\hat{C}_{\rm matter}(\vec{\m})$ to annihilate $\Ps(\vec{\m},\ph)$ for $\m_1=0$, $\m_2=0$ or $\m_3=0$.
This is an important condition that accounts for the resolution of the singularity
(see the remark in \secref{subsec:resolution of singularity}).
(More physical implications from the finiteness of the inverse triad operator are discussed
in detail in \cite{Ashtekar:2003hd}.)

\section{Dynamics: Hamiltonian Constraint ($\muzero$-Scheme)}\label{sec:dynamics}
To study the quantum dynamics, we need to obtain the Hamiltonian constraint
in the quantum theory.
We follow the same procedures used in \cite{Ashtekar:2006uz,Ashtekar:2006rx,Ashtekar:2003hd}
to construct the gravitational constraint operator in $\muzero$-scheme.
The self-adjointness of the gravitational constraint operator and its WDW limit
are also discussed.
This section focuses only on the gravitational part of the
constraint operator. The total Hamiltonian including the matter sector
and a more complete study of the analytical issues will be
explored in \secref{sec:analytical issues}.
The improved dynamics in $\mubar$-scheme is left in \secref{sec:improved dynamics}.

\subsection{Gravitational Constraint Operator}\label{subsec:grav constraint operator}
In the classicall theory, the Hamiltonian constraint in the Bianchi I model is given
by \eqnref{eqn:Hamiltonian}. However, since the corresponding operator ${\hat c}^I$ does not exist,
we cannot directly use the Hamiltonian of this form but should recast it in terms of holonomies
of the symmetric connections.
To bring out the close similarity of the regularization procedure to that in the full theory,
we start from the expression of the classical constraint
in the \emph{full} theory, which is
\ba\label{eqn:Hamiltonian full}
C_{\rm grav}&:=&\int_{{\cal V}}d^3Ne^{-1}
\left(\e_{ijk}F^i_{ab}\mbox{$\tE$}^{aj}\mbox{$\tE$}^{bk}
-2(1+\g^2){K_{[a}}^i{K_{b]}}^j{\mbox{$\tE$}^{a}}_i{\mbox{$\tE$}^{b}}_j\right)\nn\\
&=&-\g^{-2}\int_{{\cal V}}d^3x N \e_{ijk}F^i_{ab}e^{-1}\mbox{$\tE$}^{aj}\mbox{$\tE$}^{bk},
\ea
where $e:=\sqrt{|{\det\tE}|}\ {\sgn}(\det\tE)={\rm sgn}(p_1p_2p_3)\sqrt{\abs{p_1p_2p_3}}$
and the integral is restricted to the elementary cell ${\cal V}$
(of volume $V_o=\Lzero_1\Lzero_2\Lzero_3$ with respect to $\fq_{ab}$).
In the second line of \eqnref{eqn:Hamiltonian full},
we exploit the fact that for spatially flat models, the
two terms in the first line are proportional to each other (because $-\g{K_a}^i={A_a}^i$ and
$F_{ab}^i={\e^i}_{jk}{A_a}^j{A_b}^j$ when the spatial slice is flat).
Because of homogeneity, the lapse $N$ can be assumed to be constant and from now on
we set it to one.

To construct the Hamiltonian constraint operator, we first
follow the standard strategy used in gauge theories to
express the curvature component
$F^i_{ab}$ in terms of holonomies.
Given a small rectangular surface $\a_{IJ}$ parallel to the face of the cell
${\cal V}$ in the $I$-$J$ plane
(the two edges of $\a_{IJ}$ are of lengths $\muzero_I\Lzero_I$ and $\muzero_J\Lzero_J$
with respect to $\fq_{ab}$),
the ``$ab$ component'' of the curvature can be written as
\be\label{eqn:curvature component}
F^i_{ab}\t_i=\sum_{I\neq J}\fcotriad{a}{I}\,\fcotriad{b}{J}
\left(
\frac{h_{\a_{IJ}}^{(\muzero_I,\muzero_J)}-1}{\muzero_I\muzero_J\Lzero_I\Lzero_J}
+{\cal O}(c^3\,\muzero)
\right).
\ee
The holonomy around the perimeter of $\a_{IJ}$ is:
\be
h_{\a_{IJ}}^{(\muzero_I,\muzero_J)}=h_I^{(\muzero_I)}h_J^{(\muzero_J)}
(h_I^{(\muzero_I)})^{-1}(h_J^{(\muzero_J)})^{-1},
\ee
where the holonomy along the individual edges
$h_I^{(\muzero_I)}$ is given by \eqnref{eqn:hI muzero}.

Next, following the close similarity of the regularization procedure
used in the full theory \cite{Thiemann:1996ay,Thiemann:1996aw,Thiemann:1997rt},
we express the triad term $\e_{ijk}e^{-1}\mbox{$\tE$}^{aj}\mbox{$\tE$}^{bk}$
in terms of holonomies and the Poisson brackets of the holonomy and the volume function.
In the symmetry reduced phase space, this is done as:
\be\label{eqn:triad term}
\t^i\e_{ijk}e^{-1}\mbox{$\tE$}^{aj}\mbox{$\tE$}^{bk}
=-\frac{2}{8\p\g\,G}
\sum_{K}\frac{\sgn(p_1p_2p_3)}{\muzero_K\Lzero_K}
\e^{abc}\,\fcotriad{c}{K}h_K^{(\muzero_K)}
\{h_K^{(\muzero_K)-1},V\},
\ee
where $V=L_1L_2L_3=\sqrt{\abs{p_1p_2p_3}}$.
Note that \eqnref{eqn:triad term} is exact,
while \eqnref{eqn:curvature component} depends on the choice of $\muzero_I$
(``\,$\muzero$-ambiguity'').

Taking \eqnref{eqn:curvature component} and \eqnref{eqn:triad term} into \eqnref{eqn:Hamiltonian full},
we have
\ba\label{eqn:grav Hamiltonian}
C_{\rm grav}&=&-\frac{4}{8\p\g^3\,G}\frac{\sgn(p_1p_2p_3)}{\muzero_1\muzero_2\muzero_3}\nn\\
&&\ \times\sum_{IJK}\e^{IJK}{\rm tr}
\left(
h_I^{(\muzero_I)}h_J^{(\muzero_J)}h_I^{(\muzero_I)-1}h_J^{(\muzero_J)-1}
h_K^{(\muzero_K)}\{h_K^{(\muzero_K)-1},V\}
\right)
+{\cal O}(c^3\,\muzero),\quad
\ea
where the term proportional to the identity inside the parenthesis in \eqnref{eqn:curvature component}
is dropped out because of the trace and because the equations
$\e^{abc}\,\fcotriad{a}{I}\fcotriad{b}{J}\fcotriad{c}{K}=\sqrt{\fq}\,\e^{IJK}$
and $\int_{{\cal V}}d^3x\sqrt{\fq}=V_o$ are used.
Note that the limit $\muzero_I\rightarrow 0$ in \eqnref{eqn:grav Hamiltonian} exists
and it \emph{precisely} equals the classical counterpart $C_{\rm garv}$ in \eqnref{eqn:Hamiltonian}.
However, this limit is physically fictitious and only makes sense mathematically.
Since $\hat{c}$ is not well-defined in $\hilbert^S_{\rm grav}$,
the ``continuum limit'' $\muzero_I\rightarrow 0$ fails to exist and thus we cannot
remove the regulator in the quantum geometry of the reduced model.
(In the full theory, on the contrary, diffeomorphism invariance ensures
the insensitivity to the regularization procedure
\cite{Thiemann:1996ay,Thiemann:1996aw,Thiemann:1997rt}.
The dependance on the regulator
is traced back to the assumption of homogeneity.)

Since there is no natural way to express the constraint \emph{exactly}
in terms of the elementary variables, the limiting procedure is essential.
To faithfully convey the underlining geometry of the full theory,
we keep $\muzero_I$ explicitly and treat it as the
``regulator'', which reflects the ``fundamental discreteness'' of the full theory
as an imprint on the reduced theory.
(The appropriate value of $\muzero_I$ is to be determined soon.)
The resulting regulated constraint is:
\ba\label{eqn:original C grav}
&&{{\hat C}}_{{\rm grav},o}^{(\muzero)}
=
\frac{4i\ \sgn(p_1p_2p_3)}{8\p\g^3\Pl^2\muzero_1\muzero_2\muzero_3}
\sum_{IJK}\e^{IJK}{\rm tr}
\left(
{\hat h}_I^{(\muzero_I)}{\hat h}_J^{(\muzero_J)}{\hat h}_I^{(\muzero_I)-1}{\hat h}_J^{(\muzero_J)-1}
{\hat h}_K^{(\muzero_K)}[{\hat h}_K^{(\muzero_K)-1},{\hat V}]
\right)
\nn\\
&=&\!
\frac{i\ \sgn(p_1p_2p_3)}{\p\g^3\Pl^2\muzero_1\muzero_2\muzero_3}\Big\{
\sin({\muzero_1c^1})
\sin({\muzero_2c^2})
\left(\sin(\frac{\muzero_3c^3}{2}){\hat V}\cos(\frac{\muzero_3c^3}{2})
-\cos(\frac{\muzero_3c^3}{2}){\hat V}\sin(\frac{\muzero_3c^3}{2})\right)\nn\\
&&\qquad\quad+\text{ cyclic terms}
\Big\},
\ea
where ``cyclic terms''
denote the same expression as the term(s) written explicitly inside the curly bracket but with
the cyclic index replacements
$(1\rightarrow3,2\rightarrow1,3\rightarrow2)$ and
$(1\rightarrow2,2\rightarrow3,3\rightarrow1)$.
(The subscript ``$o$'' used for
${{\hat C}}_{{\rm grav},o}^{(\muzero)}$
is to indicate that the ordering ambiguity of quantization
has not been taken care of and thus ${{\hat C}}_{{\rm grav},o}^{(\muzero)}$
is not self-adjoint.)
The operator ${{\hat C}}_{{\rm grav},o}^{(\muzero)}$
acting on the eigenstates of ${\hat p}_I$ gives
\ba\label{eqn:original C grav on states}
&&{ {\hat C}}_{{\rm grav},o}^{(\muzero)}\ket{\m_1,\m_2,\m_3}
=
\frac{1}{8\p\g^3\Pl^2\muzero_1\muzero_2\muzero_3}\nn\\
&&\qquad\times\Big\{\abs{V_{\m_1,\m_2,\m_3+\muzero_3}-V_{\m_1,\m_2,\m_3-\muzero_3}}
\Big(\ket{\m_1\!+\!2\,\muzero_1,\m_2\!+\!2\,\muzero_2,\m_3}-\ket{\m_1\!-\!2\,\muzero_1,\m_2\!+\!2\,\muzero_2,\m_3}\nn\\
&&\qquad\qquad\qquad
-\ket{\m_1\!+\!2\,\muzero_1,\m_2\!-\!2\,\muzero_2,\m_3}+\ket{\m_1\!-\!2\,\muzero_1,\m_2\!-\!2\,\muzero_2,\m_3}\Big)\nn\\
&&\qquad\qquad+\text{ cyclic terms}
\Big\},
\ea
where \eqref{eqn:e cos sin} is used.

Regarded as a state in the connection representation, (every element of) the holonomy $h_I^{(\muzero_I)}$
is an eigenstate of the area operator $\hat{a}_I=\widehat{\abs{p_I}}$:
\be\label{eqn:area of h}
\hat{a}\,h_I^{(\muzero_I)}(c)=\left({4\p\g\,\muzero_I}\Pl^2\right)h_I^{(\muzero_I)}(c).
\ee
Following the same strategy ($\muzero$-scheme) in
\cite{Ashtekar:2006uz,Ashtekar:2006rx,Ashtekar:2003hd},
we demand this eigenvalue to be
be $\D=2\sqrt{3}\p\g\Pl^2$, the \emph{area gap} of the full theory
\cite{Ashtekar:1995zh,Rovelli:1994ge,Ashtekar:1996eg}.
This yields
$\muzero_I=k:=\D/(4\p\g\Pl^2)=\sqrt{3}/2$ and fixes the ``\,$\muzero$-ambiguity''.

\subsection{Self-Adjointness and the WDW Limit}\label{subsec:self adjoint and WDW limit}
In order to construct the physical sector with the physical inner product
(which will be done in \secref{sec:analytical issues}),
it is advantageous to have the Hamiltonian constraint self-adjoint.
Note that \eqnref{eqn:original C grav} is not self-adjoint. This happens simply because of
the factor ordering ambiguities common in quantum mechanics.
There are two natural ways to resolve the ordering problems. The first is to take the self-adjoint
part of \eqnref{eqn:original C grav}:
\be\label{eqn:hat C grav prime}
{{\hat C}}_{\rm grav}^{(\muzero)'}=\frac{1}{2}
\left[{ {\hat C}}_{{\rm grav},o}^{(\muzero)}+
{ {\hat C}}_{{\rm grav},o}^{(\muzero)\dagger}\right].
\ee
The second is the nature re-ordering adopted in \cite{Ashtekar:2006uz},
in which we ``re-distribute'' the $\sin(\muzero c)$ terms to get the following self-adjoint constraint:
\ba\label{eqn:hat C grav}
&&{\hat C}_{\rm grav}^{(\muzero)}
=
\frac{4i\ \sgn(p_1p_2p_3)}{8\p\g^3\Pl^2\muzero_1\muzero_2\muzero_3}\nn\\
&&\qquad\times\Big\{
\sin({\muzero_1c^1})\left(\sin(\frac{\muzero_3c^3}{2}){\hat V}\cos(\frac{\muzero_3c^3}{2})
-\cos(\frac{\muzero_3c^3}{2}){\hat V}\sin(\frac{\muzero_3c^3}{2})\right)
\sin({\muzero_2c^2})\nn\\
&&\qquad\quad+
\sin({\muzero_2c^2})\left(\sin(\frac{\muzero_3c^3}{2}){\hat V}\cos(\frac{\muzero_3c^3}{2})
-\cos(\frac{\muzero_3c^3}{2}){\hat V}\sin(\frac{\muzero_3c^3}{2})\right)\sin({\muzero_1c^1})\nn\\
&&\qquad\quad+\text{ cyclic terms}
\Big\}.
\ea
When acting on the eigenstates of ${\hat p}_I$, ${\hat C}_{\rm grav}^{(\muzero)'}$ gives
\ba\label{eqn:hat C grav prime on states}
&&{\hat C}_{\rm grav}^{(\muzero)'}\ket{\m_1,\m_2,\m_3}
=
\frac{1}{16\p\g^3\Pl^2\muzero_1\muzero_2\muzero_3}\nn\\
&&\quad \times\Big\{
\Big(\Abs{V_{\tiny \begin{array}{l}\m_1\!+\!2\muzero_1,\m_2\!+\!2\muzero_2,\\\m_3\!+\!\muzero_3\end{array}}
\!\!-V_{\tiny \begin{array}{l}\m_1\!+\!2\muzero_1,\m_2\!+\!2\muzero_2,\\\m_3\!-\!\muzero_3\end{array}}
\!\!}
+\Abs{V_{\tiny \begin{array}{l}\m_1,\,\m_2,\\\m_3\!+\!\muzero_3\end{array}}
\!\!-V_{\tiny \begin{array}{l}\m_1,\,\m_2,\\\m_3\!-\!\muzero_3\end{array}}
\!\!}\Big)
\ket{\m_1\!+\!2\,\muzero_1,\m_2\!+\!2\,\muzero_2,\m_3}\nn\\
&&\qquad \!\!-
\Big(\Abs{V_{\tiny \begin{array}{l}\m_1\!-\!2\muzero_1,\m_2\!+\!2\muzero_2,\\\m_3\!+\!\muzero_3\end{array}}
\!\!-V_{\tiny \begin{array}{l}\m_1\!-\!2\muzero_1,\m_2\!+\!2\muzero_2,\\\m_3\!-\!\muzero_3\end{array}}
\!\!}
+\Abs{V_{\tiny \begin{array}{l}\m_1,\,\m_2,\\\m_3\!+\!\muzero_3\end{array}}
\!\!-V_{\tiny \begin{array}{l}\m_1,\,\m_2,\\\m_3\!-\!\muzero_3\end{array}}
\!\!}\Big)
\ket{\m_1\!-\!2\,\muzero_1,\m_2\!+\!2\,\muzero_2,\m_3}\nn\\
&&\qquad \!\!-
\Big(\Abs{V_{\tiny \begin{array}{l}\m_1\!+\!2\muzero_1,\m_2\!-\!2\muzero_2,\\\m_3\!+\!\muzero_3\end{array}}
\!\!-V_{\tiny \begin{array}{l}\m_1\!+\!2\muzero_1,\m_2\!-\!2\muzero_2,\\\m_3\!-\!\muzero_3\end{array}}
\!\!}
+\Abs{V_{\tiny \begin{array}{l}\m_1,\,\m_2,\\\m_3\!+\!\muzero_3\end{array}}
\!\!-V_{\tiny \begin{array}{l}\m_1,\,\m_2,\\\m_3\!-\!\muzero_3\end{array}}
\!\!}\Big)
\ket{\m_1\!+\!2\,\muzero_1,\m_2\!-\!2\,\muzero_2,\m_3}\nn\\
&&\qquad\!\! +
\Big(\Abs{V_{\tiny \begin{array}{l}\m_1\!-\!2\muzero_1,\m_2\!-\!2\muzero_2,\\\m_3\!+\!\muzero_3\end{array}}
\!\!-V_{\tiny \begin{array}{l}\m_1\!-\!2\muzero_1,\m_2\!-\!2\muzero_2,\\\m_3\!-\!\muzero_3\end{array}}
\!\!}
+\Abs{V_{\tiny \begin{array}{l}\m_1,\,\m_2,\\\m_3\!+\!\muzero_3\end{array}}
\!\!-V_{\tiny \begin{array}{l}\m_1,\,\m_2,\\\m_3\!-\!\muzero_3\end{array}}
\!\!}\Big)
\ket{\m_1\!-\!2\,\muzero_1,\m_2\!-\!2\,\muzero_2,\m_3}\nn\\
&&\qquad+\text{ cyclic terms}
\Big\}
\ea
and ${\hat C}_{\rm grav}^{(\muzero)}$ gives
\ba\label{eqn:hat C grav on states}
&&{\hat C}_{\rm grav}^{(\muzero)}\ket{\m_1,\m_2,\m_3}
=
\frac{1}{16\p\g^3\Pl^2\muzero_1\muzero_2\muzero_3}\nn\\
&&\quad\times\Big\{
\Big(\Abs{V_{\tiny \begin{array}{l}\m_1\!+\!2\muzero_1,\m_2\\\m_3\!+\!\muzero_3\end{array}}
\!\!-V_{\tiny \begin{array}{l}\m_1\!+\!2\muzero_1,\m_2,\\\m_3\!-\!\muzero_3\end{array}}
\!\!}
+\Abs{V_{\tiny \begin{array}{l}\m_1,\,\m_2\!+\!2\muzero_2,\\\m_3\!+\!\muzero_3\end{array}}
\!\!-V_{\tiny \begin{array}{l}\m_1,\,\m_2\!+\!2\muzero_2,\\\m_3\!-\!\muzero_3\end{array}}
\!\!}\Big)
\ket{\m_1\!+\!2\,\muzero_1,\m_2\!+\!2\,\muzero_2,\m_3}\nn\\
&&\qquad\!\! -
\Big(\Abs{V_{\tiny \begin{array}{l}\m_1\!-\!2\muzero_1,\m_2\\\m_3\!+\!\muzero_3\end{array}}
\!\!-V_{\tiny \begin{array}{l}\m_1\!-\!2\muzero_1,\m_2,\\\m_3\!-\!\muzero_3\end{array}}
\!\!}
+\Abs{V_{\tiny \begin{array}{l}\m_1,\,\m_2\!+\!2\muzero_2,\\\m_3\!+\!\muzero_3\end{array}}
\!\!-V_{\tiny \begin{array}{l}\m_1,\,\m_2\!+\!2\muzero_2,\\\m_3\!-\!\muzero_3\end{array}}
\!\!}\Big)
\ket{\m_1\!-\!2\,\muzero_1,\m_2\!+\!2\,\muzero_2,\m_3}\nn\\
&&\qquad\!\! -
\Big(\Abs{V_{\tiny \begin{array}{l}\m_1\!+\!2\muzero_1,\m_2\\\m_3\!+\!\muzero_3\end{array}}
\!\!-V_{\tiny \begin{array}{l}\m_1\!+\!2\muzero_1,\m_2,\\\m_3\!-\!\muzero_3\end{array}}
\!\!}
+\Abs{V_{\tiny \begin{array}{l}\m_1,\,\m_2\!-\!2\muzero_2,\\\m_3\!+\!\muzero_3\end{array}}
\!\!-V_{\tiny \begin{array}{l}\m_1,\,\m_2\!-\!2\muzero_2,\\\m_3\!-\!\muzero_3\end{array}}
\!\!}\Big)
\ket{\m_1\!+\!2\,\muzero_1,\m_2\!-\!2\,\muzero_2,\m_3}\nn\\
&&\qquad\!\! +
\Big(\Abs{V_{\tiny \begin{array}{l}\m_1\!-\!2\muzero_1,\m_2\\\m_3\!+\!\muzero_3\end{array}}
\!\!-V_{\tiny \begin{array}{l}\m_1\!-\!2\muzero_1,\m_2,\\\m_3\!-\!\muzero_3\end{array}}
\!\!}
+\Abs{V_{\tiny \begin{array}{l}\m_1,\,\m_2\!-\!2\muzero_2,\\\m_3\!+\!\muzero_3\end{array}}
\!\!-V_{\tiny \begin{array}{l}\m_1,\,\m_2\!-\!2\muzero_2,\\\m_3\!-\!\muzero_3\end{array}}
\!\!}\Big)
\ket{\m_1\!-\!2\,\muzero_1,\m_2\!-\!2\,\muzero_2,\m_3}\nn\\
&&\qquad+\text{ cyclic terms}
\Big\}.
\ea

In this paper, we choose the second method of symmetrization, since
${\hat C}_{\rm grav}^{(\muzero)}$ has certain qualities better than
${\hat C}_{\rm grav}^{(\muzero)'}$. In particular, in the isotropic
cases, ${\hat C}_{\rm grav}^{(\muzero)}$ gives positive definite
$\Th$ (to be defined later), which is a pleasing feature when the
physical sector is constructed. In the anisotropic Bianchi I model,
the positive definiteness of $\Th$ no longer holds; nevertheless,
choosing ${\hat C}_{\rm grav}^{(\muzero)}$ enables us to generalize
the isotropic results in an obvious way.

The Hamiltonian constraint equation differs markedly from the
WDW equation of geometrodynamics in the Planck regime because it
crucially exploits the discreteness underlying quantum geometry.
In the continuum limit $\muzero_I\rightarrow0$ (although physically fictitious),
one expect that the LQC constraint equation will reduce to the WDW equation.
According to the analysis in \cite{Ashtekar:2006uz} (and more details in \cite{Ashtekar:2003hd}),
we can follow the arguments therein and conclude that the gravitational part
of the Hamiltonian operators reduces to:
\ba
\hat{C}_{{\rm grav},o}^{(\muzero)}\Ps(\vec{p})
&\approx&
128\p^2\Pl^4\Big\{\sgn(p_1p_2)\sqrt{\frac{\abs{p_1p_2}}{\abs{p_3}}}
\frac{\partial^2}{\partial p_1\partial p_2}+\text{cyclic terms}\Big\}\Ps(\vec{p})\qquad\qquad\qquad\qquad\nn\\
&=:&\hat{C}_{{\rm grav},o}^{\rm WDW}\Ps(\vec{p}),
\ea
\ba
\hat{C}_{{\rm grav}}^{(\muzero)'}\Ps(\vec{p})
\!\!&\approx&\!\!
64\p^2\Pl^4\Big\{\frac{\sgn(p_1p_2)}{\sqrt{\abs{p_3}}}
\left(\sqrt{\abs{p_1p_2}}\frac{\partial^2}{\partial p_1\partial p_2}
+\frac{\partial^2}{\partial p_1\partial p_2}\sqrt{\abs{p_1p_2}}\right)
+\text{cyclic terms}\Big\}\Ps(\vec{p})\nn\\
&=:&\hat{C}_{{\rm grav}}^{'{\rm WDW}}\Ps(\vec{p}),
\ea
\ba\label{eqn:WDW C}
\hat{C}_{{\rm grav}}^{(\muzero)}\Ps(\vec{p})
\!\!&\approx&\!\!
64\p^2\Pl^4\Big\{\frac{\sgn(p_1p_2)}{\sqrt{\abs{p_3}}}
\left(\frac{\partial}{\partial p_2}\sqrt{\abs{p_1p_2}}\frac{\partial}{\partial p_1}
+\frac{\partial}{\partial p_1}\sqrt{\abs{p_1p_2}}\frac{\partial}{\partial p_2}\right)
\!+\text{cyclic terms}\Big\}\Ps(\vec{p})\nn\\
&=:&\hat{C}_{{\rm grav}}^{\rm WDW}\Ps(\vec{p}).
\ea
In the above approximations, $\approx$ stands for equality modulo terms of the order
${\cal O}({}^o\!p_I)$ (where ${}^o\!p_I:=4\p\g \Pl^2\muzero_I$).
That is, in the limit when the characteristic scale is much greater than
that associated with the area gap (i.e. $\abs{p_I}\gg{}^o\!p_I$ or $\abs{\m_I}\gg\muzero_I$),
quantum geometry effects can be neglected and the LQC difference operators
reduce to the WDW differential operators.
The different ordering factors give rise to different WDW limits.
In particular, $\hat{C}_{{\rm grav}}^{(\muzero)'}$ and $\hat{C}_{{\rm grav}}^{(\muzero)}$
differ only by a factor ordering:
\be
\left(\hat{C}_{{\rm grav}}^{(\muzero)'}-\hat{C}_{{\rm grav}}^{(\muzero)}\right)\Ps({\vec p})
=\frac{48\p^2\Pl^4}{\sqrt{\abs{p_1p_2p_3}}}
\Big(\sgn(p_1p_2)+\sgn(p_1p_3)+\sgn(p_2p_3)\Big)\Ps({\vec p}).
\ee
The difference is negligible for large (volume) scales.
The operator $\hat{C}_{{\rm grav}}^{\rm WDW}$ will be used extensively in \secref{sec:WDW theory}.

\ \newline \small \underline{Note}: The results obtained in this
section can be formally ``translated'' for the isotropic case by
inspection. The translation and the related issues of isotropy
reduction from anisotropic models in $\muzero$-scheme are remarked
in \appref{subsec:first scheme for isotropic}. \hfill
$\Box$\normalsize

\section{Wheeler-DeWitt (WDW) Theory}\label{sec:WDW theory}
In this section, we make a digression to study and construct
the WDW quantum theory in the \emph{connection dynamics}.
This construction will serve two purposes:
first, it will introduce the key notions for constructing the physical space
and enable us to generalized these notions heuristically for the completion of
the program in LQC;
second, we will be able to compare and contrast the results of the WDW and LQC theories,
thereby extracting out the essential ingredient given by the quantum geometry.
In particular, the WDW theory does \emph{not} resolve the
classical singularity but the semi-classical (coherent) states in the WDW theory can be used to
build up the initial states at late times in LQC theory.

\subsection{Emergent Time and the General Solutions}\label{subsec:WDW general solutions}
Recall that the classical Hamiltonian constraint in terms of the variables $c^I$ and $p_I$ is
given by \eqnref{eqn:Hamiltonian};
that is
\ba
&&C_{\ph}=-C_{\rm grav}\nn\\
&\Rightarrow&  8\p G\,{\abs{p_1p_2p_3}^{-1/2}}p_{\ph}^2=\frac{2}{\g^2}
\left(\frac{\sgn(p_1p_2)}{\sqrt{\abs{p_3}}}\,c^1c^2\sqrt{\abs{p_1p_2}}
+\text{cyclic terms}
\right).
\ea
In WDW theory, we have the kinematic Hilbert space $\hilbert_{\rm kin}^{\rm WDW}=L^2(\mathbb{R}^2,dp\,d\ph)$.
The operators $\hat{p}_I$ and $\hat{\ph}$ act as multiplications while $\hat{c}^I$ and $\hat{p}_\ph$
act as differential operators,
represented as:
\be
\hat{c}^I\Ps=i\hbar\,8\p\g G\frac{\partial\Ps}{\partial p_I}
\quad \text{and} \quad
\hat{p}_\ph\Ps=-i\hbar\frac{\partial\Ps}{\partial\ph}.
\ee

To write down the quantum constraint operator, we have to make a choice of the factor ordering.
Since our purpose is to compare the WDW theory with LQC, it is most convenient to use the same factor
ordering that comes from the continuum limit of the LQC constraint.
Therefore, we chose $\hat{C}_{{\rm grav}}^{\rm WDW}$ defined in \eqnref{eqn:WDW C}
as the gravitational part of the WDW Hamiltonian constraint.
The total WDW Hamiltonian constraint then reads as
\ba\label{eqn:WDW master eq}
\frac{\partial^2\Ps}{\partial\ph^2}\!\!&=&\!\!8\p G\,[\underline{B}(\vec{\m})]^{-1}
\Big\{\frac{\sgn(\m_1\m_2)}{\sqrt{\abs{\m_3}}}
\left(\frac{\partial}{\partial\m_2}\sqrt{\abs{\m_1\m_2}}\frac{\partial}{\partial\m_1}
\!+\!\frac{\partial}{\partial\m_1}\sqrt{\abs{\m_1\m_2}}\frac{\partial}{\partial\m_2}\right)
\!+\text{cyclic terms}\Big\}\Ps\nn\\
&=:&-\left(\underline{\Th}_3(\m_1,\m_2)+\underline{\Th}_2(\m_1,\m_3)+\underline{\Th}_1(\m_2,\m_3)\right)\Ps
=:-\underline{\Th}(\vec{\m})\Ps,
\ea
where we define $p_I=:4\p\g\Pl^2\m_I$ and $\underline{B}(\vec{\m}):=\abs{\m_1\m_2\m_3}^{-1/2}$.
The WDW equation now has the same form as the Klein-Gordon equation, with $\ph$ playing the role of time
and $\underline{\Th}$ of the spatial Laplacian.
Therefore, the form of the Hamiltonian constraint in the WDW theory suggests that
$\ph$ can be interpreted as \emph{emergent time}.
Unlike the isotropic case, however, $\underline{\Th}$ is not elliptic and hence not positive definite.
Fortunately, this problem can be easily overcome.
We will see soon that the subspace of negative eigenvalues of $\Th$ is actually non-physical and
it turns out only $\Th_+$, the projection on the positive spectrum of $\Th$, does matter.

Note that if $\Ps(\vec{\m},\ph)$ is a solution of \eqnref{eqn:WDW
master eq}, so is $\Pi_I\Ps(\vec{\m},\ph)$ (where $\Pi_I$ are the
orientation reversal operators, which flip $\m_I$ to $-\m_I$ for
$I=1,2,3$) since $\underline{\Th}$ commutes with $\Pi_I$. Because
the standard WDW theory deals with geometry, it is completely
insensitive to the orientation of the triad and $\Pi_I$ should be
considered as large gauge transformations. Furthermore, $\Pi_I^2=1$
and there are only two eigenspaces of $\Pi_I$, one of which is the
symmetric sector and the other anti-symmetric. Therefore, it is
natural to work with the symmetric sector. Thus, the physical
Hilbert space will consist of the suitably regular solutions
$\Ps(\m,\ph)$ to \eqnref{eqn:WDW master eq} which are symmetric
under $\m_I\rightarrow-\m_I$ for $I=1,2,3$.

To obtain the general solution of \eqnref{eqn:WDW master eq}, we first note that the operator
enclosed by the curly bracket in \eqnref{eqn:WDW master eq} is self-adjoint on $L^2_S(\mathbb{R}^3,d^3\!\m)$,
the invariant subspace of $L^2(\mathbb{R}^3,d^3\!\m)$ under $\Pi_{I=1,2,3}$.
Therefore, $\underline{\Th}$ (and $\underline{\Th}_I$ as well) is self-adjoint
on $L^2_S(\mathbb{R}^3,\underline{B}(\vec{\m})\,d^3\!\m)$.
It is easy to verify that the eigenfunctions of $\underline{\Th}$
can be labeled by $\vec{k}\in\mathbb{R}^3$:
\be\label{eqn:eigenfunction ek}
\underline{e}_{\vec k}({\vec \m})=\frac{1}{(4\p)^3}\,\abs{\m_1\m_2\m_3}^{-\frac{1}{4}}\,
e^{ik_1\!\ln\abs{\m_1}}e^{ik_2\!\ln\abs{\m_2}}e^{ik_3\!\ln\abs{\m_3}}.
\ee
The corresponding eigenvalues are given by
\be\label{eqn:WDW frequency}
\underline{\Th}\,\underline{e}_{\vec k}({\vec \m})=\o^2(\vec k)\,\underline{e}_{\vec k}({\vec \m})
\quad \text{with} \quad
\o^2(\vec k)=16\p G\left(k_1k_2+k_1k_3+k_2k_3+\frac{3}{16}\right),
\ee
(where the factor $3/16$ is an artifact of the factor ordering choice).
The eigenfunctions satisfy the orthonormality relation:
\be\label{eqn:orthonormality rel}
\int d^3\!\m\,\underline{B}(\vec{\m})\,\bar{\underline{e}}_{\vec k}({\vec \m})\,\underline{e}_{{\vec k}'}({\vec \m})
=\d^3(\vec{k},\vec{k}'),
\ee
(where the right hand side is the standard Dirac distribution)
and the completeness relation:
\be\label{eqn:completeness rel}
\int d^3\!\m\,\underline{B}(\vec{\m})\,\bar{\underline{e}}_{\vec k}({\vec \m})\,\Ps(\vec{\m})=0
\quad\text{for }
\forall {\vec k},
\quad
\text{iff }
\Ps({\vec \m})=0
\ee
for any $\Ps({\vec \m})\in L^2_S(\mathbb{R}^3,\underline{B}(\vec{\m})\,d^3\!\m)$.

Unlike the case of isotropic model, however, $\underline{\Th}$ is not positive definite on
$\Ps({\vec \m})\in L^2_S(\mathbb{R}^3,\underline{B}(\vec{\m})\,d^3\!\m)$
and $\o^2$ could be negative. If $\o^2<0$, the frequency has imaginary part and the corresponding
plane wave solution to \eqnref{eqn:WDW master eq} blows up in the far future ($\ph\rightarrow\infty$)
or far past ($\ph\rightarrow-\infty$); therefore, the modes associated with negative $\o^2$ should be
discarded and only those with $\o^2>0$ are allowed. The relevant space
is the ``positive'' subspace
$L^2_S(\mathbb{R}^3,\underline{B}(\vec{\m})\,d^3\!\m;\o^2>0)$
and what matters is just the operator $\underline{\Th}_+$, obtained by projecting the action of
$\underline{\Th}$ to the positive eigenspace in its spectral decomposition.\footnote{
The physical condition $\o^2>0$ is reminiscent of
the constraint \eqnref{eqn:parameter constraint}
on the classical solution parameters $\k_1$, $\k_2$, $\k_3$ and $\k_\ph$,
which leads to $\k_\ph^2=2(\k_1\k_2+\k_1\k_3+\k_2\k_3)$. If we have the identification:
$k_{1,2,3}={\cal K}\k_{1,2,3}$, then the frequency gives
\be
\o^2(\vec{k})=8\p G\left({\cal K}^2\k_\ph^2+\frac{3}{8}\right)\approx 8\p G {\cal K}^2\k_\ph^2
\equiv \frac{p_\ph^2}{\hbar^2}> 0,
\ee
(where the approximation is valid when $p_\ph\gg\hbar\sqrt{G}$
or equivalently ${\cal K}\k_\ph\gg 1$.). This shows that
the physical condition $\o^2>0$ simply means $p_\ph$ is real.
For a semi-classical state,
the identification $k_{1,2,3}={\cal K}\k_{1,2,3}$ is sensible, since its
Fourier transformed function $\tilde{\Ps}_\pm({\vec k})$ (to be defined below)
is sharply peaked at $k_{1,2,3}={\cal K}\k_{1,2,3}$.
(See \secref{subsec:semi-classical} for more details.)}

With the eigenfunctions in hand, we can express the ``general'' solution to \eqnref{eqn:WDW master eq} as
\be\label{eqn:sol to WDW master eq}
\Ps({\vec \m},\ph)=\int_{\o^2({\vec k})>0}\!\!d^3k
\left(
\tilde{\Ps}_+({\vec k})\,\underline{e}_{\vec k}({\vec \m})\,e^{i\o\ph}+
\tilde{\Ps}_-({\vec k})\,\bar{\underline{e}}_{\vec k}({\vec \m})\,e^{-i\o\ph}
\right)
\ee
for some $\tilde{\Ps}_\pm({\vec k})$ in $L^2(\mathbb{R}^3,d^3k)$.
Following the terminology used in the Klein-Gordon theory, if $\tilde{\Ps}_\pm({\vec k})$
has support in the region of positive $k_I$, we call the solution ``incoming'' (or ``expanding'')
in the direction $I$,
while if it has support of negative $k_I$, it is said to be ``outgoing''
(or ``contracting'').\footnote{The name of
``expanding/contracting'' makes sense for semi-classical solutions, for which
$k_I$ is supported mainly in the vicinity of ${\cal K}\k_I$ as mentioned in the previous footnote.
According to \eqnref{eqn:classical sol a-phi}, if $\k_I$ is positive/negtive (and $\k_\ph>0$),
the scale factor $a_I$ in the direction $I$ is expanding/contracting with respect to $\ph$.
[Note: The nomenclature adopted here is opposite to that in \cite{Ashtekar:2006uz},
due to the minus sign used in the equation there: $p_\ph^\star=-\hbar\sqrt{3/16\p G}\,k^\star$,
which, in the anisotropic notations,
gives rise to an extra minus sign for the relation between $\k_\ph$ and $p_\ph^\star$
in \eqnref{eqn:rel of p-phi and kappa-ph}.]}

If $\tilde{\Ps}_{+/-}({\vec k})$ vanishes, the solution is said to be \emph{negative/positive frequency}.
Thus, every solution \eqnref{eqn:sol to WDW master eq} admits a natural decomposition into the positive
and negative frequency parts.
The positive (respectively negative) frequency solutions satisfy the first order (in $\ph$) equation
which can be regarded as the square-root of \eqnref{eqn:WDW master eq}:
\be
\frac{\partial \Ps_\pm}{\partial \ph}=\pm i\sqrt{\underline{\Th}_+}\ \Ps_\pm,
\ee
where $\sqrt{\underline{\Th}_+}$ is the square-root of the positive self-adjoint operator $\underline{\Th}_+$
defined via spectral decomposition on
$L^2_S(\mathbb{R}^3,\underline{B}(\vec{\m})\,d^3\!\m;\o^2>0)$.
Therefore, a general ``initial'' function
$\Ps_\pm({\vec \m},\ph_o)=f_\pm({\vec \m})\in L^2_S(\mathbb{R}^3,\underline{B}(\vec{\m})\,d^3\!\m;\o^2>0)$ at $\ph=\ph_o$
can be ``evolved'' to a solution of \eqnref{eqn:WDW master eq} by:
\be\label{eqn:WDW Schrodinger eq}
\Ps_\pm({\vec \m},\ph)=e^{\pm i \sqrt{\underline{\Th}_+}\,(\ph-\ph_o)} \Ps_\pm({\vec \m},\ph_o).
\ee

\subsection{The Physical Sector}\label{subsec:physical sector}
To endow the space of the WDW solutions with a Hilbert space structure, we follow the same strategy of
\cite{Ashtekar:2006uz}.
The idea is to introduce the operators corresponding to a complete set of Dirac observables and select
the required inner product by demanding that those Dirac operators all be self-adjoint.
In the classical theory, a complete set is given by $p_\ph$ and $\vec{\m}|_{\ph_o}$
($\vec{\m}$ at some particular instant $\ph=\ph_o$), since the set of these uniquely specifies
a classical solution and thus represents a point in the classical phase space.
In WDW theory, given a (symmetric) solution $\Ps({\vec \m},\ph)$,
since $\hat{p}_\ph$ commutes with $\underline{\Th}$,
\be
\hat{p}_\ph\Ps({\vec \m},\ph):=-i\hbar\frac{\partial\Ps}{\partial\ph}
\ee
is again a (symmetric) solution to the WDW equation. So, we can just retain this definition of $\hat{p}_\ph$ from
$\hilbert_{\rm kin}^{\rm WDW}$.
On the other hand, \eqnref{eqn:WDW Schrodinger eq} enables us to define the Dirac observables
$\widehat{{\m_I}|_{\ph_o}}$. For a given (symmetric) solution $\Ps({\vec \m},\ph)$, we decompose it
into the positive and negative parts $\Ps_\pm({\vec \m},\ph)$, freeze them at $\ph=\ph_o$,
multiply them by $\abs{\m_I}$ and evolve each of them by \eqnref{eqn:WDW Schrodinger eq}. That is
\be
\widehat{{\m_I}|_{\ph_o}}\Ps({\vec \m},\ph):=
e^{i \sqrt{\underline{\Th}_+}\,(\ph-\ph_o)}\abs{\m_I} \Ps_+({\vec \m},\ph_o)+
e^{-i \sqrt{\underline{\Th}_+}\,(\ph-\ph_o)}\abs{\m_I} \Ps_-({\vec \m},\ph_o).
\ee
The result is again a (symmetric) solution to \eqnref{eqn:WDW master eq}.

Note that both $\hat{p}_\ph$ and $\widehat{{\m_I}|_{\ph_o}}$ preserve the positive and
negative frequency sectors and hence we have \emph{super-selection} between these two sectors.
In quantum theory, we can restrict ourselves on one of them. From now on, we will
focus on the positive frequency sector and thus drop the suffix $+$ for $\Ps$.

Every WDW solution is uniquely determined by its initial condition $\Ps({\vec \m},\ph_o)$
and the Dirac observables have the actions on the initial state:
\be\label{eqn:Dirac observables}
\widehat{{\m_I}|_{\ph_o}}\Ps({\vec \m},\ph_o)=\abs{\m_I}\Ps({\vec \m},\ph_o)
\quad\text{and}\quad
\hat{p}_\ph\Ps({\vec \m},\ph_o)=\hbar\sqrt{\underline{\Th}_+}\Ps({\vec \m},\ph_o).
\ee
The unique (up to an overall scaling) inner product which makes these Dirac observalbes
self-adjoint is therefore given by
\be\label{eqn:WDW inner product}
\inner{\Ps_1}{\Ps_2}_{\rm phy}=\int\!d^3\!\m\underline{B}({\vec \m})\bar{\Ps}_1(\vec{\m},\ph_o)\Ps_2(\vec{\m},\ph_o).
\ee
(Note that the inner product is conserved, i.e., independent of the choice of the initial instant $\ph_o$.)
As a result, the physical Hilbert space $\hilbert_{\rm phy}^{\rm WDW}$ is
the space of wave functions $\Ps({\vec \m},\ph)$
which are symmetric under $\m_I\rightarrow-\m_I$ and have finite norm by \eqnref{eqn:WDW inner product}.
The representation of the complete Dirac observables on $\hilbert_{\rm phy}^{\rm WDW}$ is given by
\be\label{eqn:Dirac observables 2}
\widehat{{\m_I}|_{\ph_o}}\Ps({\vec \m},\ph)=
e^{i \sqrt{\underline{\Th}_+}\,(\ph-\ph_o)}\abs{\m_I} \Ps({\vec \m},\ph_o)
\quad\text{and}\quad
\hat{p}_\ph\Ps({\vec \m},\ph)=\hbar\sqrt{\underline{\Th}_+}\Ps({\vec \m},\ph).
\ee
With the Dirac operators as well as the physical inner product in hand,
we are able to calculate the expectation and uncertainty of the physical observables
for a given physical state and thereby to extract the physics in WDW theory.

The same representation of the algebra of Dirac observables can be
obtained by the more systematic group averaging method
\cite{Marolf:1995cn,Marolf:1994wh,Marolf:1994ss,Marolf:1994nz,Hartle:1997dc},
as has been done for the isotropic model in \cite{Ashtekar:2006uz}.
There may be some complications in anisotropic cases and we defer this
issue to future study.

\subsection{The Semi-Classical (Coherent) States}\label{subsec:semi-classical}
With the physical Hilbert space and a complete set of Dirac observables available,
we can introduce semi-classical (coherent) states and study their evolution.
We first fix the time at $\ph=\ph_o$ to construct the semi-classical state which is peaked
at $p_\ph=p_\ph^\star$ and ${\vec \m}|_{\ph_o}={\vec \m}_{\ph_o}^\star$.
To have a semi-classical state corresponding to a universe with the characteristic quantities
much larger than Planck scales, we demand that $\m_{I\ph_o}^\star\gg\muzero_I$ and $p_\ph^\star\gg\hbar\sqrt{G}$
(or equivalently ${\cal K}\k_\ph\gg1$).
At the initial time $\ph=\ph_o$, consider the coherent state:
\be\label{eqn:semi-classical state}
\Ps_{{\vec \m}_{\ph_o}^\star,p_\ph^\star}({\vec \m},\ph_o)=
\int_{\o^2({\vec k})>0}\!\!d^3k\
\tilde{\Ps}_{\vec \k}({\vec k})\
\underline{e}_{\vec k}({\vec \m})\,
e^{i\o({\vec k})(\ph_o-\ph_o^\star)},
\ee
where the Fourier amplitude $\tilde{\Ps}_{\vec \k}({\vec k})$ is given by
\be\label{eqn:Fourier amplitude}
\tilde{\Ps}_{\vec \k}({\vec k})=\frac{1}{(2\p)^{3/4}}\frac{1}{\sqrt{\s_1\s_2\s_3}}\
e^{-\frac{(k_1-{\cal K}\k_1)^2}{(2\s_1)^2}}\,
e^{-\frac{(k_2-{\cal K}\k_2)^2}{(2\s_2)^2}}\,
e^{-\frac{(k_3-{\cal K}\k_3)^2}{(2\s_3)^2}},
\ee
which is sharply (if $\s_i\ll{\cal K}\k_i$) peaked at ${\vec k}={\cal K}{\vec \k}$.
The constants ${\vec \m}_{\ph_o}^\star$ and $\ph_o^\star$ are related by
\ba\label{eqn:mu-phi0-star}
{\m_1}_{\ph_o}^\star&=&e^{\frac{1}{(2\s_1)^2}}\,e^{\sqrt{8\p G}\big(\frac{\k_2+\k_3}{\k_\ph}\big)(\ph_o^\star-\ph_o)}
=e^{\frac{1}{(2\s_1)^2}}\,e^{\sqrt{8\p G}\big(\frac{1-\k_1}{\k_\ph}\big)(\ph_o^\star-\ph_o)},\nn\\
{\m_2}_{\ph_o}^\star&=&e^{\frac{1}{(2\s_2)^2}}\,e^{\sqrt{8\p G}\big(\frac{\k_1+\k_3}{\k_\ph}\big)(\ph_o^\star-\ph_o)}
=e^{\frac{1}{(2\s_1)^2}}\,e^{\sqrt{8\p G}\big(\frac{1-\k_2}{\k_\ph}\big)(\ph_o^\star-\ph_o)},\nn\\
{\m_3}_{\ph_o}^\star&=&e^{\frac{1}{(2\s_3)^2}}\,e^{\sqrt{8\p G}\big(\frac{\k_1+\k_2}{\k_\ph}\big)(\ph_o^\star-\ph_o)}
=e^{\frac{1}{(2\s_1)^2}}\,e^{\sqrt{8\p G}\big(\frac{1-\k_3}{\k_\ph}\big)(\ph_o^\star-\ph_o)},
\ea
which are suggestive of the classical solution \eqnref{eqn:classical sol p-phi}.
Similarly, referring to \eqnref{eqn:momentum-phi}, we set
\be\label{eqn:rel of p-phi and kappa-ph}
p_\ph^\star=\hbar\sqrt{8\p G}\,{\cal K}\k_\ph.
\ee
The parameters $\k_{1,2,3}$ and $\k_\ph$ satisfy the constraint \eqnref{eqn:parameter constraint}.

As the action of the Dirac observables on the initial states and the
physical inner product are well defined in \eqnref{eqn:Dirac
observables} and \eqnref{eqn:WDW inner product}, in
\appref{sec:coherent states}, we are able to calculate the
expectations and dispersions of the Dirac observables. The results
are:
\begin{itemize}
\item
The expectation value of $\widehat{{\m_I}|_{\ph_o}}$:
\be\label{eqn:mean of mu}
\langle\,{{\m_I}|_{\ph_o}}\rangle:=
\bra{\Ps_{{\vec \m}_{\ph_o}^\star,p_\ph^\star}}
\,\widehat{{\m_I}|_{\ph_o}}\,
\ket{\Ps_{{\vec \m}_{\ph_o}^\star,p_\ph^\star}}_{\rm phy}
\approx\m_{I\ph_o}^\star.
\ee
\item
The dispersion square of $\widehat{{\m_I}|_{\ph_o}}$:
\ba\label{eqn:variance of mu}
\D^2({\m_I}|_{\ph_o})&\equiv&
\langle\,({\m_I}|_{\ph_o})^2\rangle-
\langle\,{{\m_I}|_{\ph_o}}\rangle^2
:=
\bra{\Ps_{{\vec \m}_{\ph_o}^\star,p_\ph^\star}}
\left(\widehat{{\m_I}|_{\ph_o}}\right)^2
\ket{\Ps_{{\vec \m}_{\ph_o}^\star,p_\ph^\star}}_{\rm phy}
-\langle\,{{\m_I}|_{\ph_o}}\rangle^2\nn\\
&\approx&
(e^{\frac{1}{2\s_I^2}}-1)(\m_{I\ph_o}^\star)^2.
\ea
\item
The expectation value of $\hat{p}_\ph$:
\be\label{eqn:mean of p-phi}
\langle p_\ph\rangle:=
\bra{\Ps_{{\vec \m}_{\ph_o}^\star,p_\ph^\star}}
\,\hat{p}_\ph\,
\ket{\Ps_{{\vec \m}_{\ph_o}^\star,p_\ph^\star}}_{\rm phy}
\approx p_\ph^\star.
\ee
\item
The dispersion square of $\hat{p}_\ph$:
\ba\label{eqn:variance of p-phi}
\D^2(p_\ph)&\equiv&
\langle p_\ph^2\rangle-
\langle p_\ph\rangle^2
:=
\bra{\Ps_{{\vec \m}_{\ph_o}^\star,p_\ph^\star}}
\,\hat{p}_\ph^2\,
\ket{\Ps_{{\vec \m}_{\ph_o}^\star,p_\ph^\star}}_{\rm phy}
-\langle p_\ph\rangle^2\nn\\
&\approx&
\frac{2p_\ph^{\star2}}{{\cal K}^2\k_\ph^4}
\big(\s_1^2(\k_2+\k_3)^2+\s_2^2(\k_1+\k_3)^2+\s_3^2(\k_1+\k_2)^2\big).
\ea
\end{itemize}

As the evolution of the coherent state is concerned, because of \eqnref{eqn:WDW Schrodinger eq},
the solution $\Ps_{{\vec \m}_{\ph}^\star,p_\ph^\star}({\vec \m},\ph)$
defined by the initial state $\Ps_{{\vec \m}_{\ph_o}^\star,p_\ph^\star}({\vec \m},\ph_o)$
can be obtained simply by replacing $\ph_o$ with $\ph$ in \eqnref{eqn:semi-classical state}!
This implies that when the state is evolved to the instant $\ph$, it is again a semi-classical state
peaked at
\be\label{eqn:mu-phi-star}
\langle\,{{\m_I}|_{\ph}}\rangle\approx
e^{\frac{1}{(2\s_1)^2}}\,e^{\sqrt{8\p G}\big(\frac{1-\k_I}{\k_\ph}\big)(\ph_o^\star-\ph)}
\equiv{\m_I}_{\ph}^\star
\ee
and with the uncertainty spread:
\be
\D^2({\m_I}|_{\ph_o})\approx
(e^{\frac{1}{2\s_I^2}}-1)(\m_{I\ph}^\star)^2.
\ee
Comparing \eqnref{eqn:mu-phi0-star} and \eqnref{eqn:mu-phi-star},
we have
\be
{\m_I}_{\ph}^\star=
{\m_I}_{\ph_o}^\star\,e^{\sqrt{8\p G}\big(\frac{1-\k_I}{\k_\ph}\big)(\ph-\ph_o)},
\ee
which shows that the semi-classical state continues to be peaked at the trajectory
exactly described by a
classical solution as in \eqnref{eqn:classical sol p-phi}.
On the other hand, $\langle p_\ph\rangle$ and $\D^2(p_\ph)$ are independent of time.

The fact that the semi-classical states follow the classical trajectories
implies that they are destined for the singularity (Kasner-like or not) in the backward evolution.
In this sense, WDW evolution does \emph{not}
resolve the classical singularity.

\ \newline \small \textbf{Remark:} To get the semi-classical
(coherent) state in the context of LQC, we simply replace
$\underline{e}_{\vec k}({\vec \m})$ in \eqnref{eqn:semi-classical
state} with ${e}_{\vec k}^{(s)}({\vec \m})$ (the symmetrized
eigenfunctions of $\Th$, to be defined in \secref{subsec:LQC general
solutions}). Since ${e}_{\vec k}^{(s)}({\vec
\m})\rightarrow\underline{e}_{\vec k}({\vec \m})$ for
$\abs{\m_I}\gg\muzero_I$, the wavefunction given by
\eqnref{eqn:semi-classical state} is used as the initial state at
late times to conduct the numerical analysis for LQC \cite{in
progress}. See \secref{subsec:LQC physical sector} for more details
of LQC semi-classical states. \hfill $\Box$\normalsize

\section{Analytical Issues in LQC ($\muzero$-scheme)}\label{sec:analytical issues}

In this section, we come back to the model in LQC and more detailed
analytical issues are explored. We include the matter sector to the
Hamiltonian and show that the scalar field $\ph$ can again be used
as emergent time. The total Hamiltonian constraint gives a
difference revolution equation and the general solutions to it are
closely examined. The results enable us to construct the physical
Hilbert and the Dirac observables. In \secref{subsec:resolution of
singularity}, we also prove that the state in the kinematical
Hilbert space associated with the classical singularity is
completely decoupled from the evolution, which adds evidence to
the occurrence of the big bounce.

\subsection{The Total Hamiltonian Constraint (Matter Sector Included)}\label{subsec:total Hamiltonian}
Since $C_{\ph}=8\p G\,p^2_\ph/\sqrt{\abs{p_1p_2p_3}}$ in \eqnref{eqn:Hamiltonian},
the corresponding operator $\hat{C}_\ph$ in LQC acting on the states of
$\hilbert_{\rm kin}^{\rm total}:=\hilbert^S_{\rm grav}\otimes\hilbert_{\ph}$
is given by
\be
\hat{C}_\ph\ket{\vec{\m}}\otimes\ket{\ph}=
\widehat{\left[{\frac{1}{\sqrt{\abs{p_1}}}}\right]}
\widehat{\left[{\frac{1}{\sqrt{\abs{p_2}}}}\right]}
\widehat{\left[{\frac{1}{\sqrt{\abs{p_3}}}}\right]}
\ket{\vec{\m}}\otimes8\p G\hat{p}_\ph^2\ket{\ph}.
\ee
According to \eqnref{eqn:triad operator on states}, this gives
\be
\hat{C}_\ph^{(\muzero)}\ket{\vec{\m}}\otimes\ket{\ph}=
\frac{1}{(4\p\g)^\frac{3}{2}\Pl^3}
{B}(\vec{\m})\ket{\vec{\m}}\otimes 8\p G\hat{p}_\ph^2\ket{\ph},
\ee
where
\ba
{B}(\vec{\m})&:=&\frac{1}{\muzero_1\muzero_2\muzero_3}\
\abs{\abs{\m_1+\muzero_1}^\frac{1}{2}-\abs{\m_1-\muzero_1}^\frac{1}{2}}\nn\\
&&\quad\times\abs{\abs{\m_2+\muzero_2}^\frac{1}{2}-\abs{\m_2-\muzero_2}^\frac{1}{2}}\
\abs{\abs{\m_3+\muzero_3}^\frac{1}{2}-\abs{\m_3-\muzero_3}^\frac{1}{2}}.
\ea
The total Hamiltonian constraint $\hat{C}_{\rm grav}^{(\muzero)}+\hat{C}_\ph^{(\muzero)}=0$
then yields
\be
8\p G\,B(\vec{\m})\,\hat{p}_\ph^2\,\ket{\vec{\m},\ph}+(4\p\g)^\frac{3}{2}\Pl^3
\hat{C}_{\rm grav}^{(\muzero)}\,\ket{\vec{\m},\ph}=0.
\ee

Both the classical and the WDW theories suggest that $\ph$ can be used as emergent time; i.e.,
$\ph$ is to be thought of as ``time'' while ${\vec \m}$ as the genuine, physical degrees of
freedom which evolve with respect to $\ph$.
To implement the total kinematical Hilbert space with this idea,
we choose the standard Schr\"{o}dinger representation for
$\ph$ but the ``polymer representation'' for $\vec{\m}$, which captures the quantum geometry effects;
that is
$\hilbert^{\rm total}_{\rm kin}=L^2(\mathbb{R}^3_{\rm Bohr},d^3\!\m_{\rm Bohr})\otimes L^2(\mathbb{R},d\ph)$.
In this representation, $\hat{p}_\ph=\frac{\hbar}{i}\frac{\partial}{\partial\ph}$
and the total Hamiltonian constraint reads as
\ba\label{eqn:master eq}
\frac{\partial^2}{\partial\ph^2}\Ps(\vec{\m},\ph)&=&
\frac{\p G}{2\,\muzero_1\muzero_2\muzero_3}[B(\vec{\m})]^{-1}
\Big\{
C_3^{++}(\vec{\m})\,\Ps(\m_1+2\muzero_1,\m_2+2\muzero_2,\m_3,\ph)\nn\\
&&\qquad\qquad\qquad\qquad\ -C_3^{-+}(\vec{\m})\,\Ps(\m_1-2\muzero_1,\m_2+2\muzero_2,\m_3,\ph)\nn\\
&&\qquad\qquad\qquad\qquad\ -C_3^{+-}(\vec{\m})\,\Ps(\m_1+2\muzero_1,\m_2-2\muzero_2,\m_3,\ph)\nn\\
&&\qquad\qquad\qquad\qquad\ +C_3^{--}(\vec{\m})\,\Ps(\m_1-2\muzero_1,\m_2-2\muzero_2,\m_3,\ph)\nn\\
&&\qquad\qquad\qquad\qquad+\text{\ cyclic terms}
\Big\}\nn\\
&=:&-\left(\Th_3(\m_1,\m_2)+\Th_2(\m_1,\m_3)+\Th_1(\m_2,\m_3)\right)\Ps(\vec{\m},\ph)\nn\\
&=:&-\Th({\vec \m})\Ps(\vec{\m},\ph),
\ea
where
\ba
C_3^{++}(\vec{\m})&:=&
\abs{\sqrt{\abs{(\m_1+2\muzero_1)\m_2(\m_3+\muzero_3)}}-\sqrt{\abs{(\m_1+2\muzero_1)\m_2(\m_3-\muzero_3)}}}\nn\\
&&+\abs{\sqrt{\abs{\m_1(\m_2+2\muzero_2)(\m_3+\muzero_3)}}-\sqrt{\abs{\m_1(\m_2+2\muzero_2)(\m_3-\muzero_3)}}},\nn\\
C_3^{--}(\vec{\m})&:=&
\abs{\sqrt{\abs{(\m_1-2\muzero_1)\m_2(\m_3+\muzero_3)}}-\sqrt{\abs{(\m_1-2\muzero_1)\m_2(\m_3-\muzero_3)}}}\nn\\
&&+\abs{\sqrt{\abs{\m_1(\m_2-2\muzero_2)(\m_3+\muzero_3)}}-\sqrt{\abs{\m_1(\m_2-2\muzero_2)(\m_3-\muzero_3)}}},\nn\\
C_3^{+-}(\vec{\m})&:=&
\abs{\sqrt{\abs{(\m_1+2\muzero_1)\m_2(\m_3+\muzero_3)}}-\sqrt{\abs{(\m_1+2\muzero_1)\m_2(\m_3-\muzero_3)}}}\nn\\
&&+\abs{\sqrt{\abs{\m_1(\m_2-2\muzero_2)(\m_3+\muzero_3)}}-\sqrt{\abs{\m_1(\m_2-2\muzero_2)(\m_3-\muzero_3)}}},\nn\\
C_3^{-+}(\vec{\m})&:=&
\abs{\sqrt{\abs{(\m_1-2\muzero_1)\m_2(\m_3+\muzero_3)}}-\sqrt{\abs{(\m_1-2\muzero_1)\m_2(\m_3-\muzero_3)}}}\nn\\
&&+\abs{\sqrt{\abs{\m_1(\m_2+2\muzero_2)(\m_3+\muzero_3)}}-\sqrt{\abs{\m_1(\m_2+2\muzero_2)(\m_3-\muzero_3)}}}
\ea
and similar for $C_2^{\pm\pm}$ and $C_1^{\pm\pm}$ in the cyclic manner.
The coefficients $C^{\pm\pm}_I$ have the following properties:
\ba\label{eqn:coefficients C 1}
&&C^{++}_3(0,\m_2,\m_3)=C^{+-}_3(0,\m_2,\m_3)=C^{-+}_3(0,\m_2,\m_3)=C^{--}_3(0,\m_2,\m_3),\nn\\
&&C^{++}_3(\m_1,0,\m_3)=C^{+-}_3(\m_1,0,\m_3)=C^{-+}_3(\m_1,0,\m_3)=C^{--}_3(\m_1,0,\m_3),
\ea
and
\ba\label{eqn:coefficients C 2}
&&C^{++}_3(0,0,\m_3)=C^{+-}_3(0,0,\m_3)=C^{-+}_3(0,0,\m_3)=C^{--}_3(0,0,\m_3)=0,\nn\\
&&C^{++}_3(\m_1,\m_2,0)=C^{+-}_3(\m_1,\m_2,0)=C^{-+}_3(\m_1,\m_2,0)=C^{--}_3(\m_1,\m_2,0)=0,
\ea
for $\forall\ \m_1,\m_2,\m_3$.
($C^{\pm\pm}_{1,2}$ have the similar properties in the cyclic manner.)
For later convenience, we also define
\be
\widetilde{C}_I^{\pm\pm}(\m_J,\m_K)
:=\frac{\p G}{2\,\muzero_1\muzero_2\muzero_3}[B(\vec{\m})]^{-1}C_I^{\pm\pm}(\vec{\m}),
\ee
which is independent of $\m_I$ and has the properties:
\be\label{eqn:coefficients tilde C 1}
\tilde{C}^{++}_3(-2\muzero_1,-2\muzero_2)=\tilde{C}^{+-}_3(-2\muzero_1,2\muzero_2)=
\tilde{C}^{-+}_3(2\muzero_1,-2\muzero_2)=\tilde{C}^{--}_3(2\muzero_1,2\muzero_2)=0,
\ee
\be\label{eqn:coefficients tilde C 2}
\tilde{C}^{\pm\pm}_3(0,\m_2),\;\tilde{C}^{\pm\pm}_3(\m_1,0)\rightarrow\infty,
\ee
and
\be\label{eqn:coefficients tilde C 3}
\tilde{C}^{\pm\pm}_3(\m_1,\m_2)\neq0 \text{ or } \infty,\qquad \text{elsewhere},
\ee
for $\forall\ \m_1,\m_2$.
($\tilde{C}^{\pm\pm}_{1,2}$ have the similar properties in the cyclic manner.)
Also note that $\Th_I$ is independent of $\m_I$.
It is easy to see that $\Th$ and each of $\Th_I$ individually are self-adjoint
on $L^2({\mathbb R}^3_{\rm Bohr},B({\vec \m})\,d^3\!\m_{\rm Bohr})$.

\subsection{Emergent Time and the General Solutions}\label{subsec:LQC general solutions}
The resulting Hamiltonian euqation \eqnref{eqn:master eq} is cast in the way very
similar to that of the WDW constraint \eqnref{eqn:WDW master eq}.
The key difference is that the $\ph$-independent operator $\Th$ is now a \emph{difference} operator rather
than a differential operator.
In the same spirit as for WDW constraint, the LQC quantum Hamiltonian constraint can be regarded as an ``evolution
equation'' which evolves the quantum state in the \emph{emergent time} $\ph$.

As in the isotropic model, since $\Th$ is a difference operator, the space of physical states (i.e.
appropriate solutions to the constraint equation) is naturally divided into sectors, each of which
is preserved by the ``evolution'' and by the action of the Dirac observables.
Thus, there is \emph{super-selection} among these sectors.
Let ${\cal L}_{\abs{\e_I}}$ ($\abs{\e_I}\leq\muzero_I$)
be the ``lattice'' of points $\{\abs{\e_I}+2n\muzero_I;n\in\mathbb{Z}\}$ on the $\m_I$-axis,
${\cal L}_{-\abs{\e_I}}$  be the ``lattice'' of points $\{-\abs{\e_I}+2n\muzero_I;n\in\mathbb{Z}\}$
and ${\cal L}_{\e_I}={\cal L}_{\abs{\e_I}}\bigcup{\cal L}_{-\abs{\e_I}}$.
Also let $\hilbert^{\rm grav}_{\pm\abs{\e_1},\pm\abs{\e_2},\pm\abs{\e_3}}$ (8 of them) and
$\hilbert^{\rm grav}_{\vec \e}$ denote
the physical subspaces of $L^2(\mathbb{R}^3_{\rm Bohr},B({\vec \m})\,d^3\!\m_{\rm Bohr})$
with states whose support is restricted to the lattices
${\cal L}_{\pm\abs{\e_1}}\otimes{\cal L}_{\pm\abs{\e_2}}\otimes{\cal L}_{\pm\abs{\e_3}}$
and
${\cal L}_{{\e_1}}\otimes{\cal L}_{{\e_2}}\otimes{\cal L}_{{\e_3}}$ respectively.
Each of these subspaces is mapped to itself by $\Th$. Furthermore, since $\hat{C}^{(\muzero)}_{\rm grav}$
is self-adjoint on $\hilbert^{\rm grav}_{\rm kin}=L^2(\mathbb{R}^3_{\rm Bohr},d^3\!\m_{\rm Bohr})$,
it follows that $\Th$ is self-adjoint on all these physical Hilbert subspaces.

Note however that since ${\cal L}_{\abs{\e_I}}$ and ${\cal L}_{-\abs{\e_I}}$ are mapped to each other by
the operators $\Pi_I$, none of $\hilbert^{\rm grav}_{\pm\abs{\e_1},\pm\abs{\e_2},\pm\abs{\e_3}}$
but only $\hilbert^{\rm grav}_{\vec \e}$ is left invariant by $\Pi_{1,2,3}$.
The operators $\Pi_I$ reverse the triad orientation and thus represent a large gauge transformation.
In gauge theories, we have to restrict ourselves to the sector corresponding to an eigenspace of
the group of large gauge transformations.
Since there are no fermions and no parity violating processes in our theory, we are led to choose
the symmetric sector (i.e. the subspace of $\hilbert^{\rm grav}_{\vec \e}$ in which
$\Ps({\vec \m})=\Ps(-\m_1,\m_2,\m_3)=\Ps(\m_1,-\m_2,\m_3)=\Ps(\m_1,\m_2,-\m_3)$), which will be
of our primary interest.

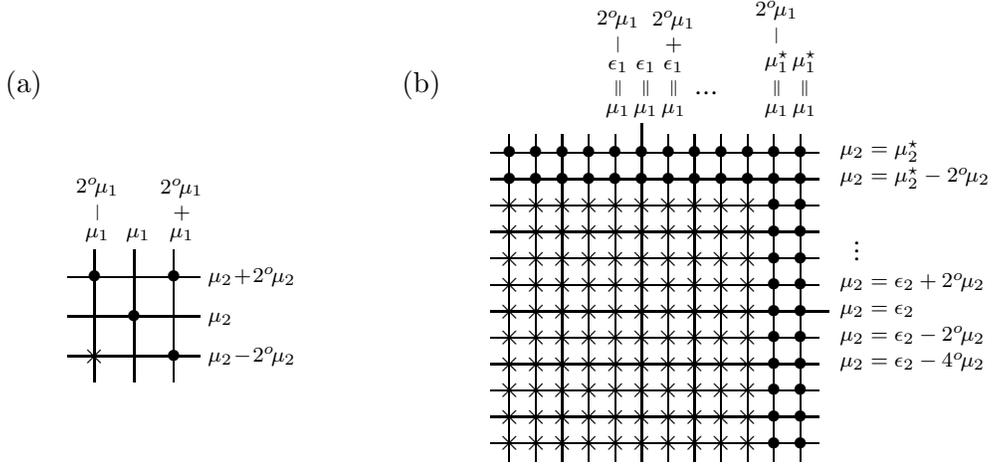
\begin{figure}
\begin{picture}(400,200)(0,0)

\put(10,150){(a)}
\put(160,150){(b)}

\put(30,40)
{
\begin{picture}(90,80)(0,0)

\multiput(0,10)(0,15){3}{\line(1,0){50}}
\multiput(10,0)(15,0){3}{\line(0,1){50}}

\multiput(10,40)(15,-15){3}{\makebox(0,0){$\bullet$}}
\put(40,40){\makebox(0,0){$\bullet$}}

\put(10,10){\makebox(0,0){$\times$}}

\put(53,8){\scriptsize $\m_2\!-\!2\muzero_2$}
\put(53,23){\scriptsize $\m_2$}
\put(53,38){\scriptsize $\m_2\!+\!2\muzero_2$}

\put(35,55){{\scriptsize \shortstack{$2\muzero_1$\\$+$\\$\m_1$}}}
\put(22,55){\scriptsize $\m_1$}
\put(3,55){{\scriptsize \shortstack{$2\muzero_1$\\{\tiny $|$}\\$\m_1$}}}


\end{picture}
}

\put(190,10)
{
\begin{picture}(200,180)(-2,-2)

\multiput(-2,5)(0,10){12}{\line(1,0){124}} 
\multiput(5,-2)(10,0){12}{\line(0,1){124}} 

\thicklines
\put(-2,55){\line(1,0){128}} 
\put(55,-2){\line(0,1){128}} 
\thinlines

\put(130,34){{\scriptsize $\m_2=\e_2-4\muzero_2$}}
\put(130,44){{\scriptsize $\m_2=\e_2-2\muzero_2$}}
\put(130,54){{\scriptsize $\m_2=\e_2$}}
\put(130,64){{\scriptsize $\m_2=\e_2+2\muzero_2$}}
\put(130,104){{\scriptsize $\m_2=\m_2^\star-2\muzero_2$}}
\put(130,114){{\scriptsize $\m_2=\m_2^\star$}}
\multiput(135,72)(0,3){3}{$\cdot$}

\put(52,130){{\scriptsize \shortstack{$\e_1$\\{\tiny $\|$}\\$\m_1$}}}
\put(59,130){{\scriptsize \shortstack{$2\muzero_1$\\$+$\\$\e_1$\\{\tiny $\|$}\\$\m_1$}}}
\put(38,130){{\scriptsize \shortstack{$2\muzero_1$\\{\tiny $|$}\\$\e_1$\\{\tiny $\|$}\\$\m_1$}}}
\put(98,130){{\scriptsize \shortstack{$2\muzero_1$\\{\tiny $|$}\\$\m_1^\star$\\{\tiny $\|$}\\$\m_1$}}}
\put(112,130){{\scriptsize \shortstack{$\m_1^\star$\\{\tiny $\|$}\\$\m_1$}}}
\multiput(75,135)(3,0){3}{$\cdot$}

\multiput(5,115)(10,0){12}{\makebox(0,0){$\bullet$}}
\multiput(5,105)(10,0){12}{\makebox(0,0){$\bullet$}}
\multiput(115,5)(0,10){10}{\makebox(0,0){$\bullet$}}
\multiput(105,5)(0,10){10}{\makebox(0,0){$\bullet$}}

\multiput(0,0)(0,10){10}{
 \multiput(5,5)(10,0){10}{\makebox(0,0){$\times$}}
}

\end{picture}
}

\end{picture}
 \caption{(\textbf{a}): Given the values of $e_{\l_{12}}$ at the lattice points marked
 $\bullet$, the value at the point marked $\times$ can be uniquely determined
 via the recursion relation \eqnref{eqn:recursion relation}.
 (\textbf{b}): If the values of $e_{\l_{12}}$ are specified on the boundary points
 (marked $\bullet$), all the values at the points marked $\times$ can be obtained
 by iterating the scheme depicted in (a).
 }\label{fig:fig1}
\end{figure}

To explore the properties of the operator $\Th$, Let us first consider a \emph{generic} $\vec{\e}$
(i.e. none of $\e_I$ equals $0$ or $\muzero_I$).
On each of the 8 Hilbert space $\hilbert^{\rm grav}_{\pm\abs{\e_1},\pm\abs{\e_2},\pm\abs{\e_3}}$,
we can solve for the eigenvalue equation
$\Th\,e_\l({\vec \m})=(\Th_1+\Th_2+\Th_3)\,e_\l({\vec \m})=\l\,e_\l({\vec \m})$.
Since $\Th_I$ is independent of $\m_I$, we solve the eigenvalue problem by
separating the variables; that is, let
$e_\l(\m_1,\m_2,\m_3)=e_{\l_1}(\m_1)\,e_{\l_2}(\m_2)\,e_{\l_3}(\m_3)$ and solve
\ba\label{eqn:eigenvalue problems}
\Th_3\,e_{\l_{12}}(\m_1,\m_2)&=&\l_{12}\,e_{\l_{12}}(\m_1,\m_2),
\quad\text{for  }
e_{\l_{12}}(\m_1,\m_2)=e_{\l_1}(\m_1)\,e_{\l_2}(\m_2),\nn\\
\Th_2\,e_{\l_{13}}(\m_1,\m_3)&=&\l_{13}\,e_{\l_{13}}(\m_1,\m_3),
\quad\text{for  }
e_{\l_{13}}(\m_1,\m_3)=e_{\l_1}(\m_1)\,e_{\l_3}(\m_3),\nn\\
\Th_1\,e_{\l_{23}}(\m_2,\m_3)&=&\l_{23}\,e_{\l_{23}}(\m_2,\m_3),
\quad\text{for  }
e_{\l_{23}}(\m_2,\m_3)=e_{\l_2}(\m_2)\,e_{\l_3}(\m_3),
\ea
with $\l=\l_{12}+\l_{13}+\l_{23}$.
Each of \eqnref{eqn:eigenvalue problems} can be solved recursively for any arbitrary $\l_{ij}$
provided that the appropriate boundary condition is given.
For example, the detail of
$\Th_3\,e_{\l_{12}}(\m_1,\m_2)=\l_{12}\,e_{\l_{12}}(\m_1,\m_2)$ reads as
\ba\label{eqn:recursion relation}
&&\widetilde{C}_3^{++}(\m_1,\m_2)\,e_{\l_{12}}(\m_1+2\muzero_1,\m_2+2\muzero_2)
-\widetilde{C}_3^{-+}(\m_1,\m_2)\,e_{\l_{12}}(\m_1-2\muzero_1,\m_2+2\muzero_2)\nn\\
&&-\widetilde{C}_3^{+-}(\m_1,\m_2)\,e_{\l_{12}}(\m_1+2\muzero_1,\m_2-2\muzero_2)
+\widetilde{C}_3^{--}(\m_1,\m_2)\,e_{\l_{12}}(\m_1-2\muzero_1,\m_2-2\muzero_2)\nn\\
&=&-\l_{12}\,e_{\l_{12}}(\m_1,\m_2).
\ea
and since the coefficients $\widetilde{C}_3^{\pm\pm}$
never vanish or go to infinity on the generic lattice
(see \eqnref{eqn:coefficients tilde C 3}),
$e_{\l_{12}}(\m_1-2\muzero_1,\m_2-2\muzero_2)$ can be obtained for any $\l_{12}$
when the values of $e_{\l_{12}}(\m_1+2\muzero_1,\m_2+2\muzero_2)$,
$e_{\l_{12}}(\m_1-2\muzero_1,\m_2+2\muzero_2)$, $e_{\l_{12}}(\m_1+2\muzero_1,\m_2-2\muzero_2)$ and
$e_{\l_{12}}(\m_1,\m_2)$ are given. (See \mbox{\figref{fig:fig1}.a}.)
Therefore, if the values of $e_{\l_{12}}(\m_1,\m_2)$ are specified on the boundary lines
$\m_1=\m_1^\star$, $\m_1=\m_1^\star-2\muzero_1$ and $\m_2=\m_2^\star$, $\m_2=\m_2^\star-2\muzero_2$
in the lattice ${\cal L}_{\pm\abs{\e_1}}\otimes{\cal L}_{\pm\abs{\e_2}}$,
$e_{\l_{12}}(\m_1,\m_2)$ can be uniquely determined via the recursion relation \eqnref{eqn:recursion relation}
for all the lattice points with $\m_1<\m_1^\star$ and $\m_2<\m_2^\star$.
(See \figref{fig:fig1}.b.)

On the other hand, the LQC operator $\Th_I$ reduces to the WDW operator $\underline{\Th}_I$
in the ``asymptotic'' regime (i.e. $\abs{\m_I}\gg\muzero_I$), and thus the LQC eigenfunction
$e_\l({\vec \m})$ approaches the WDW eigenfunction $\underline{e}_{\vec k}({\vec \m})$
given by \eqnref{eqn:eigenfunction ek} for large $\abs{\m_I}$. Therefore, we can use the same
labels $k_i$ to denote the LQC eigenfunctions (namely, $\l_{1,2,3}=k_{1,2,2}$) and similarly
decompose $\underline{e}_{\vec k}({\vec \m})$ as
$\underline{e}_{k_1}\!(\m_1)\,\underline{e}_{k_2}\!(\m_2)\,\underline{e}_{k_3}\!(\m_3)$.
It follows that $e_{k_I}(\m_I)\rightarrow\underline{e}_{k_I}(\m_I)$ for $\abs{\m_I}\gg\muzero_I$
and we have the identifications:
\be\label{eqn:lambda ij}
\l_{12}=16\p G\Big(k_1k_2+\frac{1}{16}\Big),\
\l_{13}=16\p G\Big(k_1k_3+\frac{1}{16}\Big),\
\l_{23}=16\p G\Big(k_2k_3+\frac{1}{16}\Big),
\ee
and
\be\label{eqn:lambda omega}
\l=\o^2({\vec k})=16\p G\left(k_1k_2+k_1k_3+k_2k_3+\frac{3}{16}\right).
\ee

Like $\underline{\Th}$ in the WDW theory,
the LQC operator $\Th$ is not positive definite.
For the same argument we used in \secref{subsec:WDW general solutions},
only those modes with $\o^2>0$ are
allowed and what matters is just the operator $\Th_+$,
obtained by projecting the action of $\Th$
to the positive eigenspace in its spectral decomposition.

For a given ${\vec k}$, we can find $e_{\l_{IJ}}$ as mentioned above
by appropriately specifying the boundaries
in such a way that $e_{\l_{IJ}}\!(\m_I,\m_J)\rightarrow
\underline{e}_{k_I}\!(\m_I)\,\underline{e}_{k_J}\!(\m_j)$ for
$\abs{\m_I}\gg\muzero_I$, $\abs{\m_J}\gg\muzero_J$.
However, \eqnref{eqn:eigenfunction ek} gives
$\underline{e}_{k_1}\!(\m_1)\,\underline{e}_{k_2}\!(\m_2)
=(4\p)^{-2}\abs{\m_1}^{-1/4+ik_1}\abs{\m_2}^{-1/2+ik_2}$,
which is not suitable to be used for the boundary condition, since
one of $\m_1$ and $\m_2$ could be small on the boundary
lines $\m_1=\m_1^\star$, $\m_1=\m_1^\star-2\muzero_1$ and $\m_2=\m_2^\star$, $\m_2=\m_2^\star-2\muzero_2$
and thus the WDW limit is not always valid even if both $\m_1^\star$ and $\m_1^\star$ are very big.
To reflect the effect of quantum geometry for small $\m_1$ or $\m_2$,
we follow the same strategy \`{a} la Thiemann's trick used for the
inverse triad operators in \secref{subsec:triad operators} and
designate the ``tamed'' function
\be\label{eqn:modified boundary}
\frac{1}{(4\p)^2}
\left(\frac{1}{\muzero_1}\abs{\abs{\m_1\!+\!\muzero_1}^\frac{1}{2}
\!-\!\abs{\m_1\!-\!\muzero_1}^\frac{1}{2}}\right)^{\!\!\frac{1}{2}-2ik_1}
\!\!\!\!\!\times
\left(\frac{1}{\muzero_2}\abs{\abs{\m_2\!+\!\muzero_2}^\frac{1}{2}
\!-\!\abs{\m_2\!-\!\muzero_2}^\frac{1}{2}}\right)^{\!\!\frac{1}{2}-2ik_2}
\ee
as the value of $e_{\l_{12}}(\m_1,\m_2)$
on the boundary, which again reduces to
$\underline{e}_{k_i}\!(\m_I)\,\underline{e}_{k_j}\!(\m_J)$ when both
$\abs{\m_i}\gg\muzero_i$ and $\abs{\m_j}\gg\muzero_j$.
In practice, we use \eqnref{eqn:modified boundary} to specify the boundary condition on the lines
$(\m_1=\m_1^\star,\,-\m_2^\star\leq\m_2\leq\m_2^\star)$,
$(\m_1=\m_1^\star-2\muzero_1,\,-\m_2^\star\leq\m_2\leq\m_2^\star)$,
$(-\m_1^\star\leq\m_1\leq\m_1^\star,\,\m_2=\m_2^\star)$ and
$(-\m_1^\star\leq\m_1\leq\m_1^\star,\,\m_2=\m_2^\star-2\muzero_2)$ of the lattice
for $\m_1^\star\gg\muzero_1$ and $\m_2^\star\gg\muzero_2$,
then the recursion relation \eqnref{eqn:recursion relation} uniquely gives the values
$e_{\l_{12}}\!(\m_1,\m_2)$ for all the lattice points
within the range: $-\m_1^\star\leq\m_1\leq\m_1^\star$ and $-\m_2^\star\leq\m_2\leq\m_2^\star$.
The eigenfunctions so obtained are denoted as $e^{++}_{\l_{12}}$, $e^{+-}_{\l_{12}}$ \ldots
for the lattices ${\cal L}_\abs{\e_1}\otimes{\cal L}_\abs{\e_2}$,
${\cal L}_\abs{\e_1}\otimes{\cal L}_{-\abs{\e_2}}$ \ldots
respectively. We are only interested in the symmetric sector
and have to symmetrize the eigenfunctions to get:
\ba\label{eqn:symmetrization}
e^{(s)}_{\l_{12}}(\m_1,\m_2)&=&\frac{1}{\sqrt{16}}
\Big(
e^{++}_{\l_{12}}\!(\m_1,\m_2)+e^{++}_{\l_{12}}\!(\m_1,-\m_2)
+e^{++}_{\l_{12}}\!(-\m_1,\m_2)+e^{++}_{\l_{12}}\!(-\m_1,-\m_2)\nn\\
&&\quad+\cdots\nn\\
&&\quad+
e^{--}_{\l_{12}}\!(\m_1,\m_2)+e^{--}_{\l_{12}}\!(\m_1,-\m_2)
+e^{--}_{\l_{12}}\!(-\m_1,\m_2)+e^{--}_{\l_{12}}\!(-\m_1,-\m_2)
\Big).\qquad
\ea
The same procedure is applied for $e^{(s)}_{\l_{13}}\!(\m_1,\m_3)$ and $e^{(s)}_{\l_{23}}\!(\m_2,\m_3)$.
Finally, the LQC eigenfunction $e^{(s)}_{\vec k}(\vec \m)$ in the symmetric sector
can be determined for a given $\vec{k}$ by obtaining $e_{\l_{12}}$, $e_{\l_{12}}$ and $e_{\l_{12}}$
separately.

The eigenfunctions in the symmetric sector of $\hilbert^{\rm grav}_{\vec \e}$ can be normalized such
that we have the orthonormality relation:
\be\label{eqn:LQC orthonormality rel}
\sum_{{\vec \m}\in{\cal L}_{\e_1}\otimes{\cal L}_{\e_2}\otimes{\cal L}_{\e_3}}\!\!\!\!\!
B({\vec \m})\,\bar{e}^{(s)}_{\vec k}({\vec \m})\,e^{(s)}_{{\vec k}'}({\vec \m})
=\d^3({\vec k},{\vec k}'),
\ee
(where the right hand side is the standard Dirac distribution) and the completeness relation:
\be\label{eqn:LQC completeness rel}
\sum_{{\vec \m}\in{\cal L}_{\e_1}\otimes{\cal L}_{\e_2}\otimes{\cal L}_{\e_3}}\!\!\!\!\!
B({\vec \m})\,\bar{e}^{(s)}_{\vec k}({\vec \m})\Ps({\vec \m})
=0
\quad\text{for }
\forall{\vec k},
\quad\text{iff }
\Ps({\vec \m})=0
\ee
for any symmetric $\Ps({\vec \m})\in\hilbert^{\rm grav}_{\vec \e}$.
Finally, any symmetric element $\Ps({\vec \m})$ of $\hilbert^{\rm grav}_{\vec \e}$
can be expanded as
\be
\Ps({\vec \m})=\int_{\o^2({\vec k})>0}\!\!\!\!d^3k\,\tilde{\Ps}({\vec k})\,e^{(s)}_{{\vec k}}({\vec \m})
\ee
for $e^{(s)}_{{\vec k}}({\vec \m})\in\hilbert^{\rm grav}_{\vec \e}$.

As in the WDW theory, the general symmetric solution to the LQC constraint \eqnref{eqn:master eq}
can be written as
\be\label{eqn:sol to master eq}
\Ps({\vec \m},\ph)=\int_{\o^2({\vec k})>0}\!\!d^3k
\left(
\tilde{\Ps}_+({\vec k})\,e^{(s)}_{\vec k}({\vec \m})\,e^{i\o\ph}+
\tilde{\Ps}_-({\vec k})\,\bar{e}^{(s)}_{\vec k}({\vec \m})\,e^{-i\o\ph}
\right)
\ee
where $\tilde{\Ps}_\pm({\vec k})$ are in $L^2(\mathbb{R}^3,d^3k)$.
For $\abs{\m_I}\gg\muzero_I$, \eqnref{eqn:sol to master eq}
reduces to the solution \eqnref{eqn:sol to WDW master eq}
to the WDW equation.

If $\tilde{\Ps}_{+/-}({\vec k})$ vanishes, we will say that the solution is negative/positive frequency.
Thus, every solution to \eqnref{eqn:master eq} admits a natural decomposition into positive- and
negative-frequency parts, each of which satisfies the first order differential
equation in $\ph$:
\be
\frac{\partial \Ps_\pm}{\partial \ph}=\pm i\sqrt{\Th_+}\ \Ps_\pm,
\ee
where $\sqrt{\Th_+}$ is the square-root of the positive self-adjoint operator $\Th_+$
defined via spectral decomposition.
Therefore, the solution to \eqnref{eqn:master eq} with the ``initial'' function
$\Ps_\pm({\vec \m},\ph_o)=f_\pm({\vec \m})$
is given by:
\be\label{eqn:sol given by initial function}
\Ps_\pm({\vec \m},\ph)=e^{\pm i \sqrt{\Th_+}\,(\ph-\ph_o)} \Ps_\pm({\vec \m},\ph_o).
\ee

\ \newline
\small
\textbf{Remark:}
In the above discussion, we considered a \emph{generic} $\vec{\e}$.
In the \emph{special} cases (i.e. $\e_I=0$ or $\e_I=\muzero_I$),
some differences arise because the individual lattices are
invariant under the reflection $\m_I\rightarrow-\m_I$ (i.e., the lattices
${\cal L}_\abs{\e_I}$ and ${\cal L}_{-\abs{\e_I}}$ coincide).
For example, let consider the case when $\e_1$ is generic and $\e_2$ is special.
We have ${\cal L}_\abs{\e_2}={\cal L}_{-\abs{\e_2}}={\cal L}_{\e_2}$ and $e_{\l_{12}}^{++}=e_{\l_{12}}^{+-}$.
Consequently, the symmetrization of \eqnref{eqn:symmetrization} should be modified to
\ba
e^{(s)}_{\l_{12}}(\m_1,\m_2)&=&\frac{1}{\sqrt{8}}
\Big(
e^{++}_{\l_{12}}\!(\m_1,\m_2)+e^{++}_{\l_{12}}\!(\m_1,-\m_2)
+e^{++}_{\l_{12}}\!(-\m_1,\m_2)+e^{++}_{\l_{12}}\!(-\m_1,-\m_2)\nn\\
&&\quad +e^{-+}_{\l_{12}}\!(\m_1,\m_2)+e^{-+}_{\l_{12}}\!(\m_1,-\m_2)
+e^{-+}_{\l_{12}}\!(-\m_1,\m_2)+e^{-+}_{\l_{12}}\!(-\m_1,-\m_2)
\Big).
\ea
Furthermore, if both $\e_1$ and $\e_2$ are special,
we have $e_{\l_{12}}^{++}=e_{\l_{12}}^{+-}=e_{\l_{12}}^{-+}=e_{\l_{12}}^{--}$ and
the symmetrization is performed as
\be
e^{(s)}_{\l_{12}}(\m_1,\m_2)=\frac{1}{\sqrt{4}}
\Big(
e^{++}_{\l_{12}}\!(\m_1,\m_2)+e^{++}_{\l_{12}}\!(\m_1,-\m_2)
+e^{++}_{\l_{12}}\!(-\m_1,\m_2)+e^{++}_{\l_{12}}\!(-\m_1,-\m_2)
\Big).
\ee

In the numerical computation, we can focus on the lattice of the special case
with ${\vec \e}={\vec \muzero}$ and take advantage of the fact that
${\cal L}_{+\abs{\e_I}}={\cal L}_{-\abs{\e_I}}$.
Therefore, we restrict ourselves to the positive $\m_{1,2,3}$ and
the tedious procedure of symmetrization can be avoided.
This way, the numerical computation is much more efficient.

There is a further subtlety when the recursion relation
\eqnref{eqn:recursion relation} is applied to the lattice ${\cal
L}_{\e_1}\otimes{\cal L}_{\e_2}$ with $\e_1=0$ or $\e_2=0$, because
the coefficients $\widetilde{C}_3^{\pm\pm}$ may vanish or go to
infinity at some lattice points and thus more careful consideration
is needed. This issue is closely related to the resolution of the
singularity and the detail of it is studied in
\secref{subsec:resolution of singularity}. \hfill $\Box$\normalsize

\subsection{Resolution of the Singularity and Comments on the Planar Collapse}
\label{subsec:resolution of singularity} What is the fate of the
classical singularity and planar collapse? In the classical
dynamics, the Kasner-like solutions (two of $\k_i$ positive, the
other negative) approach the Kasner-like singularity as
$\ph\rightarrow-\infty$ and the planar collapse as
$\ph\rightarrow-\infty$ (with $\k_\ph>0$), whereas the Kasner-unlike
solutions ($0<\k_1,\k_2,\k_3<1$) encounter the Kasner-unlike
singularity as $\ph\rightarrow-\infty$ and no planar collapse. Since
$1-\k_i>0$, according to \eqnref{eqn:classical sol p-phi}, the
singularity (both Kasner-like and Kasner-unlike) gives
$p_1,p_2,p_3\rightarrow0$, which corresponds to the state
$\ket{0,0,0}$ on $\hilbert_{\rm grav}^S$ and thus the point $(0,0)$
on the lattice ${\cal L}_{\e_1}\otimes{\cal L}_{\e_2}$. On the other
hand, by \eqnref{eqn:classical sol p-phi} again, the planar collapse
gives $(p_1,p_2,p_3)\rightarrow(\infty,\infty,\infty)$, which
corresponds to the state $\ket{\infty,\infty,\infty}$ on
$\hilbert_{\rm grav}^S$ and thus the point $(\infty,\infty)$ on the
lattice ${\cal L}_{\e_1}\otimes{\cal L}_{\e_2}$. We first show that
the state associated with the singularity is decoupled and thus the
quantum evolution does not break down; later, the behaviors of the
planar collapse are commented.

Unlike the classical theory, the quantum evolution does not stop at
the singularity. If one begins with a
\emph{generic} lattice, the discrete evolution determined by
\eqnref{eqn:master eq} just ``jumps'' over the states with $\m_I=0$
without encountering any subtleties. The subtleties occur only on
the \emph{special} lattice as remarked in the end of
\secref{subsec:LQC general solutions}. Thus, we revisit the
recursion relation \eqnref{eqn:recursion relation} for the two
special lattices: (i) ${\cal L}_{\e_1=0}\otimes{\cal L}_{\e_2\neq0}$
and (ii) ${\cal L}_{\e_1=0}\otimes{\cal L}_{\e_2=0}$.

On the special lattice ${\cal L}_{\e_1=0}\otimes{\cal L}_{\e_2\neq0}$,
the boundary condition given by \eqnref{eqn:modified boundary} implies that
$e_{\l_{12}}(\m_1=0,\m_2\rightarrow\infty)=0$
and thus the state $\ket{\m_1=0,\m_2\rightarrow\infty}$ on
$\hilbert_{\e_1=0,\pm\abs{\e_2}}^{\rm grav}$ (marked $\circ$ in \figref{fig:fig2})
decouples from the rest part.
On the other hand, the values of $e_{\l_{12}}(\m_1,\m_2)$
in the first and fourth quadrants
(marked $\times$ in \figref{fig:fig2}.a) can be obtained
by direct iterating \eqnref{eqn:recursion relation} while the recursion relation may not work for the
values at $\m_1=-2\muzero_1$ (points marked $\diamond$ in \figref{fig:fig2}.b)
since the coefficients $\tilde{C}^{\pm\pm}_3(0,\m_2)$ blow up by
\eqnref{eqn:coefficients tilde C 2}.
However, this problem is merely artificial.
To faithfully express the recursion relation,
\eqnref{eqn:recursion relation} can be rewritten as
\ba
&&{C}_3^{++}(\vec{\m})\,e_{\l_{12}}(\m_1+2\muzero_1,\m_2+2\muzero_2)
-{C}_3^{-+}(\vec{\m})\,e_{\l_{12}}(\m_1-2\muzero_1,\m_2+2\muzero_2)\nn\\
&&-{C}_3^{+-}(\vec{\m})\,e_{\l_{12}}(\m_1+2\muzero_1,\m_2-2\muzero_2)
+{C}_3^{--}(\vec{\m})\,e_{\l_{12}}(\m_1-2\muzero_1,\m_2-2\muzero_2)\nn\\
&=&-B({\vec{\m}})\l_{12}\,e_{\l_{12}}(\m_1,\m_2).
\ea
Provided $B(0,\m_2,\m_3)\l_{12}=0$
[Note: $B(0,\m_2,\m_3)=B(\m_1,0,\m_3)=B(\m_1,\m_2,0)=0$ by its nature],
the relation in \eqnref{eqn:coefficients C 1} then gives
\ba\label{eqn:revised recursion rel}
e_{\l_{12}}(2\muzero_1,\m_2+2\muzero_2)&=&
e_{\l_{12}}(2\muzero_1,\m_2-2\muzero_2)
+e_{\l_{12}}(2\muzero_1,\m_2+2\muzero_2)\nn\\
&&-e_{\l_{12}}(-2\muzero_1,\m_2+2\muzero_2),
\ea
which can be used to determine the values of $e_{\l_{12}}$ at the points marked $\diamond$
without difficulty.
Furthermore, if the boundary condition is given symmetric under $\m_1\rightarrow-\m_1$,
we then have $e_{\l_{12}}(2\muzero_1,\m_2+2\muzero_2)=e_{\l_{12}}(-2\muzero_1,\m_2+2\muzero_2)$
and \eqnref{eqn:revised recursion rel} yields
$e_{\l_{12}}(2\muzero_1,\m_2+2\muzero_2)=e_{\l_{12}}(2\muzero_1,\m_2-2\muzero_2)$,
which, as expected, tells that
the values at $\diamond$ are the exact mirrors (reflected by the line $\m_1=0$) of the lattice points
in the line $\m_1=2\muzero_1$.
Once the values at $\diamond$ are all obtained, the iteration can be continued to all
unmarked points in \figref{fig:fig2}.a.

\begin{figure}
\begin{picture}(460,200)(0,0)
\put(10,160){(a)} \put(245,160){(b)}

\put(25,10) {
\begin{picture}(200,180)(-2,-2)

\multiput(-2,5)(0,10){12}{\line(1,0){124}} 
\multiput(5,-2)(10,0){12}{\line(0,1){124}} 

\thicklines
\put(-2,55){\line(1,0){128}} 
\put(55,-2){\line(0,1){128}} 
\thinlines

\put(130,34){{\scriptsize $\m_2=\e_2-4\muzero_2$}}
\put(130,44){{\scriptsize $\m_2=\e_2-2\muzero_2$}}
\put(130,54){{\scriptsize $\m_2=\e_2$}} \put(130,64){{\scriptsize
$\m_2=\e_2+2\muzero_2$}} \put(130,104){{\scriptsize
$\m_2=\m_2^\star-2\muzero_2$}} \put(130,114){{\scriptsize
$\m_2=\m_2^\star$}} \multiput(135,72)(0,3){3}{$\cdot$}

\put(52,130){{\scriptsize \shortstack{$0$\\{\tiny $\|$}\\$\m_1$}}}
\put(56,130){{\scriptsize \shortstack{$\ \ 2\muzero_1$\\{\tiny
$\|$}\\$\m_1$}}} \put(30,130){{\scriptsize \shortstack{$-2\muzero_1\
\ $\\{\tiny $\|$}\\$\m_1$}}} \put(98,130){{\scriptsize
\shortstack{$2\muzero_1$\\{\tiny $|$}\\$\m_1^\star$\\{\tiny
$\|$}\\$\m_1$}}} \put(112,130){{\scriptsize
\shortstack{$\m_1^\star$\\{\tiny $\|$}\\$\m_1$}}}
\multiput(75,135)(3,0){3}{$\cdot$}

\multiput(5,115)(10,0){5}{\makebox(0,0){$\bullet$}}
\multiput(5,105)(10,0){5}{\makebox(0,0){$\bullet$}}
\multiput(65,115)(10,0){6}{\makebox(0,0){$\bullet$}}
\multiput(65,105)(10,0){6}{\makebox(0,0){$\bullet$}}
\multiput(115,5)(0,10){10}{\makebox(0,0){$\bullet$}}
\multiput(105,5)(0,10){10}{\makebox(0,0){$\bullet$}}

\put(55,105){\makebox(0,0){$\circ$}}
\put(55,115){\makebox(0,0){$\circ$}}

\multiput(50,0)(0,10){10}{
 \multiput(5,5)(10,0){5}{\makebox(0,0){$\times$}}
}

\multiput(45,5)(0,10){10}{\makebox(0,0){$\diamond$}}

\end{picture}
}

\put(260,10) {
\begin{picture}(200,180)(-2,-2)

\multiput(-2,5)(0,10){12}{\line(1,0){124}} 
\multiput(5,-2)(10,0){12}{\line(0,1){124}} 

\thicklines
\put(-2,55){\line(1,0){128}} 
\put(55,-2){\line(0,1){128}} 
\thinlines

\put(130,34){{\scriptsize $\m_2=\e_2-4\muzero_2$}}
\put(130,44){{\scriptsize $\m_2=-2\muzero_2$}}
\put(130,54){{\scriptsize $\m_2=0$}} \put(130,64){{\scriptsize
$\m_2=2\muzero_2$}} \put(130,104){{\scriptsize
$\m_2=\m_2^\star-2\muzero_2$}} \put(130,114){{\scriptsize
$\m_2=\m_2^\star$}} \multiput(135,72)(0,3){3}{$\cdot$}

\put(52,130){{\scriptsize \shortstack{$0$\\{\tiny $\|$}\\$\m_1$}}}
\put(56,130){{\scriptsize \shortstack{$\ \ 2\muzero_1$\\{\tiny
$\|$}\\$\m_1$}}} \put(30,130){{\scriptsize \shortstack{$-2\muzero_1\
\ $\\{\tiny $\|$}\\$\m_1$}}} \put(98,130){{\scriptsize
\shortstack{$2\muzero_1$\\{\tiny $|$}\\$\m_1^\star$\\{\tiny
$\|$}\\$\m_1$}}} \put(112,130){{\scriptsize
\shortstack{$\m_1^\star$\\{\tiny $\|$}\\$\m_1$}}}
\multiput(75,135)(3,0){3}{$\cdot$}

\multiput(5,115)(10,0){5}{\makebox(0,0){$\bullet$}}
\multiput(5,105)(10,0){5}{\makebox(0,0){$\bullet$}}
\multiput(65,115)(10,0){6}{\makebox(0,0){$\bullet$}}
\multiput(65,105)(10,0){6}{\makebox(0,0){$\bullet$}}
\multiput(115,5)(0,10){5}{\makebox(0,0){$\bullet$}}
\multiput(105,5)(0,10){5}{\makebox(0,0){$\bullet$}}
\multiput(115,65)(0,10){4}{\makebox(0,0){$\bullet$}}
\multiput(105,65)(0,10){4}{\makebox(0,0){$\bullet$}}

\put(55,105){\makebox(0,0){$\circ$}}
\put(55,115){\makebox(0,0){$\circ$}}
\put(105,55){\makebox(0,0){$\circ$}}
\put(115,55){\makebox(0,0){$\circ$}}

\multiput(60,0)(0,10){10}{
 \multiput(5,5)(10,0){4}{\makebox(0,0){$\times$}}
} \multiput(0,60)(0,10){4}{
 \multiput(5,5)(10,0){6}{\makebox(0,0){$\times$}}
} \put(55,45){\makebox(0,0){$\times$}}
\put(45,55){\makebox(0,0){$\times$}}

\put(35,35){\makebox(0,0){\tiny $\square$}}
\put(45,45){\makebox(0,0){\tiny $\square$}}
\put(35,55){\makebox(0,0){\tiny $\square$}}
\put(55,35){\makebox(0,0){\tiny $\square$}}

\put(55,55){\makebox(0,0){$\otimes$}}

\end{picture}
}

\end{picture}
 \caption{(\textbf{a}): The lattice ${\cal L}_{\e_1=0}\otimes{\cal L}_{\e_2\neq0}$.
 (\textbf{b}): The lattice ${\cal L}_{\e_1=0}\otimes{\cal L}_{\e_2=0}$.
 The points marked $\bullet$ are the boundary lines with given values; in particular,
 $\circ$ denotes the boundary points at which the value of $e_{\l_{12}}$ vanishes.
 The points marked $\times$ are the places whose values can be easily obtained,
 while those with $\diamond$, {\tiny $\square$} and $\otimes$ require more careful consideration.
 (The largeness of $\m_1^\star$ and $\m_2^\star$ is assumed
 but not properly presented in this figure.)
 }\label{fig:fig2}
\end{figure}

Next, let us study the special lattice ${\cal
L}_{\e_1=0}\otimes{\cal L}_{\e_2=0}$. Again, we have the states
$\ket{0,\infty}$ and $\ket{\infty,0}$ decoupled on the distant
boundary (marked $\circ$ in \figref{fig:fig2}.b). Furthermore, in
the iterating procedure, the values of $e_{\l_{12}}$ can be obtained
only to the first, second and fourth quadrants plus two extra points
(all marked $\times$). When the recursion relation is applied for
the value at the origin ($\m_1=\m_2=0$, marked $\otimes$), the
vanishing of $\tilde{C}^{--}_3(2\muzero_1,2\muzero_2)$ by
\eqnref{eqn:coefficients tilde C 1} makes \eqnref{eqn:recursion
relation} a consistency relation:
\ba\label{eqn:consistency relation}
&&\widetilde{C}_3^{++}(2\muzero_1,2\muzero_2)\,e_{\l_{12}}(4\muzero_1,4\muzero_2)
-\widetilde{C}_3^{-+}(2\muzero_1,2\muzero_2)\,e_{\l_{12}}(0,4\muzero_2)\nn\\
&&-\widetilde{C}_3^{+-}(2\muzero_1,2\muzero_2)\,e_{\l_{12}}(4\muzero_1,0)
+\l_{12}\,e_{\l_{12}}(2\muzero_1,2\muzero_2)
=0
\ea
and leaves the value of $e_{\l_{12}}(0,0)$ \emph{undetermined}.
Nevertheless,
thanks to \eqnref{eqn:coefficients tilde C 1} again,
the values of $e_{\l_{12}}$ at the points marked {\tiny $\square$}
can still be determined without knowing the value at $\otimes$.
Moreover, if the Hilbert space $\hilbert_{\e_1=0,\e_2=0}^{\rm grav}$ is restricted to the symmetric sector,
given the symmetric boundary condition,
the consistency relation \eqnref{eqn:consistency relation} is the mirror
(reflected by the origin) of
the relation among the four points marked {\tiny $\square$}
and therefore it is automatically satisfied.
As a result, the point $\otimes$ is \emph{completely decoupled}
and the iteration ``jumps'' over the origin to those points marked {\tiny $\square$}
and consequently to all the unmarked points in \figref{fig:fig2}.b.

Put all together, it follows that the states associated with the
point marked $\circ$ and $\otimes$ in \figref{fig:fig2} are
irrelevant. The same arguments can be easily applied for the special
lattices in $(\m_1,\m_3)$- and $(\m_2,\m_3)$-planes in the cyclic
manner. In the end, combining $e_{\l_{12}}(\m_1,\m_2)$,
$e_{\l_{13}}(\m_1,\m_3)$ and $e_{\l_{23}}(\m_2,\m_3)$ to get
$e_{\vec k}^{(s)}(\vec{\m})$, we conclude that the sate
$\ket{0,0,0}$ associated with the singularity on $\hilbert^S_{\rm
grav}$ is \emph{completely decoupled} from the difference evolution
and the evolution remains \emph{deterministic} across the deep
Planck regime. Consequently, the classical singularity (both
Kasner-like and Kasner-unlike) is expected to be \emph{resolved} in
the quantum evolution. (The absence of the Kasner singularity has
been shown in \cite{Date:2005nn} using an effective classical
Hamiltonian obtained from loop quantization of vacuum Bianchi I
model. Here, inclusion of the scalar field $\ph$ allows us to prove
the absence of the singularity on the \emph{exact} evolution
equation as $\ph$ being the emergent time.)

When the matter sector is included to the model, the
\emph{directional factor} defined as
$\varrho_I:=p_\ph^{2}/\abs{p_I}^3$ for any of the diagonal
directions with the constant of motion $p_\ph$ plays the same role
as the matter density $\r:=\frac{1}{2}p_\ph^2/\abs{p_1p_2p_3}$ does
in the isotropic case. The fact that the problematic state is
decoupled suggests that the directional factor in any diagonal
direction cannot go to infinity when the quantum discreteness is
imposed. Therefore, the classical singularity (which gives infinite
directional factors) is expected to be resolved and replaced by the
big bounce in LQC.\footnote{Note that, unlike the isotropic case, it
is the \emph{directional factor} $\varrho_I$ (not the \emph{matter
density} $\r$) that indicates the happening of the big bounce. Near
the epoch of the singularity, the three directional factors
$\varrho_I$ may approach their critical values at slightly different
instants. Therefore, the detailed occurrence of the big bounce may
happen as many as three times, not just once.} However, following
the lesson we learned for the isotropic model in
\cite{Ashtekar:2006uz}, the critical value of the directional factor
can be arbitrarily small by increasing $p_\ph$ since we are now
adopting $\muzero$-scheme. Later in \secref{sec:improved dynamics},
the dynamics will be refined in $\mubar$-scheme, by which this
problem is supposed to be fixed and the critical value for
$\varrho_I$ is expected to be ${\cal O}(\hbar\,\Pl^{-4})$,
independent of $p_\ph$. (This has been shown for the effective
dynamics in \cite{to appear}.)

The argument here is essentially the generalization of that used for
the isotropic case in \cite{Ashtekar:2003hd}. The result solidifies
the assertion that the key features responsible for the resolution
of the singularity are robust \cite{Bojowald:2002ny} and persist in
more complicated cosmological models \cite{Bojowald:2003md}.
However, the proof presented here only shows the decoupling of the
state in the \emph{kinematical} Hilbert space associated with the
classical singularity, but details of the \emph{dynamical} behaviors
of the solutions have yet to be investigated. To draw the definitive
conclusion that the singularity is resolved and replaced by the
bounce for those states which are semi-classical at late times, we
have to figure out the \emph{physical} Hilbert space, Dirac
observables and semi-classical states, (which will be the topic of
\secref{subsec:LQC physical sector}) and afterwards a thorough
numerical investigation can be initiated \cite{in progress}.

\ \newline \small \textbf{Remark:} The proof in this subsection is
rather general, insensitive to the details of the matter sector. As
far as the matter sector is concerned, the only proviso to make the
above argument valid is:
$B(0,\m_2,\m_3)\l_{12}=B(\m_1,0,\m_3)\l_{12}=0$, etc. By
\eqnref{eqn:lambda ij} and \eqnref{eqn:lambda omega}, this condition
is equivalent to: $-B({\vec \m})\Th({\vec \m})=\hat{C}_{\rm
matter}({\vec \m})\rightarrow0$, when any of $\m_I$ goes to zero.
That is, $\hat{C}_{\rm matter}({\vec \m})$ annihilates
$\Ps(\vec{\m},\ph)$ ($\ph$ represents the matter field in general)
for $\m_1=0$, $\m_2=0$ or $\m_3=0$, which is the case for the
minimally coupled matter source
\cite{Bojowald:2002gz,Bojowald:2001vw}.

Without the discreteness imposed for the inverse triad operator,
$B(\vec{\m})$ goes to infinity when $\m_I\rightarrow0$ and the above
proviso no longer holds. The finiteness of the eigenvalue of
$\widehat{1/\sqrt{p_I}}$ mentioned in \secref{subsec:triad
operators} therefore plays an important role for the decoupling of
the singularity. However, the results of both effective dynamics and
the detailed numerical analysis in isotropic models tell us: it is
the ``non-locality'' of the gravity (i.e, $\hat{c}_I$ replaced by
holonomy), not the finiteness of the inverse triad operator, that
accounts for the occurrence of the big bounce and in fact the bounce
takes place much earlier before the discreteness correction on the
inverse triad operator becomes significant.

Nevertheless, the fact that the problematic state is decoupled
kinematically is still important, giving the lowest bound at the
deep Planck regime, sometime before which the bounce should take
place. For the semi-classsical states at late times, however, the
exact point at which the bounce occurs depends on the detailed
dynamics and it is very likely to happen much earlier before the
lowest bound as in the case of the isotropic model. Likewise, in the
anisotropic model, we also anticipate the same situation, and
accordingly expect the directional factor $\varrho_I$ as the
indication of happening of the bounce \cite{to appear}.\hfill $\Box$
\normalsize

\ \newline \indent Contrary to the resolution of the singularity
shown above, the state $\ket{\infty,\infty,\infty}$ on
$\hilbert_{\rm grav}^S$ associated with the planar collapse is
\emph{not} decoupled but persists in the evolution equation. This
fact is somewhat undesirable since it means one of the length scale
factors $a_I$ continues the vanishing behavior in the forward
evolution and thus eventually enters the deep Planck length scale
without a stop. On the other hand, however, this is expected as the
classical solutions yield $\muzero_Ic_I\rightarrow0$ toward the
planar collapse and therefore the quantum corrections become more
and more negligible. (This remains true in $\mubar$-scheme since
$\mubar_Ic_I\rightarrow0$. For more details of the effectiveness of
quantum corrections, see \cite{to appear}.) It can also be
understood as follows. The planar collapse corresponds to the point
$(\infty,\infty,\infty)$ in the lattice ${\cal L}_{\e_1}\otimes{\cal
L}_{\e_2}\otimes{\cal L}_{\e_3}$, and for the points with large
values of $p_I$ the lattice discreteness can be ignored. In terms
of $p_I$, nothing is really collapsing and the classical solution
toward the planar collapse is perfectly well-behaved if we treat
$p_I$ as the fundamental variables.

The result suggests that smallness of $p_I$ (not of $a_I$) is the
indication for the significance of quantum effect and thus also for
the occurrence of bounces. While $a_I$ are the \emph{length} scale
factors, $p_I$ can be interpreted as the \emph{area} scale factors
according to \eqnref{eqn:p and a}. This seems to support the
speculation: ``area is more fundamental than length in LQG'',
although whether this is simply a technical issue or reflects some
deeper facts is still unclear. (See Section VII.B of
\cite{Rovelli:1997yv} for some comments on this aspect and
\cite{Rovelli:1993vu} for more discussions.) Meanwhile, whereas the
length operator has been shown to also have a discrete spectrum
\cite{Thiemann:1996at}, the fact that the vanishing of the length
scale factor is not resolved seems to contradict the discreteness of
the length spectrum. Whether we have missed some important
ingredients when imposing the discreteness corrections from the full
theory or indeed area is more essential than length remains an open
question for further investigation.

\subsection{The Physical Sector and Semi-Classical States}\label{subsec:LQC physical sector}
Results of \secref{subsec:total Hamiltonian} and \secref{subsec:LQC
general solutions} show that while the LQC operator $\Th$ differs
from the WDW operator $\underline{\Th}$ in an interesting way, the
structural forms of these two Hamiltonian constraint equations are
quite the same. Therefore, in the LQC theory, apart from the issue
of super-selection, we can introduce the Dirac observables and
identify the physical Hilbert space with the inner product by
demanding that the Dirac observables be self-adjoint (or by carrying
out group averaging method) in the way closely analogous to what we did in
WDW theory.

The super-selection sector of the physical Hilbert space
$\hilbert_{\rm phy}^{\vec \e}$ labeled by ${\vec
\e}\in[0,\muzero_1]\times[0,\muzero_2]\times[0,\muzero_3]$ consists
of the positive frequency solutions $\Ps({\vec \m},\ph)$ to
\eqnref{eqn:master eq} with initial data $\Ps({\vec \m},\ph_o)$ in
the symmetric sector of $\hilbert^{\rm grav}_{\vec \e}$. By
\eqnref{eqn:sol to master eq}, the explicit expression of $\Ps({\vec \m},\ph)$
in terms of
the symmetric eigenfunctions $e_{\vec k}^{(s)}$ reads as
\be\label{eqn:LQC physical functions}
\Ps({\vec \m},\ph)=\int_{\o^2({\vec k})>0}\!\!d^3k\,
\tilde{\Ps}({\vec k})\,e^{(s)}_{\vec k}({\vec \m})\,e^{i\o\ph},
\ee
where, as in the WDW theory, $\o^2({\vec k})=16\p G(k_1k_2+k_1k_3+k_2k_3+3/16)$.
The physical inner product is given by
\be\label{eqn:LQC inner product}
\inner{\Ps_1}{\Ps_2}_{\rm phy}^{\vec \e}=\sum_{{\vec \m}
\in{\cal L}_{\e_1}\otimes{\cal L}_{\e_2}\otimes{\cal L}_{\e_3}}
\!\!\!\!\!\!B({\vec \m})\,\bar{\Ps}_1({\vec \m},\ph_o)\Ps_2({\vec \m},\ph_o).
\ee
The action of the Dirac observables is independent of ${\vec \e}$ and has the same form as in the WDW theory:
\be\label{eqn:LQC Dirac observables}
\widehat{\m_I|_{\ph_o}}\Ps({\vec \m},\ph)=e^{i \sqrt{\Th_+}\,(\ph-\ph_o)}\abs{\m_I} \Ps({\vec \m},\ph_o)
\quad\text{and}\quad
\hat{p}_\ph\Ps({\vec \m},\ph)=-i\hbar\frac{\partial\Ps({\vec \m},\ph)}{\partial\ph}.
\ee

The kinematical Hilbert space $\hilbert_{\rm kin}^{\rm total}$ is
non-separable but, because of super-selection, each physical sector
$\hilbert_{\rm phy}^{\vec \e}$ is separable. The eigenvalues of the
Dirac observable $\widehat{\m_I|_{\ph_o}}$ constitute a discrete
subset of the real line in each sector. In $\hilbert_{\rm
kin}^{\rm grav}$, on the contrary, the spectrum of $\hat{p}_I$ is
discrete in a subtler sense: every real value is allowed but the
spectrum space has discrete topology. Therefore, the more delicate
discreteness of the spectrum of $\hat{p}_I$ on $\hilbert_{\rm
kin}^{\rm grav}$ descends to the standard type of discreteness of the
Dirac observables on each physical sector $\hilbert_{\rm phys}^{\vec{\e}}$.

Furthermore, comparing \eqnref{eqn:LQC physical functions} with the
WDW semi-classical state given by \eqnref{eqn:semi-classical state},
by analogy, we can construct the LQC semi-classical state at the time
$\ph=\ph_o$:
\be\label{eqn:LQC semi-classical state}
\Ps_{{\vec
\m}_{\ph_o}^\star,p_\ph^\star}({\vec \m},\ph_o)= \int_{\o^2({\vec
k})>0}\!\!d^3k\ \tilde{\Ps}({\vec k})\ {e}^{(s)}_{\vec k}({\vec
\m})\, e^{i\o({\vec k})\ph_o}, \ee where the Fourier amplitude
$\tilde{\Ps}({\vec k})$ is given by \be \tilde{\Ps}({\vec k})=
\tilde{\Ps}_{\vec \k}({\vec k})\, e^{-i\o({\vec k})\ph_o^\star}
\ee
with the same $\tilde{\Ps}_{\vec \k}({\vec k})$ and $\ph_o^\star$ as
defined in \eqnref{eqn:Fourier amplitude} and
\eqnref{eqn:mu-phi0-star}. Since ${e}_{\vec k}^{(s)}({\vec
\m})\rightarrow\underline{e}_{\vec k}({\vec \m})$ for
$\abs{\m_I}\gg\muzero_I$, the wavefunction given by \eqnref{eqn:LQC
semi-classical state} is very close to that in
\eqnref{eqn:semi-classical state} for large $\abs{\m_I}$. Therefore,
the LQC semi-classical state is peaked at the classical trajectory
for the region far away from the singularity and thus satisfies the
\emph{pre-classicality} \cite{Bojowald:2001xa,Bojowald:2003ij} (at
least for one asymptotic side of the evolution).

It has been argued in \cite{Cartin:2005an} that only the zero
solution satisfies the pre-classical condition for both the vacuum
case and the addition of a cosmological constant in the Bianchi I
model. Inclusion of matter changes the situation as suggested in
\cite{Cartin:2005an}; here, with $\ph$ being the emergent time, the
LQC semi-classical sates are constructed and shown to be
pre-classical.

We have shown in \secref{subsec:resolution of singularity} that the
quantum evolution is deterministic, but the question whether the
solution is pre-classical again on the other side across the
singularity or descends into a quantum foam needs more
investigation. Although the thorough numerical analysis is awaited
to be done \cite{in progress}, the occurrence of the big bounce has
been shown in \cite{to appear} and it turns out that the
semi-classical states are pre-classical on both asymptotic sides of
the evolution. (For the issues of pre-classicality and time
asymmetry on either side of the evolution in the vacuum Bianchi I
model, see \cite{Cartin:2005sm}.)

\section{Improved Dynamics ($\mubar$-Scheme)}\label{sec:improved dynamics}
\secref{sec:dynamics} -- \secref{sec:analytical issues} complete the
analytical investigation for the precursor strategy
($\muzero$-scheme). In this section, we repeat the whole analysis
for the improved dynamics ($\mubar$-scheme), which was first
proposed and studied for the isotropic model in
\cite{Ashtekar:2006wn}. The prescription to endow the fundamental
discreteness from the full theory is modified and implemented in a
more sophisticated way, thus more directly conveying the underlying
physics of quantum geometry. The corresponding WDW theory is
unchanged but recast in \secref{subsec:improved WDW theory},
featuring the use of affine variables (to be defined soon).

\subsection{Gravitational Constraint Operator}\label{subsec:improved grav constraint operator}
Fixing the $\muzero$-ambiguity by setting $\muzero$ to the finite
number can be understood as the imprint of the fundamental
discreteness of the full theory. In the full theory of LQG, a proper
regularization procedure is taken to define the Hamiltonian
operator: the 3d coordinate space is partitioned into small regions
of coordinate volume $\e^3$. When acting on the diffeomorphism
equivalence classes, the dependence on $\e$ is gone when $\e$ is
smaller than some particular value because of the diffeomorphism
invariance. Therefore, the Hamiltonian constraint is independent of
the regulator in the end, simply because making the partition finer
cannot change anything below the Planck scale
\cite{Thiemann:1996ay,Thiemann:1996aw,Thiemann:1997rt}.

In the symmetry reduced theory, since the diffeomorphism invariance
is gauge fixed in the beginning, the independence of the regulator
is no longer true and we have to fix the regulator by hand. However,
the strategy used in $\muzero$-scheme is rather heuristic: the
holonomy $h_I^{(\muzero_I)}$ in \eqnref{eqn:area of h} is not a
``state'' of the Hilbert space and thus setting its ``eigenvalue''
to the area gap is not completely physical. To reflect the
fundamental discreteness more rigourously, we introduce the improved
strategy: ``$\mubar$-scheme''.

In $\cyl$ of the full theory, the \emph{area operator}
$\hat{A}({\cal S})$ for a coordinate surface yields quantized
spectrum when acting on the spin network states
$\ket{\a,\textbf{j},\textbf{I}}$. To carry the parallel from the
full theory to the reduced theory, first, recall that the states
${\cal N}_{\vec \m}=\ket{{\vec \m}}$ in $\cyl_S$ are analogous to
the spin network states ${\cal
N}_{\a,\textbf{j},\textbf{I}}=\ket{\a,\textbf{j},\textbf{I}}$ in
$\cyl$. Second, note that the area operator $\hat{A}({\cal S})$ in
$\cyl$ is replaced by $E(S_I):=E(S_I,f_I=1)$ in $\cyl_S$ defined in
\eqnref{eqn:area operator}.

Let $S^{(\mubar_3)}$ be the square surface normal to the $z$-direction (with respect to $\fq$)
with the side lengths $\mubar_3\Lzero_1$ and $\mubar_3\Lzero_2$ (with respect to $\fq$)
in $x$- and $y$-directions respectively. According to \eqnref{eqn:area operator}, when acting on the state
$\ket{{\vec \m}}$, the area operator associated with $S^{(\mubar_3)}$ gives
\be
E(S^{(\mubar_3)})\ket{{\vec \m}}=4\p\g\Pl^2\,\mubar_3^2\mu_3\ket{{\vec \m}}.
\ee
If $S^{(\mubar_3)}$ is the smallest area we can shrink to,
we should then have $4\p\g\Pl^2\,\mubar_3^2\mu_3=\pm\D$ or
\be
\abs{\mubar_3^2\mu_3}=k:=\D/(4\p\g\Pl^2)=
\sqrt{3}/2.
\ee
In the other two directions, we set $S^{(\mubar_1)}$ and $S^{(\mubar_2)}$ to the smallest area
and have the similar results: $\abs{\mubar_1^2\mu_1}=\abs{\mubar_2^2\mu_2}=k$.
Therefore, to faithfully reflect the area gap in the full theory, instead of the $\muzero$ -scheme
used earlier, we should set
\be\label{eqn:mu bar}
\mubar_1=\sqrt{\frac{k}{\abs{\m_1}}},
\qquad
\mubar_2=\sqrt{\frac{k}{\abs{\m_2}}},
\qquad
\mubar_3=\sqrt{\frac{k}{\abs{\m_3}}}.
\ee
Unlike $\muzero$-scheme, in which $\muzero_I$ are fixed as constant, in the improved strategy,
the values $\mubar_I$ to be set to fix the regulator depend \emph{adaptively}
on the state $\ket{{\vec \m}}$. This is due to the presence of the operator $p_I$
in \eqnref{eqn:area operator}.

To figure out what the action of the operator $e^{\pm\frac{i}{2}\mubar_{(I)}c^{(I)}}$ is when acting on $\ket{{\vec \m}}$,
we first recall that the first line of \eqnref{eqn:e cos sin} corresponds to
$e^{\mp\muzero_1\frac{\partial}{\partial \m_1}}\ps(\m_1,\m_2,\m_3)=\ps(\m_1\mp\muzero_1,\m_2,\m_3)$;
that is, $c^I=2i\frac{\partial}{\partial\m_I}$ in the ${\vec \m}$-representation.
By the same spirit, we define the affine parameter $v_I$ such that
\be
\mubar_{(I)}\frac{\partial}{\partial\m_{(I)}}=\sqrt{\frac{k}{\abs{\m_I}}}\,\frac{\partial}{\partial\m_I}
:=\frac{\partial}{\partial v_I},
\ee
which leads to
\be\label{eqn:affine variables}
v_I=\sgn(\m_I)\frac{2}{3\sqrt{k}}\,\abs{\m_I}^{3/2}.
\ee
Changing the coordinates $\m_1,\m_2,\m_3$ to $v_1,v_2,v_3$, we then have\footnote{In the ordinary
quantum mechanics, $p=-\frac{i}{\hbar}\frac{d}{dx}$ and $e^{iap/\hbar}\ps(x)=\ps(x+a)$ but we cannot simply take
$e^{if(x)p/\hbar}=e^{f(x)\frac{d}{dx}}$ because it is not unitary in general. Here, on the contrary,
since $\cyl_S$ is in the ``polymer representation'' and the right hand side of \eqnref{eqn:inner product}
is the Kronecker delta (not the Dirac distribution), the same problem does not occur.
In fact, $\hilbert_{\rm kin}^{\rm grav}=
L^2(\mathbb{R}^3_{\rm Bohr},d^3\!\m_{\rm Bohr})=L^2(\mathbb{R}^3_{\rm Bohr},d^3v_{\rm Bohr})$
and $e^{\pm\frac{i}{2}\mubar_{(I)}c^{(I)}}$ is unitary.}
\be
e^{\mp\mubar_1\frac{\partial}{\partial\m_1}}\ps(\m_1,\m_2,\m_3)
=e^{\mp\frac{\partial}{\partial v_1}}\ps(v_1,v_2,v_3)
=\ps(v_1\mp 1,v_2,v_3),
\ee
or equivalently
\be
e^{\pm\frac{i}{2}\mubar_1 c^1}\ket{\m_1,\m_2,\m_3}
=\ket{v_1\pm 1,v_2,v_3}.
\ee
In terms of the new variables, that is
\ba
e^{\pm\frac{i}{2}\mubar_1c^1}\ket{\vec v}&=&\ket{v_1\!\pm1\!,v_2,v_3},\nn\\
e^{\pm\frac{i}{2}\mubar_2c^2}\ket{\vec v}&=&\ket{v_1,v_2\!\pm1\!,v_3},\nn\\
e^{\pm\frac{i}{2}\mubar_3c^3}\ket{\vec v}&=&\ket{v_1,v_2,v_3\!\pm\!1}.
\ea

In $\mubar$-scheme, the new operators corresponding to
\eqnref{eqn:hat C grav prime} and \eqnref{eqn:hat C grav}
are obtained by replacing $\muzero_I$ with $\mubar_I$.
Consequently, when acting on the states labeled by $v_1,v_2,v_3$,
\eqnref{eqn:hat C grav prime on states} is modified to
\ba\label{eqn:improved hat C grav prime on states}
&&{\hat C}^{({\mubar})'}_{\rm grav}\ket{v_1,v_2,v_3}
=
\frac{1}{16\p\g^3\Pl^2\mubar_1\mubar_2\mubar_3}\nn\\
&&\quad\times\Big\{
\Big(\abs{V_{v_1+2,v_2+2,v_3+1}
-V_{v_1+2,v_2+2,v_3-1}}
+\abs{V_{v_1,v_2,v_3+1}-V_{v_1,v_2,v_3-1}}\Big)
\ket{v_1+2,v_2+2,v_3}\nn\\
&&\qquad\!\!-
\Big(\abs{V_{v_1-2,v_2+2,v_3+1}
-V_{v_1-2,v_2+2,v_3-1}}
+\abs{V_{v_1,v_2,v_3+1}-V_{v_1,v_2,v_3-1}}\Big)
\ket{v_1-2,v_2+2,v_3}\nn\\
&&\qquad\!\!-
\Big(\abs{V_{v_1+2,v_2-2,v_3+1}
-V_{v_1+2,v_2-2,v_3-1}}
+\abs{V_{v_1,v_2,v_3+1}-V_{v_1,v_2,v_3-1}}\Big)
\ket{v_1+2,v_2-2,v_3}\nn\\
&&\qquad\!\!+
\Big(\abs{V_{v_1-2,v_2-2,\m_3+1}
-V_{v_1-2,v_2-2,v_3-1}}
+\abs{V_{v_1,v_2,v_3+1}-V_{v_1,v_2,v_3-1}}\Big)
\ket{v_1-2,v_2-2,v_3}\nn\\
&&\qquad+\text{ cyclic terms}
\Big\}
\ea
and \eqnref{eqn:hat C grav on states} is modified to
\ba\label{eqn:improved hat C grav on states}
&&{\hat C}^{({\mubar})}_{\rm grav}\ket{v_1,v_2,v_3}
=
\frac{1}{16\p\g^3\Pl^2\mubar_1\mubar_2\mubar_3}\nn\\
&&\quad\times\Big\{
\Big(\abs{V_{v_1+2,v_2,v_3+1}-V_{v_1+2,v_2,v_3-1}}
+\abs{V_{v_1,v_2+2,v_3+1}-V_{v_1,v_2+2,v_3-1}}\Big)
\ket{v_1+2,v_2+2,v_3}\nn\\
&&\qquad\!\! -
\Big(\abs{V_{v_1-2,v_2,v_3+1}-V_{v_1-2,v_2,v_3-1}}
+\abs{V_{v_1,v_2+2,v_3+1}-V_{v_1,v_2+2,v_3-1}}\Big)
\ket{v_1-2,v_2+2,v_3}\nn\\
&&\qquad\!\! -
\Big(\abs{V_{v_1+2,v_2,v_3+1}+V_{v_1+2,v_2,v_3-1}}
+\abs{V_{v_1,v_2-2,v_3+1}-V_{v_1,v_2-2,v_3-1}}\Big)
\ket{v_1+2,v_2-2,v_3}\nn\\
&&\qquad\!\! +
\Big(\abs{V_{v_1-2,v_2,v_3+1}-V_{v_1-2,v_2,v_3-1}}
+\abs{V_{v_1,v_2-2,v_3+1}-V_{v_1,v_2-2,v_3-1}}\Big)
\ket{v_1-2,v_2-2,v_3}\nn\\
&&\qquad+\text{ cyclic terms}
\Big\},
\ea
where the eigenvalues of the volume operator are given by
\be
V_{v_1,v_2,v_3}:=\left(4\p\g\right)^{3/2}\frac{3\sqrt{k}}{2}\ \abs{v_1v_2v_3}^{1/3}\,\Pl^3.
\ee
In terms of the new notation, it is equivalent to
\ba\label{eqn:improved hat C grav on states in new notation}
{\hat C}^{({\mubar})}_{\rm grav}\ket{v_1,v_2,v_3}
&=&
\frac{\sqrt{\p}\,\Pl}{2\g^\frac{3}{2}\mubar_1\mubar_2\mubar_3}
\Big\{
C_3^{++}(\vec{v})\,
\ket{v_1+2,v_2+2,v_3}-
C_3^{-+}(\vec{v})\,
\ket{v_1-2,v_2+2,v_3}\nn\\
&&\qquad\qquad\quad\! -
C_3^{+-}(\vec{v})\,
\ket{v_1+2,v_2-2,v_3}+
C_3^{--}(\vec{v})\,
\ket{v_1-2,v_2-2,v_3}\nn\\
&&\qquad\qquad\quad\;+\text{ cyclic terms}
\Big\},
\ea
where
\ba
C_3^{++}(\vec{v})&:=&\frac{3\sqrt{k}}{2}\,\Big(\Abs{
|(v_1+2)v_2(v_3+1)|^{1/3}
-|(v_1+2)v_2(v_3-1)|^{1/3}}\nn\\
&&\qquad\quad+
\Abs{
|v_1(v_2+2)(v_3+1)|^{1/3}
-|v_1(v_2+2)(v_3-1)|^{1/3}}\Big),\nn\\
C_3^{--}(\vec{v})&:=&\frac{3\sqrt{k}}{2}\,\Big(\Abs{
|(v_1-2)v_2(v_3+1)|^{1/3}
-|(v_1-2)v_2(v_3-1)|^{1/3}}\nn\\
&&\qquad\quad+
\Abs{
|v_1(v_2-2)(v_3+1)|^{1/3}
-|v_1(v_2-2)(v_3-1)|^{1/3}}\Big),\nn\\
C_3^{+-}(\vec{v})&:=&\frac{3\sqrt{k}}{2}\,\Big(\Abs{
|(v_1+2)v_2(v_3+1)|^{1/3}
-|(v_1+2)v_2(v_3-1)|^{1/3}}\nn\\
&&\qquad\quad+
\Abs{
|v_1(v_2-2)(v_3+1)|^{1/3}
-|v_1(v_2-2)(v_3-1)|^{1/3}}\Big),\nn\\
C_3^{-+}(\vec{v})&:=&\frac{3\sqrt{k}}{2}\,\Big(\Abs{
|(v_1-2)v_2(v_3+1)|^{1/3}
-|(v_1-2)v_2(v_3-1)|^{1/3}}\nn\\
&&\qquad\quad+
\Abs{
|v_1(v_2+2)(v_3+1)|^{1/3}
-|v_1(v_2+2)(v_3-1)|^{1/3}}\Big)
\ea
and similar for $C_2^{\pm\pm}$ and $C_1^{\pm\pm}$ in the cyclic manner.

By inspection, it is clear that $\hat{C}^{({\mubar})'}_{\rm grav}$ and
$\hat{C}^{({\mubar})}_{\rm grav}$ commute with
the parity operators $\Pi_{1,2,3}$:
\be
[\hat{C}^{({\mubar})'}_{\rm grav},\, \Pi_I]=0,
\qquad
[\hat{C}^{({\mubar})}_{\rm grav},\, \Pi_I]=0,
\ee
where $\Pi_1\ket{{\vec v}}=\ket{-v_1,v_2,v_3}$
(or equivalently, $\Pi_1\ket{{\vec \m}}=\ket{-\m_1,\m_2,\m_3}$)
and so on,
which flip the orientation of the triad.

Note that, unlike $\hat{C}^{({\muzero})'}_{\rm grav}$ and
$\hat{C}^{({\muzero})}_{\rm grav}$,
because of the presence of nonconstant $\mubar_I$ in
\eqnref{eqn:improved hat C grav prime on states} and
\eqnref{eqn:improved hat C grav on states}, the operators
$\hat{C}_{\rm grav}^{(\mubar)'}$ and $\hat{C}_{\rm
grav}^{(\mubar)'}$ are \emph{not} self-adjoint in $\hilbert_{\rm
grav}^S$, unless further re-ordering is performed. Nevertheless, the operator
$\Th$ to be defined in \eqnref{eqn:improved master eq} is
self-adjoint in $L^2(\mathbb{R}^3_{\rm
Bohr},\tilde{B}(\vec{v})\,d^3v_{\rm Bohr})$ with proper measure
$\tilde{B}(\vec{v})$ chosen. Therefore, the procedures used in
$\muzero$-scheme to construct the physical sector can still be
applied for the improved scheme. (Also see the remark in
\appref{subsec:second scheme for isotropic} for the issue of
re-ordering in the isotropic case.)

\ \newline \small \textbf{Remark:} To impose the ``adaptive''
regularization, a seemingly direct but in fact misconceived method
would be to set $S^{(\mubar_1,\mubar_2)}$ (the rectangular surface
normal to the $z$-direction with the side lengths $\mubar_3\Lzero_1$
and $\mubar_3\Lzero_2$ in $x$- and $y$-directions) to the smallest
area $\D$ and so on for the other two directions. Instead of
\eqnref{eqn:mu bar}, this method gives
$\mubar_1=\sqrt{\frac{\abs{\m_1}k}{\abs{\m_2\m_3}}}$,
$\mubar_2=\sqrt{\frac{\abs{\m_2}k}{\abs{\m_1\m_3}}}$, and
$\mubar_3=\sqrt{\frac{\abs{\m_3}k}{\abs{\m_1\m_2}}}$, which mingle
the three diagonal directions and complicate the Hamiltonian
constraint (in particular, the evolution constraint is no longer a
difference equation). A more careful consideration urges that we
should use the ``square'' surface $S^{(\mubar_3)}$ rather than the
generic ``rectangular'' surface $S^{(\mubar_1,\mubar_2)}$ to reflect
the discreteness imprinted from the full theory. In the full theory,
the discretized area is associated with the edge of the spin network
which penetrates the surface and carries the $su(2)$ label $j$.
Therefore, in the symmetry reduced theory, to properly impose the
discreteness, the role of $j$ is replaced by the \emph{single} label
$\mubar_I$ in the normal direction to describe the area of the
associated surface.\footnote{The author thanks Ashtekar and
Pawlowski for clarifying this confusion.} \hfill $\Box$\normalsize

\subsection{Inverse Triad Operators and the Total Hamiltonian Constraint}\label{subsec:improved total hamiltonian}
In $\mubar$-scheme, by the same spirit as discussed in the previous subsection,
the inverse triad operators defined in \eqnref{eqn:triad operator}
should be changed as well by replacing $\muzero_I$ with $\mubar_I$,
and correspondingly, when acting on the states, \eqnref{eqn:triad operator on states} is modified as
\ba\label{eqn:improved triad operator on states}
\widehat{\left[{\frac{1}{\sqrt{\abs{p_I}}}}\right]}\ket{{\vec v}}
&=&\frac{1}{4\p\g\Pl^2\mubar_1}\Abs{L_{v_I+1}-L_{v_I-1}}\,\ket{{\vec v}}\nn\\
&=&(4\p\g k)^{-\frac{1}{2}}\Pl^{-1}
\abs{v_I}^\frac{1}{3}\,\Abs{\abs{v_I+1}^\frac{1}{3}-\abs{v_I-1}^\frac{1}{3}}\,\ket{{\vec v}},
\ea
where $L_I$ is the eigenvalue of the length operators:
\be
L_I=\left(4\p\g\right)^{1/2}\left(\frac{3\sqrt{k}}{2}\right)^{\!\!1/3}\abs{v_I}^{1/3}\,\Pl.
\ee
Note that as in $\muzero$-scheme, the inverse triad operators in $\mubar$-scheme
commute with $\hat{p}_I$ and are self-adjoint in $\hilbert_{\rm grav}^S$ and bounded above
as well.
The upper bound is given at $v_I=\pm 1$:
\be
\abs{p}_{\rm max}^{-\frac{1}{2}}=2^{1/3}(4\p\g k)^{-1/2}\,\Pl^{-1}.
\ee
Furthermore, the eigenvalue of $\widehat{1/\sqrt{p_I}}$ goes to zero when $\m_I\rightarrow0$
and this fact is important for the resolution of the singularity
as discussed in \secref{subsec:triad operators}.

Since $C_{\ph}=8\p G\,p^2_\ph/\sqrt{\abs{p_1p_2p_3}}$ in \eqnref{eqn:Hamiltonian},
the corresponding operator $\hat{C}^{({\mubar})}_\ph$ acting on the states of
$\hilbert_{\rm kin}^{\rm total}:=\hilbert^S_{\rm grav}\otimes\hilbert_{\ph}$
yields
\be
\hat{C}^{({\mubar})}_\ph\ket{\vec{v}}\otimes\ket{\ph}=
\frac{1}{(4\p\g)^\frac{3}{2}\Pl^3}{B}(\vec{v})\ket{\vec{v}}\otimes8\p G\hat{p}_\ph^2\ket{\ph},
\ee
where
\ba\label{eqn:B of v}
{B}(\vec{v})&:=&\frac{3\sqrt{k}}{2}\,
\frac{1}{\mubar_1\mubar_2\mubar_3}\,\Abs{\abs{v_1+1}^{1/3}-\abs{v_1-1}^{1/3}}
\,
\Abs{\abs{v_2+1}^{1/3}-\abs{v_2-1}^{1/3}}\nn\\
&&\qquad\qquad\qquad\times
\Abs{\abs{v_3+1}^{1/3}-\abs{v_2-1}^{1/3}}\nn\\
&=&\frac{9}{4\sqrt{k}}\,
\abs{v_1v_2v_3}^{1/3}\,\Abs{\abs{v_1+1}^{1/3}-\abs{v_1-1}^{1/3}}
\,
\Abs{\abs{v_2+1}^{1/3}-\abs{v_2-1}^{1/3}}\nn\\
&&\qquad\qquad\qquad\times
\Abs{\abs{v_3+1}^{1/3}-\abs{v_2-1}^{1/3}}.
\ea
The total Hamiltonian constraint $\hat{C}^{({\mubar})}_{\rm grav}+\hat{C}^{({\mubar})}_\ph=0$ then gives
\be
8\p G\,B(\vec{v})\,\hat{p}_\ph^2\,\ket{\vec{v},\ph}+(4\p\g)^\frac{3}{2}\Pl^3
\hat{C}^{({\mubar})}_{\rm grav}\,\ket{\vec{v},\ph}=0.
\ee

Again, we choose the standard Schr\"{o}dinger representation for
$\ph$ but the ``polymer representation'' for $\vec{v}$ (or equivalently for $\vec{\m}$); that is
$\hilbert^{\rm total}_{\rm kin}=L^2(\mathbb{R}^3_{\rm Bohr},d^3v_{\rm Bohr})\otimes L^2(\mathbb{R},d\ph)$.
In this representation, $\hat{p}_\ph=\frac{\hbar}{i}\frac{\partial}{\partial\ph}$
and the total Hamiltonian constraint read as
\ba\label{eqn:improved master eq}
&&\frac{\partial^2}{\partial\ph^2}\Ps(\vec{v},\ph)\nn\\
&=&\frac{\p G}{2}[\tilde{B}(\vec{v})]^{-1}
\Big\{
C_3^{++}(\vec{v})\,\Ps(v_1+2,v_2+2,v_3,\ph)
-C_3^{-+}(\vec{v})\,\Ps(v_1-2,v_2+2,v_3,\ph)\nn\\
&&\qquad\qquad\quad\,-C_3^{+-}(\vec{v})\,\Ps(v_1+2,v_2-2,v_3,\ph)
+C_3^{--}(\vec{v})\,\Ps(v_1-2,v_2-2,v_3,\ph)\nn\\
&&\qquad\qquad\qquad+\text{ cyclic terms}
\Big\}\nn\\
&=:&-\left(\Th_3(v_1,v_2)+\Th_2(v_1,v_3)+\Th_1(v_2,v_3)\right)
\Ps(\vec{v},\ph)
=:-\Th(\vec{v})\,\Ps(\vec{v},\ph),
\ea
where we define
\ba\label{eqn:tilde B of v}
&&\tilde{B}({\vec v}):=\mubar_1\mubar_2\mubar_3B({\vec v})\nn\\
&=&
\frac{3\sqrt{k}}{2}\,
\Abs{\abs{v_1\!+\!1}^{1/3}\!-\!\abs{v_1\!-\!1}^{1/3}}
\,
\Abs{\abs{v_2\!+\!1}^{1/3}\!-\!\abs{v_2\!-\!1}^{1/3}}
\Abs{\abs{v_3\!+\!1}^{1/3}\!-\!\abs{v_2\!-\!1}^{1/3}}.\qquad
\ea
By inspection, we can easily see that $\Th$ and each of $\Th_{1,2,3}$ individually are self-adjoint
on $L^2({\mathbb R}^3_{\rm Bohr},\tilde{B}({\vec v})\,d^3v_{\rm Bohr})$.

\ \newline \small \underline{Note}: The total Hamiltonian constraint
in $\mubar$-scheme for the isotropic case can be obtained by
formally ``translating'' the anisotropic results found in this
section. This is shown in \appref{subsec:second scheme for
isotropic}, where the different factor orderings for $\hat{C}^{\rm
WDW}_{\rm grav}$ and $\hat{C}^{(\mubar)}_{\rm grav}$ used in this
paper and in \cite{Ashtekar:2006wn} are also remarked. \hfill
$\Box$\normalsize

\subsection{WDW Theory (Revisited)}\label{subsec:improved WDW theory}
When the improved dynamics ($\mubar$-scheme) is adopted in LQC, the
corresponding WDW theory remains the same. In order to compare the
WDW theory with LQC in the new scheme, we recast the results in
\secref{sec:WDW theory} in terms of the affine variables $v_I$
defined in \eqnref{eqn:affine variables}. This change of variables
is straightforward and we only list the results.

The total WDW Hamiltonian constraint in \eqnref{eqn:WDW master eq} becomes
\ba
\frac{\partial^2\Ps}{\partial\ph^2}&=&8\p G\,[\underline{B}(\vec{v})]^{-1}
\Big\{\frac{9}{4}\left(\frac{3\sqrt{k}}{2}\right)^{\!\!-\frac{5}{3}}
\frac{\sgn(v_1v_2)}{\abs{v_3}^{1/3}}\nn\\
&&\qquad\qquad\quad\times
\left(\abs{v_2}^\frac{1}{3}\frac{\partial}{\partial v_2}
\left(\abs{v_2}^\frac{1}{3}\abs{v_1}^\frac{2}{3}\right)
\frac{\partial}{\partial v_1}
+
\abs{v_1}^\frac{1}{3}\frac{\partial}{\partial v_1}
\left(\abs{v_1}^\frac{1}{3}\abs{v_2}^\frac{2}{3}\right)
\frac{\partial}{\partial v_2}
\right)\nn\\
&&\qquad\qquad\qquad+\text{cyclic terms}\Big\}\Ps\nn\\
&=&8\p G\,[\tilde{\underline{B}}(\vec{v})]^{-1}
\frac{2}{3}\left(\frac{3\sqrt{k}}{2}\right)^{\!\!\frac{1}{3}}
\Big\{\frac{\sgn(v_1v_2)}{\abs{v_3}^{2/3}}
\left(\frac{\partial}{\partial v_2}
\abs{v_1v_2}^\frac{1}{3}
\frac{\partial}{\partial v_1}
+
\frac{\partial}{\partial v_1}
\abs{v_1v_2}^\frac{1}{3}
\frac{\partial}{\partial v_2}
\right)\nn\\
&&\qquad\qquad\qquad+\text{cyclic terms}\Big\}\Ps\nn\\
&=:&-(\underline{\Th}_3(v_1,v_2)+\underline{\Th}_2(v_1,v_3)+\underline{\Th}_1(v_1,v_3))\Ps
=:-\underline{\Th}(\vec{v})\Ps,
\ea
where $\underline{B}(\vec{v})=\underline{B}(\vec{\m})=(3\sqrt{k}/2)^{-1}\abs{v_1v_2v_3}^{-1/3}$
and we also define
\be
\tilde{\underline{B}}(\vec{v}):=\left(\frac{2}{3}\right)^{\!\!3}\left(\frac{3\sqrt{k}}{2}\right)^{\!\!2}
\abs{v_1v_2v_3}^{-1/3}\underline{B}(\vec{v})
=\left(\frac{2}{3}\right)^{\!\!3}\left(\frac{3\sqrt{k}}{2}\right)
\abs{v_1v_2v_3}^{-2/3},
\ee
which gives
\be
\underline{B}(\vec{\m})\,d^3\!\m=\tilde{\underline{B}}(\vec{v})\,d^3v.
\ee
Note that in LQC theory, $B({\vec v})$ and $\tilde{B}({\vec v})$ defined
in \eqnref{eqn:B of v} and \eqnref{eqn:tilde B of v} give the expected asymptotic behaviors:
$B({\vec v})\rightarrow\underline{B}({\vec v})$ and
$\tilde{B}({\vec v})\rightarrow\tilde{\underline{B}}({\vec v})$
when $\abs{v_{1,2,3}}\gg1$.

The operators $\underline{\Th}$ and each of $\underline{\Th}_{1,2,3}$ individually are self-adjoint
on $L^2(\mathbb{R}^3,\tilde{\underline{B}}(\vec{v})\,d^3v)$.
The (symmetric) eigenfunctions of $\underline{\Th}$ in \eqnref{eqn:eigenfunction ek}
now reads as
\ba\label{eqn:eigenfunction ek improved}
\underline{e}_{\vec k}({\vec v})=
\underline{e}_{\vec k}({\vec \m})=
\frac{\big({3\sqrt{k}}/{2}\big)^{\!-\frac{1}{2}+\frac{2i}{3}(k_1\!+\!k_2\!+\!k_3)}}{(4\p)^3}
\ \abs{v_1v_2v_3}^{-\frac{1}{6}}\,
e^{i\frac{2}{3}k_1\!\ln\abs{v_1}}\,e^{i\frac{2}{3}k_2\!\ln\abs{v_2}}\,e^{i\frac{2}{3}k_3\!\ln\abs{v_3}}.
\ea
The corresponding eigenvalues are unchanged and given by \eqnref{eqn:WDW frequency}.
The orthonormality relation \eqnref{eqn:orthonormality rel} is rewritten as
\be
\int d^3v\,\tilde{\underline{B}}(\vec{v})\,\bar{\underline{e}}_{\vec k}({\vec v})\,\underline{e}_{{\vec k}'}({\vec v})
=\d^3(\vec{k},\vec{k}'),
\ee
and the completeness relation \eqnref{eqn:completeness rel} gives
\be
\int d^3v\,\tilde{\underline{B}}(\vec{v})\,\bar{\underline{e}}_{\vec k}({\vec v})\,\Ps(\vec{v})=0
\quad\text{for }
\forall {\vec k},
\quad
\text{iff }
\Ps({\vec v})=0
\ee
for any $\Ps({\vec v})\in L^2_S(\mathbb{R}^3,\tilde{\underline{B}}(\vec{v})\,d^3v)$.

The complete set of Dirac observables is comprised of $\hat{p}_\ph$ and $\widehat{{\vec v}|_{\ph_o}}$
($\hat{{\vec v}}$ at some particular instant $\ph=\ph_o$).
The representation of the Dirac observables on $\hilbert_{\rm phy}^{\rm WDW}$ is given by
(cf. \eqnref{eqn:Dirac observables 2}):
\be
\widehat{v_I|_{\ph_o}}\Ps({\vec v},\ph)=e^{i \sqrt{\underline{\Th}_+}\,(\ph-\ph_o)}\abs{v_I} \Ps({\vec v},\ph_o)
\quad\text{and}\quad
\hat{p}_\ph\Ps({\vec v},\ph)=\hbar\sqrt{\underline{\Th}_+}\Ps({\vec v},\ph).
\ee
Both $\hat{p}_\ph$ and $\widehat{{\vec v}|_{\ph_o}}$ are self-adjoint on the physical Hilbert space
$\hilbert_{\rm phy}^{\rm WDW}$ endowed with the inner product (cf. \eqnref{eqn:WDW inner product}):
\be
\inner{\Ps_1}{\Ps_2}_{\rm phy}=\int\!\! d^3v\,
\tilde{\underline{B}}({\vec v})\bar{\Ps}_1(\vec{v},\ph_o)\Ps_2(\vec{v},\ph_o).
\ee

The semi-classical (coherent) states are the same as discussed in \secref{subsec:semi-classical}.
The expectation value of $\widehat{{\vec v}|_{\ph_o}}$ is given by (cf. \eqnref{eqn:mean of mu}):
\ba
\langle\,{{v_I}|_{\ph_o}}\rangle&=&
\sgn(\m_I)\frac{2}{3\sqrt{k}}\langle\Abs{\m_I|_{\ph_o}}^{3/2}\rangle
\approx\sgn(\m_I)\frac{2}{3\sqrt{k}}\abs{\langle\m_I|_{\ph_o}\rangle}^{3/2}\nn\\
&\approx&\sgn(\m_I)\frac{2}{3\sqrt{k}}\abs{\m_{I\ph_o}^\star}^{3/2}
\ea
and the dispersion square of $\widehat{{\vec v}|_{\ph_o}}$ is (cf. \eqnref{eqn:variance of mu}):
\be
\D^2({v_I}|_{\ph_o})\approx
\left(\frac{2}{3\sqrt{k}}\right)^{\!\!2}\abs{\D({\m_I}|_{\ph_o})}^3
\approx
\left(\frac{2}{3\sqrt{k}}\right)^{\!\!2}
\left(e^{\frac{1}{2\s_I^2}}-1\right)^{\!\!3/2}\abs{\m_{I\ph_o}^\star}^3.
\ee
The approximations used here are valid
if the semi-classical states are sharply peaked at the classical values.

The WDW theory recast in terms of the affine variables gives us the asymptotic behavior
($\abs{v_{1,2,3}}\gg1$) of the general solutions and the physical sector of LQC, which will be discussed
in the next subsection.

\subsection{Emergent Time, General Solutions and the Physical Sector}\label{subsec:improved general solutions}
Just like the Hamiltonian equation \eqnref{eqn:master eq} in
$\muzero$-scheme, the improved Hamiltonian equation
\eqnref{eqn:improved master eq} in $\mubar$-scheme has the similar
structure. All the results studied in \secref{subsec:LQC general
solutions} -- \secref{subsec:LQC physical sector} can be easily
reproduced and modified for the $\mubar$-scheme in the obvious
manner (using the affine variables $v_I$). Again,
\eqnref{eqn:improved master eq} can be regarded as an evolution
equation with respect to the emergent time $\ph$ and the
$\ph$-independent operator $\Th({\vec v})$ is a difference operator.

The space of physical states is naturally divided into sectors, each of which
is preserved by the evolution and by the action of the Dirac observables.
This again implies the super-selection for different sectors.
Let ${\cal L}_{\abs{\e_I}}$ ($\abs{\e_I}\leq1$)
be the ``lattice'' of points $\{\abs{\e_I}+2n;n\in\mathbb{Z}\}$ on the $v_I$-axis,
${\cal L}_{-\abs{\e_I}}$  be the ``lattice'' of points $\{-\abs{\e_I}+2n;n\in\mathbb{Z}\}$
and ${\cal L}_{\e_I}={\cal L}_{\abs{\e_I}}\bigcup{\cal L}_{-\abs{\e_I}}$.
Also let $\hilbert^{\rm grav}_{\pm\abs{\e_1},\pm\abs{\e_2},\pm\abs{\e_3}}$ (8 of them) and
$\hilbert^{\rm grav}_{\vec \e}$ denote
the physical subspaces of $L^2(\mathbb{R}^3_{\rm Bohr},\tilde{B}({\vec v})\,d^3v_{\rm Bohr})$
with states whose support is restricted to the lattices
${\cal L}_{\pm\abs{\e_1}}\otimes{\cal L}_{\pm\abs{\e_2}}\otimes{\cal L}_{\pm\abs{\e_3}}$
and
${\cal L}_{{\e_1}}\otimes{\cal L}_{{\e_2}}\otimes{\cal L}_{{\e_3}}$ respectively.
Each of these subspaces is mapped to itself by $\Th$. Furthermore, since $\hat{C}^{(\mubar)}_{\rm grav}$
is self-adjoint on $\hilbert^{\rm grav}_{\rm kin}=L^2(\mathbb{R}^3_{\rm Bohr},d^3v_{\rm Bohr})$,
it follows that $\Th$ is self-adjoint on all these physical Hilbert subspaces.

As in the $\muzero$-scheme, since ${\cal L}_{\abs{\e_I}}$
and ${\cal L}_{-\abs{\e_I}}$ are mapped to each other by
the operator $\Pi_I$,
we restrict ourselves to
the symmetric sector (i.e. the subspace of $\hilbert^{\rm grav}_{\vec \e}$ in which
$\Ps({\vec v})=\Ps(-v_1,v_2,v_3)=\Ps(v_1,-v_2,v_3)=\Ps(v_1,v_2,-v_3)$).

For a generic $\vec{\e}$ (i.e., none of $\e_I$ equals to $0$ or $1$),
we can solve for the eigenvalue equation
$\Th\,e_\l({\vec v})=(\Th_1+\Th_2+\Th_3)\,e_\l({\vec v})=\l\,e_\l({\vec v})$
on each of the 8 Hilbert space $\hilbert^{\rm grav}_{\pm\abs{\e_1},\pm\abs{\e_2},\pm\abs{\e_3}}$.
Let
$e_\l(v_1,v_2,v_3)=e_{\l_1}(v_1)\,e_{\l_2}(v_2)\,e_{\l_3}(v_3)$ and solve
(cf. \eqnref{eqn:eigenvalue problems}):
\ba\label{eqn:eigenvalue problems improved}
\Th_3\,e_{\l_{12}}(v_1,v_2)&=&\l_{12}\,e_{\l_{12}}(v_1,v_2),
\quad\text{for  }
e_{\l_{12}}(v_1,v_2)=e_{\l_1}(v_1)\,e_{\l_2}(v_2),\nn\\
\Th_2\,e_{\l_{13}}(v_1,v_3)&=&\l_{13}\,e_{\l_{13}}(v_1,v_3),
\quad\text{for  }
e_{\l_{13}}(v_1,v_3)=e_{\l_1}(v_1)\,e_{\l_3}(v_3),\nn\\
\Th_1\,e_{\l_{23}}(v_2,v_3)&=&\l_{23}\,e_{\l_{23}}(v_2,v_3),
\quad\text{for  }
e_{\l_{23}}(v_2,v_3)=e_{\l_2}(v_2)\,e_{\l_3}(v_3),
\ea
with $\l=\l_{12}+\l_{13}+\l_{23}$. (Note that $\Th_I$ is independent of $v_I$.)
Each of \eqnref{eqn:eigenvalue problems improved} can be solved recursively for any arbitrary $\l_{ij}$
provided that the appropriate boundary condition is given.
For example, $\Th_3\,e_{\l_{12}}(v_1,v_2)=\l_{12}\,e_{\l_{12}}(v_1,v_2)$ reads as
(cf. \eqnref{eqn:recursion relation}):
\ba\label{eqn:recursion relation improved}
&&\widetilde{C}_3^{++}(v_1,v_2)\,e_{\l_{12}}(v_1+2,v_2+2)
-\widetilde{C}_3^{-+}(v_1,v_2)\,e_{\l_{12}}(v_1-2,v_2+2)\nn\\
&&-\widetilde{C}_3^{+-}(v_1,v_2)\,e_{\l_{12}}(v_1+2,v_2-2_2)
+\widetilde{C}_3^{--}(v_1,v_2)\,e_{\l_{12}}(v_1-2,v_2-2)\nn\\
&=&-\l_{12}\,e_{\l_{12}}(v_1,v_2),
\ea
where we define
\be
\widetilde{C}_I^{\pm\pm}(v_J,v_K):=\frac{\p G}{2}[\tilde{B}(\vec{v})]^{-1}C_I^{\pm\pm}(\vec{v}),
\ee
which is independent of $\m_I$ and has the same properties as $\widetilde{C}_I^{\pm\pm}(\m_J,\m_K)$
does in \eqnref{eqn:coefficients tilde C 1}--\eqnref{eqn:coefficients tilde C 3}.
Since the coefficients $\widetilde{C}_3^{\pm\pm}$ never vanish or go to infinity on the generic lattice,
$e_{\l_{12}}(v_1-2,v_2-2)$ can be obtained for any $\l_{12}$
when the values of $e_{\l_{12}}(v_1+2,v_2+2)$,
$e_{\l_{12}}(v_1-2,v_2+2)$, $e_{\l_{12}}(v_1+2,v_2-2)$ and
$e_{\l_{12}}(v_1,v_2)$ are given.
As a result, if the values of $e_{\l_{12}}(v_1,v_2)$ are specified on the boundary lines
$v_1=v_1^\star$, $v_1=v_1^\star-2$ and $v_2=v_2^\star$, $v_2=v_2^\star-2$
in the lattice ${\cal L}_{\pm\abs{\e_1}}\otimes{\cal L}_{\pm\abs{\e_2}}$,
$e_{\l_{12}}(v_1,v_2)$ can be uniquely determined via the recursion relation
\eqnref{eqn:recursion relation improved}
for all the lattice points with $v_1<v_1^\star$ and $v_2<v_2^\star$.

On the other hand, the LQC operator $\Th_I$ reduces to the WDW operator $\underline{\Th}_I$
in the asymptotic regime (i.e. $\abs{v_I}\gg1$), and thus the LQC eigenfunction
$e_\l({\vec v})$ approaches the WDW eigenfunction $\underline{e}_{\vec k}({\vec v})$
given by \eqnref{eqn:eigenfunction ek improved} for large $\abs{v_I}$. Therefore, we can use the same
labels $k_i$ to denote the LQC eigenfunctions (i.e., $\l_{1,2,3}=k_{1,2,2}$) and similarly
decompose $\underline{e}_{\vec k}({\vec v})$ as
$\underline{e}_{k_1}\!(v_1)\,\underline{e}_{k_2}\!(v_2)\,\underline{e}_{k_3}\!(v_3)$.
It follows that $e_{k_I}(v_I)\rightarrow\underline{e}_{k_I}(v_I)$ for $\abs{v_I}\gg1$
and we have the identifications \eqnref{eqn:lambda ij} and \eqnref{eqn:lambda omega} unchanged.
Again, the LQC operator $\Th$ is not positive definite
and only those modes with $\o^2>0$ are allowed.
What matters is the operator $\Th_+$, obtained by projecting the action of $\Th$
to the positive eigenspace in its spectral decomposition.

For a given ${\vec k}$, $e_{\l_{IJ}}$ can be obtained as mentioned above by appropriately specifying the boundaries
in such a way that $e_{\l_{IJ}}\!(v_I,v_J)\rightarrow\underline{e}_{k_I}\!(v_J)\,\underline{e}_{k_J}\!(v_J)$ for
$\abs{v_I}\gg1$, $\abs{v_J}\gg1$.
To be used for the boundary condition, however, the asymptotic function
$\underline{e}_{k_1}\!(v_1)\,\underline{e}_{k_2}\!(v_2)=
(4\p)^{-2}(\frac{3\sqrt{k}}{2})^{-\frac{1}{3}+\frac{2i}{3}(k_1+k_2)}
\abs{v_1}^{-\frac{1}{6}+\frac{2i}{3}k_1}\abs{v_2}^{-\frac{1}{6}+\frac{2i}{3}k_2}$
given by \eqnref{eqn:eigenfunction ek improved} should be further ``tamed''.
Following the same spirit used for $\muzero$-scheme,
the value of $e_{\l_{12}}\!(\m_1,\m_2)$ on the boundary is given by
(cf. \eqnref{eqn:modified boundary}):
\ba\label{eqn:modified boundary improved}
\frac{\big({3\sqrt{k}}/{2}\big)^{-\frac{1}{3}+\frac{2i}{3}(k_1\!+\!k_2)}}{(4\p)^2}
\left(\frac{3}{2}\abs{\abs{v_1\!+\!1}^\frac{1}{3}
\!-\!\abs{v_1\!-\!1}^\frac{1}{3}}\right)^{\!\!\frac{1}{4}-ik_1}
\!\!\!\!\!\!\!\times\!
\left(\frac{3}{2}\abs{\abs{v_2\!+\!1}^\frac{1}{3}
\!-\!\abs{v_2\!-\!1}^\frac{1}{3}}\right)^{\!\!\frac{1}{4}-ik_2},\quad
\ea
which again reduces to $\underline{e}_{k_I}\!(v_J)\,\underline{e}_{k_I}\!(v_J)$ when both
$\abs{v_I}\gg1$ and $\abs{v_J}\gg1$.
In practice, we use \eqnref{eqn:modified boundary improved} to specify the boundary condition on the lines
$(v_1=v_1^\star,\,-v_2^\star\leq v_2\leq v_2^\star)$, $(v_1=v_1^\star-2,\,-v_2^\star\leq v_2\leq v_2^\star)$,
$(-v_1^\star\leq v_1\leq v_1^\star,\,v_2=v_2^\star)$ and
$(-v_1^\star\leq v_1\leq v_1^\star,\,v_2=v_2^\star-2)$ of the lattice
for $v_1^\star\gg1$ and $v_2^\star\gg1$,
then the recursion relation \eqnref{eqn:recursion relation improved} uniquely gives the values
$e_{\l_{12}}\!(v_1,v_2)$ for all the lattice points
within the range $-v_1^\star\leq v_1\leq v_1^\star$ and $-v_2^\star\leq v_2\leq v_2^\star$.

Similarly, all the other results and remarks in \secref{subsec:LQC general solutions} --
\secref{subsec:LQC physical sector} can be easily modified for the $\mubar$-scheme
by formally taking the replacements: $\m_I\rightarrow v_I$
and $B(\vec{\m})\rightarrow\tilde{B}(\vec{v})$.
In particular, the orthonormality relation for
the eigenfunctions in the symmetric sector of $\hilbert^{\rm grav}_{\vec \e}$ is given by
(cf. \eqnref{eqn:LQC orthonormality rel}):
\be
\sum_{{\vec v}\in{\cal L}_{\e_1}\otimes{\cal L}_{\e_2}\otimes{\cal L}_{\e_3}}\!\!\!\!\!
\tilde{B}({\vec v})\,\bar{e}^{(s)}_{\vec k}({\vec v})\,e^{(s)}_{{\vec k}'}({\vec v})
=\d^3({\vec k},{\vec k}')
\ee
and the completeness relation reads as (cf. \eqnref{eqn:LQC completeness rel}):
\be
\sum_{{\vec v}\in{\cal L}_{\e_1}\otimes{\cal L}_{\e_2}\otimes{\cal L}_{\e_3}}\!\!\!\!\!
\tilde{B}({\vec v})\,\bar{e}^{(s)}_{\vec k}({\vec v})\Ps({\vec v})
=0
\quad\text{for }
\forall{\vec k},
\quad\text{iff }
\Ps({\vec v})=0
\ee
for any symmetric $\Ps({\vec v})\in\hilbert^{\rm grav}_{\vec \e}$.
The general symmetric solution to the improved LQC constraint \eqnref{eqn:improved master eq}
can be written as (cf. \eqnref{eqn:sol to master eq}):
\be\label{eqn:sol to improved master eq}
\Ps({\vec v},\ph)=\int_{\o^2({\vec k})>0}\!\!d^3k
\left(
\tilde{\Ps}_+({\vec k})\,e^{(s)}_{\vec k}({\vec v})\,e^{i\o\ph}+
\tilde{\Ps}_-({\vec k})\,\bar{e}^{(s)}_{\vec k}({\vec v})\,e^{-i\o\ph}
\right)
\ee
where $\tilde{\Ps}_\pm({\vec k})$ are in $L^2(\mathbb{R}^3,d^3k)$.
The solution to \eqnref{eqn:improved master eq} with the ``initial'' function
$\Ps_\pm({\vec v},\ph_o)=f_\pm({\vec v})$
is (cf. \eqnref{eqn:sol given by initial function}):
\be
\Ps_\pm({\vec v},\ph)=e^{\pm i \sqrt{\Th_+}\,(\ph-\ph_o)} \Ps_\pm({\vec v},\ph_o),
\ee
where $\sqrt{\Th_+}$ is the square-root of the positive self-adjoint operator $\Th_+$
defined via spectral decomposition.

Finally, the super-selection sector of the physical Hilbert space $\hilbert_{\rm phy}^{\vec \e}$ labeled by
${\vec \e}\in[0,1]\times[0,1]\times[0,1]$ consists of the positive frequency solutions
$\Ps({\vec v},\ph)$ to \eqnref{eqn:improved master eq} with initial data $\Ps({\vec v},\ph_o)$ in the symmetric sector
of $\hilbert^{\rm grav}_{\vec \e}$. By \eqnref{eqn:sol to improved master eq},
the explicit expression in terms of the
symmetric eigenfunctions $e_{\vec k}^{(s)}$ is given by (cf. \eqnref{eqn:LQC physical functions}):
\be
\Ps({\vec v},\ph)=\int_{\o^2({\vec k})>0}\!\!\!\!\!d^3k\,
\tilde{\Ps}({\vec k})\,e^{(s)}_{\vec k}({\vec v})\,e^{i\o\ph},
\ee
where $\o^2({\vec k})=16\p G(k_1k_2+k_1k_3+k_2k_3+3/16)$.
The physical inner product on the physical sector
$\hilbert_{\rm phys}^{\vec \e}$
is given by (cf. \eqnref{eqn:LQC inner product}):
\be
\inner{\Ps_1}{\Ps_2}_{\rm phy}^{\vec \e}=\sum_{{\vec v}\in{\cal L}_{\e_1}\otimes{\cal L}_{\e_2}\otimes{\cal L}_{\e_3}}
\!\!\!\!\!\!\tilde{B}({\vec v})\,\bar{\Ps}_1({\vec v},\ph_o)\Ps_2({\vec v},\ph_o).
\ee
The action of the Dirac observables is independent of ${\vec \e}$ and has the same form as in the WDW theory
(cf. \eqnref{eqn:LQC Dirac observables}):
\be
\widehat{v_I|_{\ph_o}}\Ps({\vec v},\ph)=e^{i \sqrt{\Th_+}\,(\ph-\ph_o)}\abs{v_I} \Ps({\vec v},\ph_o)
\quad\text{and}\quad
\hat{p}_\ph\Ps({\vec v},\ph)=-i\hbar\frac{\partial\Ps({\vec v},\ph)}{\partial\ph}.
\ee

Just like the case in $\muzero$-scheme, because of super-selection,
each physical sector $\hilbert_{\rm phy}^{\vec \e}$ is separable
while the kinematical Hilbert space $\hilbert_{\rm kin}^{\rm total}$
is not. The standard type of discreteness of the Dirac observables
$\widehat{v_I|_{\ph_o}}$ on each physical sector is descended from
the more delicate discreteness of the spectrum of $\hat{v}_I$ on
$\hilbert_{\rm kin}^{\rm grav}=L^2(\mathbb{R}^3_{\rm Bohr},d^3v_{\rm
Bohr})$.

Finally, it is easy to check that the coefficients
$C^{\pm\pm}_I({\vec v})$ and $\tilde{C}^{\pm\pm}_I(v_J,v_K)$ have
the properties exactly the same as those of $C^{\pm\pm}_I({\vec
\m})$ and $\tilde{C}^{\pm\pm}_I(\m_J,\m_K)$ listed in
\eqnref{eqn:coefficients C 1}, \eqnref{eqn:coefficients C 2} and
\eqnref{eqn:coefficients tilde C 1}--\eqnref{eqn:coefficients tilde
C 3}. Together with the fact that
$\tilde{B}(0,v_2,v_3)=\tilde{B}(v_1,0,v_3)=\tilde{B}(v_1,v_2,0)=0$,
the argument used in \secref{subsec:resolution of singularity} can
be applied to prove the resolution of the classical singularity in
$\mubar$-scheme and to suggest the occurrence of the big bounce.
Moreover, by the lesson learned in \cite{Ashtekar:2006wn}, in
$\mubar$-scheme, the bounces are expected to happen when the
directional factor $\varrho_I:=p_\ph^{2}/\abs{p_I}^3$ enters the
Planck regime of ${\cal O}(\hbar\,\Pl^{-4})$, independent of the
values of $p_\ph$. Meanwhile, the planar collapse is not resolved
but persists in $\mubar$-scheme, yielding the same comments as made
in the end of \secref{subsec:resolution of singularity}.

\section{Discussion}\label{sec:discussion}
The analytical investigation in this paper is the direct extension
of \cite{Ashtekar:2006uz} and \cite{Ashtekar:2006wn}, developing a
comprehensive LQC framework for the anisotropic Bianchi I model.
Both the precursor ($\muzero$-) and the improved ($\mubar$-)
strategies are successfully deployed. The key features responsible
for the desired results remain the same: (i) the kinematical
framework of LQC forces us to replace the connection variables by
the holonomies, and (ii) the underlying quantum geometry gives rise
to the discreteness in the spectrum of the triad operators, as the
imprint of the fundamental ``area gap'' in the full theory.
Furthermore, following the $\mubar$-scheme newly proposed in
\cite{Ashtekar:2006wn}, a more sophisticated prescription is
introduced to impose the discreteness, thereby capturing the
underlying physics of quantum geometry in a more sensible way.

The success of this extension affirms the robustness of the methods
in \cite{Ashtekar:2006uz,Ashtekar:2006wn} and thus enlarges their
domain of validity. It also shows the fact that the singularity
persists in the WDW theory is rather generic, even in the less
symmetric model. The main difference with anisotropy is that the
operator $\Th$ is no longer positive definite. Nevertheless, the
detailed analysis shows that the only relevant part is $\Th_+$ (the
projection of $\Th$ on its positive eigenspace) and therefore the
strategies can be readily carried through without a big
modification. Consequently, the expected results for the Bianchi I
model are established: (i) the scalar field $\ph$ again serves as an
internal clock and is treated as emergent time; (ii) the total
Hamiltonian constraint is derived in both $\muzero$- and
$\mubar$-schemes and gives the evolution as a difference equation;
and (iii) the physical Hilbert space, Dirac observables and
semi-classical states are constructed rigorously.

On the other hand, the variety of anisotropy does give new features,
notably distinct from those in isotropic models. Unlike the
isotropic case, the Bianchi I model with a massless scalar field
admits both Kasner-like and Kasner-unlike solutions in the classical
dynamics. In the quantum theory, we anticipate that the singularity
(both Kasner-like and Kasner-unlike) is resolved. This is evidenced
by the observation: in the kinematical Hilbert space, \emph{the
state associated with the classical singularity is completely
decoupled} from the other states in the difference evolution
equation, and therefore \emph{the evolution is deterministic} across
the deep Planck regime.

Meanwhile, we have the somewhat surprising observation that the
planar collapse for Kasner-like solutions is not resolved and thus
the vanishing behavior of the length scale factors $a_I$ is not
prevented by the quantum effect. This suggests that either area is
more fundamental than length in LQG or some important ingredient is
missing when we impose the discreteness correction from the full
theory. Either case requires further studies.

To actually figure out the behaviors of the quantum evolution, the
analytical investigation alone is not enough. A complete numerical
analysis or at least the study of effective dynamics is in demand
\cite{in progress, to appear}. We expect that the \emph{big bounce}
occurs whenever one of the area scales $p_I$ undergoes the vanishing
behavior. It may happens as many as three times. Furthermore, in the
improved dynamics ($\mubar$-scheme), the bounce takes place at the
moment when the directional factor $\varrho_I$ in any of the three
diagonal directions comes close to the critical value in the Planck
regime, regardless of the values of $p_\ph$.

Other possible extensions have been discussed and commented in
Discussion of \cite{Ashtekar:2006uz}. In particular, in order to
understand the BKL scenario in the context of LQC, the next step
would be to explore Bianchi IX models based on the same formulation.
Meanwhile, as both the isotropic and Bianchi I models have been
analyzed in great detail, another feasible research direction is to
study the emergence of isotropy in the arena of Bianchi I models,
which could shed some light on the symmetry reduction for more
complicated situations. As remarked in the end of
\appref{subsec:first scheme for isotropic}, this issue is not
trivial and demands more detailed investigation.

\begin{acknowledgments} I am grateful for the warm hospitality during
my current visit at IGPG and the helpful discussions with Golam
Hossain, Tomasz Pawlowski, Parampreet Singh, Kevin Vandersloot and
especially Abhay Ashtekar, who introduced me the field of LQC and
initiated this research. It is also greatly appreciated that Ori
Ganor encouraged me to explore LQG. This work was supported in part
by the Director, Office of Science, Office of High Energy and
Nuclear Physics, of the U.S. Department of Energy under Contract
DE-AC03-76SF00098.
\end{acknowledgments}

\appendix

\section{WDW Semi-Classical (Coherent) States}\label{sec:coherent states}
In this appendix, we study more details about the WDW semi-classical states described in
\secref{subsec:semi-classical}.
We first look closer at the semi-classical state to better understand its structure
and then compute the physical quantities corresponding to the Dirac observables.

The semi-classical states in WDW theory are constructed in
\eqnref{eqn:semi-classical state} with $\tilde{\Ps}_{\vec \k}({\vec
k})$ sharply peaked at ${\vec k}={\cal K}{\vec \k}$. Since
$\tilde{\Ps}_{\vec \k}({\vec k})$ is sharply distributed, only those
${\vec k}$ close to ${\cal K}{\vec \k}$ contribute. For ${\cal
K}\k_\ph\gg1$, $\o^2({\vec k})\approx 16\p
G(k_1k_2+k_1k_3+k_2k_3)\approx 8\p G {\cal K}^2\k_\ph^2$ in the vicinity of
${\vec k}={\cal K}{\vec \k}$. As a result, the physical condition
$\o^2>0$ is automatically satisfied for the relevant ${\vec k}$.
Therefore, when a sharply-peaked semi-classical state is concerned,
the restriction $\o^2>0$ for the integral range in
\eqnref{eqn:semi-classical state} can be ignored. We will make use
of this fact to simplify the following calculations.

With the inner product defined in \eqnref{eqn:WDW inner product}, we have
\ba
&&\inner{\Ps_{{\vec \m}_{\ph_o}^\star,p_\ph^\star}}{\Ps_{{\vec \m}_{\ph_o}^\star,p_\ph^\star}}_{\rm phy}=
\int_{\o^2({\vec k}),\,\o^2({\vec k}')>0}
\!\!\!\!\!\!\!\!\!\!\!\!\!d^3k'd^3k\ \d^3({\vec k},{\vec k}')
\bar{\tilde{\Ps}}_{\vec \k}({\vec k}') {\tilde{\Ps}}_{\vec \k}({\vec k})\,
e^{i(\o({\vec k})-\o({\vec k}'))(\ph_o-\ph_o^\star)}\nn\\
&\approx& \frac{1}{(2\p)^{3/2}\s_1\s_2\s_3}\int_{-\infty}^\infty\!\int_{-\infty}^\infty\!\int_{-\infty}^\infty
dk_1dk_2dk_3\
e^{-\frac{(k_1-{\cal K}\k_1)^2}{2\s_1^2}}
e^{-\frac{(k_2-{\cal K}\k_2)^2}{2\s_2^2}}
e^{-\frac{(k_3-{\cal K}\k_3)^2}{2\s_3^2}}
=1,
\ea
where the identity \eqnref{eqn:orthonormality rel} is used and the restriction $\o^2>0$ is released.
The overall factors in \eqnref{eqn:Fourier amplitude} has been chosen to make the physical norm of the
state normalized.

Before computing the expectation value and dispersion for $\widehat{{\vec \m}|_{\ph_o}}$,
first note that by $e^{a\frac{d}{dx}}f(x)=f(x+a)$ and \eqnref{eqn:eigenfunction ek}
we have
\be\label{eqn:trick on ek}
e^{-i\a\frac{\partial}{\partial k_I}}\underline{e}_{{\vec k}}({\vec \m})=
\abs{\m_I}^\a\underline{e}_{{\vec k}}({\vec \m})
\ee
and consequently
\be
\int d^3\!\m\,\underline{B}(\vec{\m})\,\bar{\underline{e}}_{\vec k}({\vec \m})\,
\abs{\m_I}^\a
\underline{e}_{{\vec k}'}({\vec \m})
=e^{-i\a\frac{\partial}{\partial k_I}}\d^3(\vec{k},\vec{k}').
\ee

The derivatives of $\d^3({\vec k},{\vec k}')$ can be handled with integral by part. This then
leads to
\ba
\langle\,{{\m_I}|_{\ph_o}}\rangle&:=&
\bra{\Ps_{{\vec \m}_{\ph_o}^\star,p_\ph^\star}}
\,\widehat{{\m_I}|_{\ph_o}}\,
\ket{\Ps_{{\vec \m}_{\ph_o}^\star,p_\ph^\star}}_{\rm phy}\nn\\
&\approx&\int d^3k\,d^3k'\ \d^3({\vec k},{\vec k}')\,
\bar{\tilde{\Ps}}_{\vec \k}({\vec k}')\,e^{-i\o({\vec k}')(\ph_o-\ph_o^\star)}\,
e^{i\frac{\partial}{\partial k_I}}
\left(
\tilde{\Ps}_{\vec \k}({\vec k})\,e^{i\o({\vec k})(\ph_o-\ph_o^\star)}
\right).
\ea
Applying $e^{a\frac{d}{dx}}f(x)=f(x+a)$ again, we have (taking $I=1$)
\ba
&&e^{i\frac{\partial}{\partial k_1}}
\left(
\tilde{\Ps}_{\vec \k}({\vec k})\,e^{i\o({\vec k})(\ph_o-\ph_o^\star)}
\right)=
\tilde{\Ps}_{\vec \k}(k_1+i,k_2,k_3)\,e^{i\o(k_1+i,k_2,k_3)(\ph_o-\ph_o^\star)}\nn\\
&=&
e^{\frac{-2i(k_1-{\cal K}\k_1)+1}{(2\s_1)^2}}
e^{\sqrt{4\p G}\left(\frac{k_2+k_3}{\sqrt{k_1k_2+k_1k_2+k_2k_3}}+{\cal O}(k^{-1})\right)(\ph_o^\star-\ph_o)}\
\tilde{\Ps}_{\vec \k}({\vec k})\,e^{i\o({\vec k})(\ph_o-\ph_o^\star)}.
\ea
This follows
\ba
\langle\,{{\m_1}|_{\ph_o}}\rangle
&\approx&
\int d^3k\
e^{\frac{-2i(k_1-{\cal K}\k_1)+1}{(2\s_1)^2}}
e^{\sqrt{4\p G}\left(\frac{k_2+k_3}{\sqrt{k_1k_2+k_1k_3+k_2k_3}}\right)(\ph_o^\star-\ph_o)}\,
\bar{\tilde{\Ps}}_{\vec \k}({\vec k})
\tilde{\Ps}_{\vec \k}({\vec k})\nn\\
&\approx&
e^{\frac{1}{(2\s_1)^2}}
e^{\sqrt{8\p G}\big(\frac{\k_2+\k_3}{\k_\ph}\big)(\ph_o^\star-\ph_o)}\!\!
\int d^3k\,\bar{\tilde{\Ps}}_{\vec \k}({\vec k})
\tilde{\Ps}_{\vec \k}({\vec k})\nn\\
&=&e^{\frac{1}{(2\s_1)^2}}
e^{\sqrt{8\p G}\big(\frac{\k_2+\k_3}{\k_\ph}\big)(\ph_o^\star-\ph_o)}\equiv\m_{1\ph_o}^\star,
\ea
where in the second line, we simply assign the peaked value to the integrand.
This is the result in \eqnref{eqn:mean of mu}.

Similarly, taking $\a=2$ in \eqnref{eqn:trick on ek} and following the same procedure,
we have
\ba
\langle({{\m_1}|_{\ph_o}})^2\rangle
&\approx&
\int d^3k\
e^{\frac{-4i(k_1-{\cal K}\k_1)+4}{(2\s_1)^2}}
e^{\sqrt{16\p G}\left(\frac{k_2+k_3}{\sqrt{k_1k_2+k_1k_3+k_2k_3}}\right)(\ph_o^\star-\ph_o)}\,
\bar{\tilde{\Ps}}_{\vec \k}({\vec k})
\tilde{\Ps}_{\vec \k}({\vec k})\nn\\
&\approx&
e^{\frac{4}{(2\s_1)^2}}
e^{2\sqrt{8\p G}\big(\frac{\k_2+\k_3}{\k_\ph}\big)(\ph_o^\star-\ph_o)}\!\!
\int d^3k\,\bar{\tilde{\Ps}}_{\vec \k}({\vec k})
\tilde{\Ps}_{\vec \k}({\vec k})\nn\\
&=&e^{\frac{4}{(2\s_1)^2}}
e^{2\sqrt{8\p G}\big(\frac{\k_2+\k_3}{\k_\ph}\big)(\ph_o^\star-\ph_o)}
\equiv e^{\frac{1}{2\s_1^2}}(\m_{1\ph_o}^\star)^2,
\ea
which leads to the variance in \eqnref{eqn:variance of mu}.

Next, to compute the expectation value for $p_\ph$, we recall the action of $\hat{p}_\ph$ defined in
\eqnref{eqn:Dirac observables} and use the fact
$\sqrt{\underline{\Th}_+}\,\underline{e}_{\vec k}({\vec \m})=\o({\vec k})\,\underline{e}_{\vec k}({\vec \m})$.
It follows
\ba\label{eqn:computing p-ph}
\langle p_\ph\rangle&=&
\int_{\o^2({\vec k})>0}\!\!d^3k\,\hbar\,\o({\vec k})\,\bar{\tilde{\Ps}}_{\vec \k}({\vec k})
\tilde{\Ps}_{\vec \k}({\vec k})
\approx
\hbar\,\o({\cal K}{\vec \k})\int d^3k\,\bar{\tilde{\Ps}}_{\vec \k}({\vec k})
\tilde{\Ps}_{\vec \k}({\vec k})\nn\\
&=&\hbar\sqrt{8\p G}{\cal K}\k_\ph\equiv p_\ph^\star,
\ea
where again we ignore the restriction for the integral range and directly put the peaked value
to the integrand. This is the result in \eqnref{eqn:mean of p-phi}.

Following the same procedure, we have
\ba\label{eqn:computing p-ph square}
\langle p_\ph^2\rangle&=&
\int_{\o^2({\vec k})>0}\!\!d^3k\,\hbar^2\,\o^2({\vec k})\,\bar{\tilde{\Ps}}_{\vec \k}({\vec k})
\tilde{\Ps}_{\vec \k}({\vec k})\nn\\
&\approx&
\frac{16\p G\hbar^2}{(2\p)^{3/2}\s_1\s_2\s_3}\!\int\!d^3k(k_1k_2\!+\!k_1k_3\!+\!k_2k_3)\,
e^{-\frac{(k_1-{\cal K}\k_1)^2}{2\s_1^2}}
e^{-\frac{(k_2-{\cal K}\k_2)^2}{2\s_2^2}}
e^{-\frac{(k_3-{\cal K}\k_2)^2}{2\s_3^2}}\nn\\
&=&16\p G\hbar^2{\cal K}^2(\k_1\k_2+\k_1\k_3+\k_2\k_3)\equiv p_\ph^{\star2},
\ea
which leads to $\D^2(p_\ph)=\langle p_\ph^2\rangle-\langle p_\ph\rangle^2=0$.
However, the vanishing of $\D(p_\ph)$ is artificial, resulting merely from the approximations we have used in
\eqnref{eqn:computing p-ph} and \eqnref{eqn:computing p-ph square},
which neglect the variation in $\D(p_\ph)$ in the order of
$p_\ph^\star\,{\cal O}\!\left((\frac{\s}{{\cal K}\k})^{n\geq1}\right)$.
In order to recover the correct estimate of $\D(p_\ph)$ for the desired order
while keep the integration tractable,
we assume $\langle p_\ph^2\rangle\approx\langle p_\ph^4\rangle^{1/2}$
(based on the assumption that $\tilde{\Ps}_{\vec \k}({\vec k})$ is sharply peaked)
and compute $\langle p_\ph^4\rangle$ instead:
\ba
\langle p_\ph^4\rangle&=&
\int_{\o^2({\vec k})>0}\!\!d^3k\,\hbar^2\,\o^4({\vec k})\,\bar{\tilde{\Ps}}_{\vec \k}({\vec k})
\tilde{\Ps}_{\vec \k}({\vec k})\nn\\
&\approx&
\frac{(16\p G)^2\hbar^4}{(2\p)^{3/2}\s_1\s_2\s_3}\!\int\!d^3k(k_1k_2\!+\!k_1k_3\!+\!k_2k_3)^2\,
e^{-\frac{(k_1-{\cal K}\k_1)^2}{2\s_1^2}}
e^{-\frac{(k_2-{\cal K}\k_2)^2}{2\s_2^2}}
e^{-\frac{(k_3-{\cal K}\k_2)^2}{2\s_3^2}}\nn\\
&=&(16\p G)^2\hbar^4
\Big\{
{\cal K}^2\big(\s_1^2(\k_2+\k_3)^2+\s_2^2(\k_1+\k_3)^2+\s_3^2(\k_1+\k_2)^2\big)\nn\\
&&\quad
+(\s_1^2\s_2^2+\s_1^2\s_3^2+\s_2^2\s_3^2)
+{\cal K}^4(\k_1\k_2+\k_1\k_3+\k_2\k_3)^2
\Big\}\nn\\
&\equiv&
p_\ph^{\star4}
\Big\{
1+\frac{4}{{\cal K}^2\k_\ph^4}\big(\s_1^2(\k_2+\k_3)^2+\s_2^2(\k_1+\k_3)^2+\s_3^2(\k_1+\k_2)^2\big)\nn\\
&&\qquad\quad
+\frac{4}{{\cal K}^4\k_\ph^4}(\s_1^2\s_2^2+\s_1^2\s_3^2+\s_2^2\s_3^2)
\Big\}.
\ea
This then gives
\ba
\langle p_\ph^2\rangle\approx\langle p_\ph^4\rangle^{1/2}
&=&p_\ph^{\star2}
\Big\{
1+\frac{2}{{\cal K}^2\k_\ph^4}\big(\s_1^2(\k_2+\k_3)^2+\s_2^2(\k_1+\k_3)^2+\s_3^2(\k_1+\k_2)^2\big)\nn\\
&&\qquad\quad+{\cal O}\left(\big(\frac{\s}{{\cal K}\k}\big)^4\right)
\Big\}
\ea
and therefore
\be
\D^2(p_\ph)=\langle p_\ph^2\rangle-\langle p_\ph\rangle^2
\approx
\frac{2p_\ph^{\star2}}{{\cal K}^2\k_\ph^4}
\big(\s_1^2(\k_2+\k_3)^2+\s_2^2(\k_1+\k_3)^2+\s_3^2(\k_1+\k_2)^2\big),
\ee
which is the result shown in \eqnref{eqn:variance of p-phi}.

\section{Remarks on the Isotropic Model}\label{sec:isotropic model}
\subsection{Precursor Strategy ($\m_o$-Scheme)}\label{subsec:first scheme for isotropic}
In this subsection, we discuss how to read off the gravitational
constraint operator for the isotropic case from the results obtained
in \secref{subsec:grav constraint operator} and \secref{subsec:self
adjoint and WDW limit} for the anisotropic cases. In the end, we
comment on the symmetry reduction from anisotropy to isotropy.

In the isotropic model, the classical phase space is given by the variables $c$ and $p$ with
the fundamental Poisson bracket (note the difference of the factor 3):
\be
\{c,p\}=\frac{8\p\g G}{3}.
\ee
In quantum kinematics, the eigenstates of $\hat{p}$ are the basis states $\ket{\m}$:
\be
\hat{p}\ket{\m}=\frac{4\p\g\Pl^2}{3}\m\ket{\m}.
\ee
The basis states are also the eigenstates of the volume operator $\hat{V}$:
\be
\hat{V}\ket{\m}=V_\m\ket{\m}:=\left(\frac{4\p\g}{3}\abs{\m}\right)^{\frac{3}{2}}\Pl^3\ket{\m}.
\ee

To reproduce the isotropic result from the anisotropic Bianchi I models, we can simply
take the replacement:
\be
c^1,c^2,c^3\longrightarrow c,
\qquad
p_1,p_2,p_3\longrightarrow p.
\ee
In LQC, correspondingly, we take
\be
\ket{\m_1+\a\,\muzero_1,\m_2+\b\,\muzero_2,m_3+\g\,\muzero_3}
\longrightarrow
\ket{\m+(\a+\b+\g)\m_o}
\ee
and
\be
V_{\m_1+\a\,\muzero_1,\m_2+\b\,\muzero_2,\m_3+\g\,\muzero_3}
\longrightarrow
V_{\m+(\a+\b+\g)\m_o},
\ee
where in the end we also impose
\be
\m_o=3\,\muzero_I=3k\equiv3\sqrt{3}/2.
\ee
The replacement rules then ``translate''
\eqnref{eqn:original C grav} \eqnref{eqn:hat C grav prime}, \eqnref{eqn:hat C grav}
and \eqnref{eqn:original C grav on states} \eqnref{eqn:hat C grav prime on states} \eqnref{eqn:hat C grav on states}
to their counterparts in the isotropic model:
\be
\hat{C}^{(\m_o)}_{{\rm grav},o}=\frac{24i\ \sgn(p)}{8\p\g^3\Pl^2\m_o^3}
\left[
\sin^2({\m_o c})\left(\sin(\frac{\m_o c}{2}){\hat V}\cos(\frac{\m_o c}{2})
-\cos(\frac{\m_o c}{2}){\hat V}\sin(\frac{\m_o c}{2})\right)
\right],
\ee
\be
\hat{C}^{(\m_o)'}_{\rm grav}=
\frac{1}{2}\left[\hat{C}^{(\m_o)}_{{\rm grav},o}+\hat{C}^{(\m_o)\dagger}_{{\rm grav},o}\right],
\ee
\be
\hat{C}^{(\m_o)}_{\rm grav}=\frac{24i\ \sgn(p)}{8\p\g^3\Pl^2\m_o^3}
\left[
\sin({\m_o c})\left(\sin(\frac{\m_o c}{2}){\hat V}\cos(\frac{\m_o c}{2})
-\cos(\frac{\m_o c}{2}){\hat V}\sin(\frac{\m_o c}{2})\right)
\sin({\mu_o c})\right]
\ee
and
\be
\hat{C}^{(\m_o)}_{{\rm grav},o}\ket{\m}=
\frac{3}{8\p\g^3\Pl^2\m_o^3}\Big\{
\abs{V_{\m+\m_o}-V_{\m-\m_o}}\left(\ket{\m+4\m_o}-2\ket{\m}+\ket{\m-4\m_o}\right)
\Big\},
\ee
\ba
&&\hat{C}^{(\m_o)'}_{\rm grav}\ket{\m}
=
\frac{3}{16\p\g^3\Pl^2\m_o^3}\Big\{
\big(\abs{V_{\m+5\m_o}-V_{\m+3\m_o}}+\abs{V_{\m+\m_o}-V_{\m-\m_o}}\big)\ket{\m+4\m_o}\nn\\
&&\qquad+\big(\abs{V_{\m-3\m_o}-V_{\m-5\m_o}}+\abs{V_{\m+\m_o}-V_{\m-\m_o}}\big)\ket{\m-4\m_o}
-4\abs{V_{\m+\m_o}-V_{\m-\m_o}}\ket{\m}
\Big\},\qquad
\ea
\ba\label{eqn:hat C grav isotropic}
{\hat C}_{\rm grav}^{(\m_o)}\ket{\m}
&=&
\frac{6}{16\p\g^3\Pl^2\m_o^3}\Big\{
\abs{V_{\m+3\m_o}-V_{\m+\m_o}}\ket{\m+4\m_o}
+\abs{V_{\m-\m_o}-V_{\m-3\m_o}}\ket{\m-4\m_o}\nn\\
&&\qquad\qquad\qquad-\big(\abs{V_{\m+3\m_o}-V_{\m+\m_o}}+\abs{V_{\m-\m_o}-V_{\m-3\m_o}}\big)\ket{\m}
\Big\}.
\ea
The last one with $\m_o=3k$ is the master equation used in
\cite{Ashtekar:2006uz}.

\ \newline \small \textbf{Remark:} According to \eqnref{eqn:hat C
grav on states}, the subspace spanned by $\ket{\m_1,\m_2,\m_3}$ with
$\m_1=\m_2=\m_2$ is not invariant under $\hat{C}^{(\muzero)}_{\rm
grav}$. Hence, by restricting the initial states to the ``isotropic
sector'', the total Hamiltonian constraint $\hat{C}^{(\muzero)}_{\rm
grav}+\hat{C}^{(\muzero)}_\ph=0$ does \emph{not} yield the solution
which lays on the isotropic sector for all the time of the
evolution. To systematically study how the isotropy appears in the
anisotropic framework, more conceptual considerations and technical
tools for a better notion of symmetry reduction such as developed in
\cite{Bojowald:1999eh,Engle:2005ca} may be needed for the future
research. \hfill $\Box$\normalsize

\subsection{Improved Strategy ($\mubar$-Scheme)}\label{subsec:second scheme for isotropic}
In this subsection, we follow the ideas used in \secref{subsec:improved grav constraint operator}
and \secref{subsec:improved total hamiltonian} to construct the total Hamiltonian constraint
in $\mubar$-scheme for the isotropic case. In the end, the issue of the different
factor orderings for $\hat{C}^{\rm WDW}_{\rm grav}$ and $\hat{C}^{(\mubar)}_{\rm grav}$
used in this paper and in \cite{Ashtekar:2006wn} are also remarked.

In the isotropic model, the area operator in \eqnref{eqn:area operator} is replaced by
\be
E(S,f)=pV_o^{-\frac{2}{3}}A_{S,f},
\ee
with $A_{S,f}$ the area of $S$ measured by $\fq_{ab}$ times the factor $f$.
Consider $S^{(\mubar)}$ to be a square surface with side length $\mubar V_o^{-1/3}$ (measured by $\fq_{ab}$),
which gives
\be
E(S^{(\mubar)})\ket{\m}=p\,\mubar^2\,\ket{\m}=\frac{4\p\g}{3}\Pl^2\,\mubar^2\m\,\ket{\m}.
\ee
When the $\mubar$-scheme is applied to the isotropic case, again, we require the smallest area to be
$4\p\g\Pl^2\mubar^2\m/3=\pm\D$ and thus we set
\be
\mubar=\sqrt{{3k}/{\abs{\m}}}.
\ee

Following the same procedure for the anisotropic case, we solve the
affine parameter \be
\mubar\frac{\partial}{\partial\m}:=\frac{\partial}{\partial v}
\quad\Rightarrow\quad v=\sgn(\m)\frac{2}{3\sqrt{3k}}\,\abs{\m}^{3/2}
\ee and it leads to \be
e^{\mp\mubar\frac{\partial}{\partial\m}}\ps(\m)=
e^{\mp\frac{\partial}{\partial v}}\ps(v)=\ps(v\mp1) \ee or
equivalently \be e^{\pm\frac{i}{2}\mubar c}\,\ket{\m}=\ket{v\pm1}.
\ee Therefore, \eqnref{eqn:hat C grav isotropic} is altered to
\be\label{eqn:hat C grav isotropic on states} \hat{C}_{\rm
grav}^{(\mubar)}\,\ket{v} =
\frac{\sqrt{\p}\,\Pl}{\sqrt{3}\,\g^\frac{3}{2}\mubar^3} \Big\{
C^+(v)\ket{v+4}+C^-(v)\ket{v-4}-C^o(v)\ket{v} \Big\}, \ee where \ba
C^+(v)&:=&\frac{3\sqrt{3k}}{2}\Abs{\abs{v+3}-\abs{v+1}},\nn\\
C^-(v)&:=&\frac{3\sqrt{3k}}{2}\Abs{\abs{v-1}-\abs{v-3}},\nn\\
C^o(v)&:=&\frac{3\sqrt{3k}}{2}
\left(\Abs{\abs{v+3}-\abs{v+1}\big|+\big|\abs{v-1}-\abs{v-3}}\right).
\ea

As in the $\m_o$-scheme shown in \appref{subsec:first scheme for
isotropic}, the result of \eqnref{eqn:hat C grav isotropic on
states} can be easily obtained from the anisotropic $\hat{C}_{\rm
grav}^{(\mubar)}$ in \eqnref{eqn:improved hat C grav on states}
simply by taking the formal replacements:
\ba
&&\qquad\ket{v_1+\a,v_2+\b,v_3+\g}\longrightarrow\ket{v+(\a+\b+\g)},\nn\\
&&\mubar_{1,2,3}\longrightarrow3^{-{1}/{3}}\,\mubar
\quad\text{and}\quad
V_{v_1+\a,v_2+\b,v_3+\g}\longrightarrow V_{v+(\a+\b+\g)},
\ea
where $V_v$ is the eigenvalues of the volume operator:
\be
V_v=\left(\frac{4\p\g}{3}\right)^{\!\!3/2}\left(\frac{3\sqrt{3k}}{2}\right)\abs{v}\,\Pl^3.
\ee

Similarly, the inverse triad operator in the new scheme is given by
\be
\widehat{\left[{\frac{{\sgn}(p)}{\sqrt{\abs{p}}}}\right]}\ket{v}
=\frac{3}{4\p\g\Pl^2\mubar}(V^{1/3}_{v+1}-V^{1/3}_{v-1})\,\ket{v}.
\ee
Classically, $C_{\ph}=8\p G\,p^2_\ph/\abs{p}^{3/2}$
and this is quantized to the LQC operator $\hat{C}_\ph^{({\mubar})}$, which when
acting on the states of $\hilbert^{\rm total}_{\rm kin}$
yields
\be
\hat{C}_\ph^{({\mubar})}\ket{v}\otimes\ket{\ph}=
\left(\frac{3}{4\p\g}\right)^{\!\!3/2}\frac{1}{\Pl^3}{B}(v)\ket{v}\otimes8\p G\hat{p}_\ph^2\ket{\ph}
\ee
with
\be
{B}(v):=\frac{3\sqrt{3k}}{2}\,\frac{1}{\mubar^3}\,\abs{\abs{v+1}^{1/3}-\abs{v-1}^{1/3}}^3.
\ee
Consequently, the total Hamiltonian constraint leads to
\ba\label{eqn:improved master eq for isotropic case}
\frac{\partial^2}{\partial\ph^2}\ps(v,\ph)
&=&\frac{\p G}{9}[\tilde{B}(v)]^{-1}
\Big\{
C^+(v)\Ps(v+4,\ph)+C^-(v)\Ps(v-4,\ph)-C^o(v)\Ps(v,\ph)
\Big\}\nn\\
&=&\frac{\p G}{9}\,\abs{\abs{v+1}^{1/3}-\abs{v-1}^{1/3}}^{-3}\nn\\
&&\ \times
\Big\{
\Abs{\abs{v+3}-\abs{v+1}}\,\ps(v+4,\ph)+\Abs{\abs{v-1}-\abs{v-3}}\,\ps(v-4,\ph)\nn\\
&&\qquad
-\left(\Abs{\abs{v+3}-\abs{v+1}}+\Abs{\abs{v-1}-\abs{v-3}}\right)\ps(v,\ph)
\Big\}\nn\\
&=:&\Th(v)\Ps(v,\ph),
\ea
where we define
\be
\tilde{B}(v):=\mubar^3B(v)
=\frac{3\sqrt{3k}}{2}\,\abs{\abs{v+1}^{1/3}-\abs{v-1}^{1/3}}^3.
\ee
It can be easily inspected that the operator $\Th$ is self-adjoint
in $L^2({\mathbb R}_{\rm Bohr},\tilde{B}(v)\,dv_{\rm Bohr})$.

\ \newline
\small
\textbf{Remark:} The total Hamiltonian constraint given in
\eqnref{eqn:improved master eq for isotropic case} is almost the same as
the master equation used in \cite{Ashtekar:2006wn}. However, there is a subtle difference.
In \cite{Ashtekar:2006wn}, $\tilde{B}(v)$ is not introduced and the physical Hilbert space is
identified as $L_S^2({\mathbb R}_{\rm Bohr},B(v)\,dv_{\rm Bohr})$.
Here, on the contrary, the physical Hilbert space is
$L_S^2({\mathbb R}_{\rm Bohr},\tilde{B}(v)\,dv_{\rm Bohr})$.

This difference corresponds to different structures in the WDW
theory. In \cite{Ashtekar:2006wn}, the physical Hilbert space in the
corresponding WDW theory is $L_S^2({\mathbb
R},\underline{B}(v)\,dv)$, which is \emph{different} from
$L_S^2({\mathbb R},\underline{B}(\m)\,d\m)$. [Note:
$\underline{B}(v)=\underline{B}(\m):=1/\abs{\m}^{3/2}$ and
$B(v)\rightarrow\underline{B}(v)$ when $\abs{v}\gg1$.] Here, on the
other hand, the physical Hilbert space in the corresponding WDW
theory is taken to be $L_S^2({\mathbb
R},\tilde{\underline{B}}(v)\,dv)=L_S^2({\mathbb
R},\underline{B}(\m)\,d\m)$. [Note: $\underline{B}(v)$ is defined
such that $\underline{B}(v)\,dv=B(\m)\,d\mu$ and we have
$\tilde{B}(v)\rightarrow\tilde{\underline{B}}(v)$ when
$\abs{v}\gg1$.] This difference can be understood as two different
choices of factor ordering both in LQC and WDW theories. Both
choices of factor ordering have their own advantages.

The approach used in this paper by introducing $\tilde{B}(v)$ (and
$\tilde{B}(\vec{v})$ for the anisotropic case) is to insist the same
WDW theory for both $\muzero$-scheme (or say $\m_o$-scheme for the
isotropic case) and $\mubar$-scheme, even though the LQC theory is
modified from $\muzero$-scheme to $\mubar$-scheme. The advantage of
this choice is two-fold. Firstly, the expression of the constraint
equation is simpler. Secondly, since the WDW theory is unchanged,
all we have to do in $\mubar$-scheme is changing variables from
$\m_I$ to $v_I$. Therefore, we can quite straightforwardly
``translate'' the results we obtained in \secref{sec:WDW theory} and
\secref{sec:analytical issues} to those in $\mubar$-scheme as what
we did in \secref{subsec:improved WDW theory} and
\secref{subsec:improved general solutions}. However, the drawback of
this factor ordering is that the gravitational part of the
Hamiltonian constraint $\hat{C}_{\rm grav}^{(\mubar)}$ is not
self-adjoint in $\hilbert_{\rm grav}^S$.\footnote{However, this is
not an obstacle to constructing the physical sector with the desired
inner product (such that the Dirac operators are self-adjoint),
since the operator $\Th$ is still self-adjoint in $L^2({\mathbb
R}_{\rm Bohr},\tilde{B}(v)\,dv_{\rm Bohr})$ and thus the procedures
use for $\m_o$-scheme can be easily applied.}

On the contrary, the choice made in \cite{Ashtekar:2006wn} insists
$\hat{C}_{\rm grav}^{(\mubar)}$ to be self-adjoint in $\hilbert_{\rm
grav}^S$ and yields the ``natural'' factor ordering of the WDW
theory in accordance with the use of Laplace-Beltrami operator in
geometrodynamics (see \cite{Ashtekar:2006wn} for details). \hfill
$\Box$\normalsize


\end{document}